\documentclass[12pt,preprint]{aastex}
\usepackage{psfig,epsfig,amsfonts,amsmath,amssymb,graphicx,morefloats,appendix,tensor}
\usepackage{color}
\usepackage{natbib}
\usepackage{hyperref}
\usepackage{subfigure}
\renewcommand{\dataset}[1]{doi:{#1}, \url{https://doi.org/#1}}

\begin{document}

\slugcomment{Accepted: Sept. 21, 2023}

\title{Characterizing the Near-infrared Spectra of Flares from TRAPPIST-1 During JWST Transit Spectroscopy Observations}

\author{Ward S. Howard\altaffilmark{1,2}, Adam F. Kowalski\altaffilmark{1,3,4}, Laura Flagg\altaffilmark{5}, Meredith A. MacGregor\altaffilmark{6}, Olivia Lim\altaffilmark{7}, Michael Radica\altaffilmark{7}, Caroline Piaulet\altaffilmark{7}, Pierre-Alexis Roy\altaffilmark{7}, David Lafreni{\`e}re\altaffilmark{7}, Bj{\"o}rn Benneke\altaffilmark{7}, Alexander Brown\altaffilmark{8}, N{\'e}stor Espinoza\altaffilmark{6,9}, Ren{\'e} Doyon\altaffilmark{7, 10}, Louis-Philippe Coulombe\altaffilmark{7}, Doug Johnstone\altaffilmark{11, 12}, Nicolas B. Cowan\altaffilmark{13,14}, Ray Jayawardhana \altaffilmark{5, 6}, Jake D. Turner\altaffilmark{2,5}, Lisa Dang\altaffilmark{7}}

\altaffiltext{1}{Department of Astrophysical and Planetary Sciences, University of Colorado, 2000 Colorado Avenue, Boulder, CO 80309, USA}
\altaffiltext{2}{NASA Hubble Fellowship Program Sagan Fellow}

\altaffiltext{3}{National Solar Observatory, University of Colorado Boulder, 3665 Discovery Drive, Boulder, CO 80303, USA}
\altaffiltext{4}{Laboratory for Atmospheric and Space Physics, University of Colorado Boulder, 3665 Discovery Drive, Boulder, CO 80303, USA.}

\altaffiltext{5}{Department of Astronomy and Carl Sagan Institute, Cornell University, 122 Sciences Drive, Ithaca, NY 14853, USA}
\altaffiltext{6}{Department of Physics and Astronomy, Johns Hopkins University, 3400 N Charles St, Baltimore, MD 21218, USA}
\altaffiltext{7}{Institut Trottier de recherche sur les exoplan{\`e}tes and D{\'e}partement de Physique, Universit{\'e} de Montr{\'e}al, 1375 Avenue Th{\'e}r{\`e}se-Lavoie-Roux, Montr{\'e}al, QC, H2V 0B3, Canada}
\altaffiltext{8}{Center for Astrophysics and Space Astronomy, University of Colorado, 385 UCB, Boulder, CO 80309}
\altaffiltext{9}{Space Telescope Science Institute, 3700 San Martin Drive, Baltimore, MD 21218, USA}
\altaffiltext{10}{Observatoire du Mont-M{\'e}gantic, Universit{\'e} de Montr{\'e}al, Montr{\'e}al, QC, H3C 3J7, Canada}
\altaffiltext{11}{NRC Herzberg Astronomy and Astrophysics, 5071 West Saanich Rd, Victoria, BC, V9E 2E7, Canada}
\altaffiltext{12}{Department of Physics and Astronomy, University of Victoria, Victoria, BC, V8P 5C2, Canada}
\altaffiltext{13}{Department of Physics, McGill University, Montr\'{e}al, QC, Canada}
\altaffiltext{14}{Department of Earth and Planetary Sciences, McGill University, Montr\'{e}al, QC, Canada}

\begin{abstract}
We present the first analysis of JWST near-infrared spectroscopy of stellar flares from TRAPPIST-1 during transits of rocky exoplanets. Four flares were observed from 0.6--2.8~$\mu$m with NIRISS and 0.6--3.5~$\mu$m with NIRSpec during transits of TRAPPIST-1b, f, and g. We discover P$\alpha$ and Br$\beta$ line emission and characterize flare continuum at wavelengths from 1--3.5$\mu$m for the first time. Observed lines include H$\alpha$, P$\alpha$-P$\epsilon$, Br$\beta$, He I $\lambda$0.7062$\mu$m, two Ca II infrared triplet (IRT) lines, and the He I IRT. We observe a reversed Paschen decrement from P$\alpha$-P$\gamma$ alongside changes in the light curve shapes of these lines. The continuum of all four flares is well-described by blackbody emission with an effective temperature below 5300 K, lower than temperatures typically observed at optical wavelengths. The 0.6--1~$\mu$m spectra were convolved with the TESS response, enabling us to measure the flare rate of TRAPPIST-1 in the TESS bandpass. We find flares of 10$^{30}$ erg large enough to impact transit spectra occur at a rate of 3.6$\substack{+2.1 \\ -1.3}$ flare d$^{-1}$, $\sim$10$\times$ higher than previous predictions from \textit{K2}. We measure the amount of flare contamination at 2~$\mu$m for the TRAPPIST-1b and f transits to be 500$\pm$450 and 2100$\pm$400 ppm, respectively. We find up to 80\% of flare contamination can be removed, with mitigation most effective from 1.0--2.4~$\mu$m. These results suggest transits affected by flares may still be useful for atmospheric characterization efforts.
\end{abstract}


\section{Introduction}\label{sec:intro}
The majority of rocky planets suitable for atmospheric characterization with JWST orbit M-dwarfs \citep{Kempton:2018}. M-dwarfs make up 75\% of stars in the solar neighborhood \citep{Henry:2006} and have high occurrence rates of rocky planets \citep{Dressing:2015, Hardegree:2019}. Many terrestrial planets orbit within the so-called ``habitable zone" where water oceans can exist in the presence of an atmosphere \citep{Kopparapu:2013}, while others orbit interior to the HZ as potential Venus-analogs \citep{Kane:2014}. Many of these rocky planets belong to multi-planet systems \citep{Sandford:2019, Zhu:2020}, providing an ideal opportunity to investigate the effect of decreasing stellar radiation on the atmospheric compositions of these worlds \citep{Barclay:2023}. A number of rocky planets in nearby multi-planet systems are high-priority targets for JWST, including three transiting planets around L 98-59 \citep{Kostov:2019}, two transiting planets around LTT 1445A \citep{Winters:2022}, and seven transiting planets in the TRAPPIST-1 system \citep{Gillon:2017}.

High levels of stellar activity are observed from M-dwarfs in the form of starspots, faculae, plages, and flares (e.g. \citealt{Delfosse:1998, Berdyugina:2005, Astudillo-Defru:2017, Medina:2020}), complicating the characterization of exoplanet atmospheres \citep{Rackham:2023}. Transmission spectroscopy assumes that an out-of-transit, disk-integrated spectrum of the host star can be subtracted to isolate the spectrum of the planet \citep{McCullough:2014, Rackham:2017, Rackham:2019}. Stellar activity that varies in time and both occulted and unocculted surface inhomogeneities therefore contaminate the transmission spectrum, altering the transit depth as a function of wavelength \citep{Ballerini:2012}. Contamination from flares in the near-infrared (NIR) has received less attention than slowly-varying noise sources such as spots (\citealt{Rackham:2023} and references therein). While flares peak at near-ultraviolet (NUV) and blue optical wavelengths \citep{Kowalski:2013}, they emit across the entire electromagnetic spectrum \citep{MacGregor:2021} and must be considered as a key source of unpredictable stellar contamination.

Flares occur when magnetic reconnection events in the corona accelerate electrons along field lines toward the photosphere, heating the plasma. Particles of different energies brake at different depths in the chromosphere to the upper photosphere, producing emission at different wavelengths from the FUV to the NIR \citep{Fuhrmeister:2008, Kowalski:2013, Klein_Dalla:2017}. Flare spectra are well-described by the superposition of line emission onto a blackbody-like spectrum that evolves in time. Flare light curves generally occur in two phases, with a rapid period of prompt emission characterized by an effective temperature of 9000--16,000 K, followed by a more gradual phase with a lower effective temperature. A blackbody spectrum with an effective temperature of 9000 K would place the NIR in the Rayleigh-Jeans tail \citep{Fuhrmeister:2008} and produce a flux enhancement of 700 ppm at 2~$\mu$m for even a small flare of 10$^{30}$ erg. On the Sun, such flares occur on a monthly basis near solar maximum \citep{Youngblood:2017}. Multiple flares of 10$^{30}$ erg are emitted from active M-dwarfs on a daily basis \citep{Lacy:1976}. Although the NIR enhancement is low compared to the total energy of the flare, such signals are substantial compared to the $\sim$20--200 ppm signatures of atmospheric features in transmission spectra of rocky exoplanets (e.g. \citealt{Moran:2023, Lustig-Yaeger:2023, Barclay:2023}) and far larger than pre-launch estimates of the JWST noise floor near 10 ppm \citep{Schlawin:2020, Schlawin:2021, Feinstein:2022b}. Furthermore, NIR spectra provide some of the strongest available constraints on cooler sources of red optical emission during the long decay phase of flares \citep{Kowalski:2016}.

Continuum emission is the dominant source of flare contamination at wavelengths from 1--3.5~$\mu$m, assuming a 9000 K blackbody \citep{Fuhrmeister:2008}. Nevertheless, flare continuum has not yet been observed at all beyond 1.7~$\mu$m, and has not been characterized in spectra at wavelengths longer than 1~$\mu$m. Near-infrared spectroscopy from 1.1--1.7$\mu$m was obtained for two small flares from TRAPPIST-1 with the HST Wide Field Camera 3 (WFC3) by \citet{Zhang:2018}. However, \citet{Zhang:2018} note the spectrum of each flare cannot be meaningfully constrained, as the variation in the spectrum is mostly at the 1-2$\sigma$ level. Characterization of the continuum does exist at NIR wavelengths below 1~$\mu$m for several flares. Continuum was detected in a large flare from Wolf 359 out to 1.01~$\mu$m with the red arm of the Ultraviolet-Visual Echelle Spectrograph (UVES) on the ESO Kueyen telescope \citep{Fuhrmeister:2008}. Flare continuum has also been detected out to 0.76~$\mu$m in an event from the M9.5 dwarf 2MASS J0149090+295613 with the Low Resolution Imaging Spectrograph on Keck II \citep{Liebert:1999} and out to 0.94~$\mu$m in a large flare from the M7 dwarf 2MASS J1028404-143843 \citep{Schmidt:2007}. The lack of spectral coverage or sensitivity at longer wavelengths during each of these flares makes it difficult to characterize contamination at the wavelengths used for JWST transit spectroscopy.

The lower order hydrogen I Paschen and Brackett lines, 1.0833~$\mu$m helium I infrared triplet (IRT), and several metal lines such as the Ca II IRT (0.85 to 0.87~$\mu$m) each contribute to flare contamination at NIR wavelengths. The cumulative effect of the forest of weak lines near the Paschen jump (0.82~$\mu$m) and the Brackett jump (1.46~$\mu$m) may also induce contamination in transit spectra. While P$\alpha$ has not been detected yet during an M-dwarf flare, the higher-order Paschen lines have been detected in several flares. High-resolution flare spectra obtained in the NIR with CARMENES contain 57 and 30 detections of P$\beta$ and P$\gamma$, respectively \citep{Fuhrmeister:2023}. However, spectra of each star are typically obtained several days apart, making it difficult to know if the flares are observed during the peak or decay phase. P$\beta$ has been detected with spectral time series in three flares \citep{Schmidt:2012} using the 0.95--2.45~$\mu$m TripleSpec near-infrared spectrograph on the 3.5 m ARC telescope at APO, and P$\gamma$ and the He I IRT have been detected in five flares with TripleSpec and the 0.810--1.280~$\mu$m Habitable-zone Planet Finder high-resolution spectrograph on the 10 m Hobby-Eberly Telescope \citep{Schmidt:2012, Kanodia:2022}. Evidence for He I IRT emission in flares was also obtained in CARMENES spectra \citep{Fuhrmeister:2020}. A wealth of detections for P$\delta$ (1.005~$\mu$m) and higher order Paschen lines during flares exists in the literature below 1.1~$\mu$m (e.g. \citealt{Liebert:1999, Fuhrmeister:2008, Schmidt:2012, Kanodia:2022}). A single detection of Brackett emission (Br$\gamma$) was reported by \citet{Schmidt:2012}. However, the lack of observations of the lowest energy transitions limits our understanding of hydrogen line emission in the NIR more broadly since these lines are likely to dominate the energetics of their respective series.

TRAPPIST-1 is an M8 dwarf that hosts a system of seven terrestrial planets, three of which orbit in the habitable zone \citep{Gillon:2017}. At a distance of only 12.43$\pm$0.02 pc, TRAPPIST-1 is the subject of numerous JWST programs to obtain spectroscopic time series for its rocky planets (e.g. GTO 1177; PI: Greene, GTO 1331; PI: Lewis, GO 1981; PI: Stevenson, GO 2304; PI: Kreidberg, GO 2420; PI: Rathcke). In particular, the NIRISS Exploration of the Atmospheric diversity of Transiting exoplanets (NEAT) collaboration is observing TRAPPIST-1b, c, and g as part of JWST Cycle 1 GO 2589 (PI: Lim) and TRAPPIST-1d and f as part of Cycle 1 GTO 1201 (PI: Lafreni{\`e}re). TRAPPIST-1 is fairly active given its apparent 3.3 d rotation period \citep{Luger:2017, Roettenbacher_Kane:2017} and 7.6$\pm$2.2 Gyr age \citep{Burgasser_Mamajek:2017}, and exhibits spot modulation, flares, and a L$_\mathrm{H\alpha}$/L$_\mathrm{bol}$ activity index typical of a star undergoing spin-down \citep{Roettenbacher_Kane:2017, Burgasser_Mamajek:2017}.

Flares observed by \textit{Spitzer} to partially overlap transits of TRAPPIST-1 planets in \citet{Gillon:2017} have raised concerns about flare contamination \citep{Davenport:2017}. However, the exact flare rate of the star is uncertain \citep{Wilson:2021}. K2 observed TRAPPIST-1 for 70.6 days and observed 39 flares with energies from 2.0$\times$10$^{29}$ to 2.3$\times$10$^{32}$ erg \citep{Vida:2017, Paudel:2018}. The flare frequency distribution (FFD) derived from the K2 data predicts a flare of 10$^{30}$ erg occurs at a rate of 0.3 flares d$^{-1}$ when considering only flares above the 100\% completeness limit as determined by injection testing \citep{Paudel:2018}. The average flare rate of TRAPPIST-1 analog stars was measured in TESS data and corrected for missing flares using an incompleteness curve and uncertainties measured from injection testing \citep{Seli:2021}. The resulting average flare rate of the TRAPPIST-1 analogs corrected for incompleteness predicts $\sim$1--5 flares d$^{-1}$ of 10$^{30}$ erg. The NEAT programs will obtain a total of 75 hr of JWST observations across the entire sample, including 14.3 hr in transit. Many of these observations should occur during flares, with 1--15 flares of 10$^{30}$ erg expected based on the flare rate. These flares provide a serendipitous opportunity to characterize both the continuum and line emission of flares from 0.6--3.5~$\mu$m at unprecedented precision. In turn, the flare observations can enable empirical insights to be developed for the mitigation of flare contamination, increasing both the yield of usable transits and sensitivity to spectral features from the planet atmospheres in spectra impacted by flares.

In Section \ref{neat_sample_obs}, we describe the JWST transit observations. In Section \ref{spec_reductions}, we describe the spectral reduction methods for each instrument and the extraction of calibrated spectra and light curves. In Section \ref{flare_line_methods}, we describe the methods used for the identification and characterization of flare lines. In Section \ref{flare_continuum_methods}, we describe the methods used to measure the continuum properties. In Section \ref{flare_params}, we describe how the amplitude, energy, and shape of the flare light curves are quantified. In Section \ref{results}, we present the detection of flare lines and continuum emission and discuss the properties of the flares both individually and as a sample. In Section \ref{x_rel_results}, we model the luminosity emitted by the flare lines using radiative hydrodynamic flare models. In Section \ref{planet_results}, we mitigate flare contamination in the TRAPPIST-1b and f transits. In Section \ref{discuss_conclude}, we discuss the implications of our findings, and conclude.

\begin{figure*}
	\centering
	{
		\includegraphics[width=0.98\textwidth]{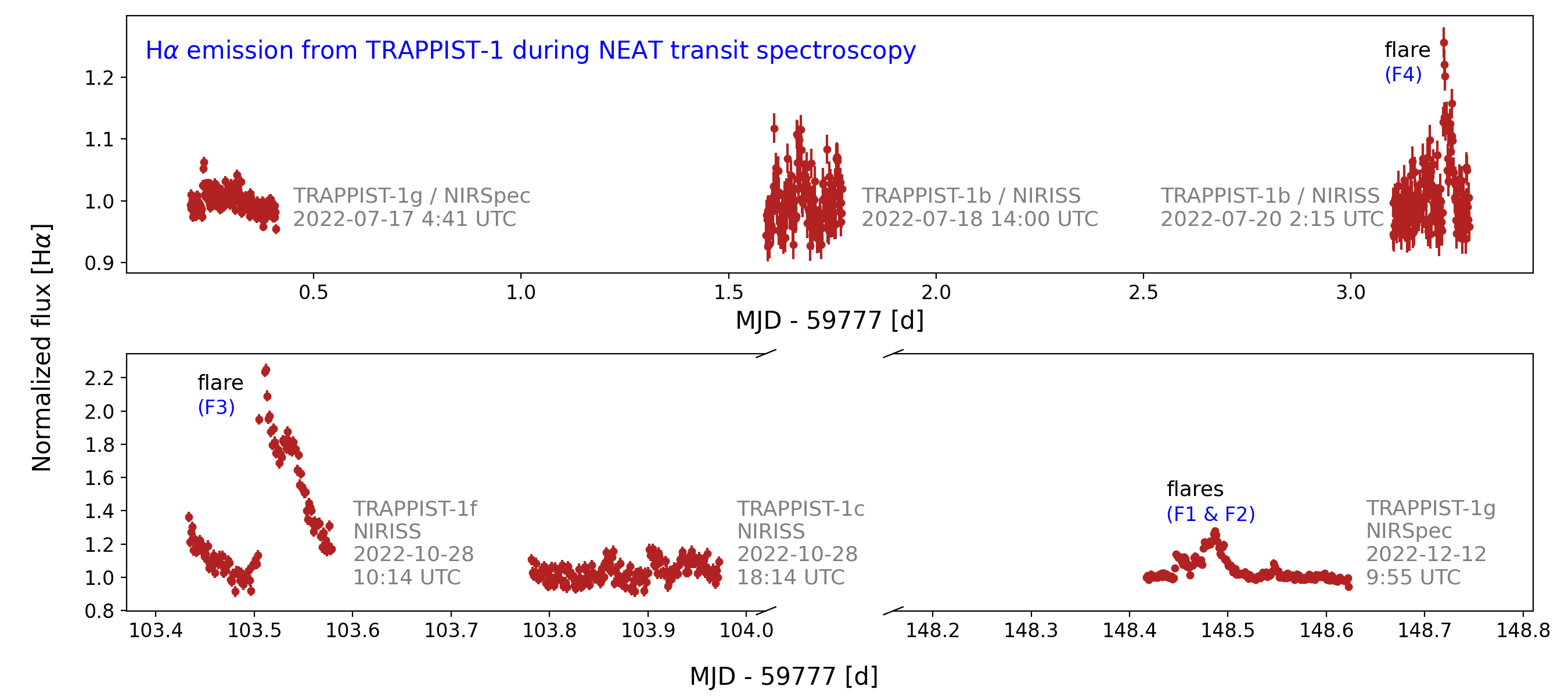}
	}
	\caption{Overview of stellar activity in the H$\alpha$ line during transit spectroscopy observations of TRAPPIST-1 used in this work. During the JWST observations, the H$\alpha$ line shows considerable variability, including flares as described in \S \ref{create_lcvs} and \S \ref{identify_lines}. Flares with high enough signal for multi-wavelength characterization (approximately $E_\mathrm{H\alpha}$=10$^{29}$ erg in the H$\alpha$ line) are numbered by TESS band energy. Energies in these bands are given in Tables \ref{tab:line_table} and \ref{tab:continuum_table}. The x-axis time scales of the top and bottom panels are fixed within each panel to better observe changes in activity between observations. The final transit of TRAPPIST-1g is 44.4 d after the previous observation, so a break in the x-axis (but still at the same time scale) is used to include this data in the figure. NIRSpec observations are binned to the same cadence as NIRISS and have a lower photometric uncertainty.}
	\label{fig:activity_overview}
\end{figure*}

\section{Observations of TRAPPIST-1 with JWST}\label{neat_sample_obs}
We describe the transit spectroscopy observations from the NEAT program used in this work.

\subsection{The NEAT program and Target Selection}\label{neat_targets}
The NEAT program is designed to measure molecular abundances in planetary atmospheres as functions of the planet mass and radiation environment in order to explore the processes sculpting the formation of close-in planets. Most targets are observed with the Single Object Slitless Spectroscopy (SOSS; \citealt{Albert:2023}) mode of the Near Infrared Imager and Slitless Spectrograph (NIRISS; \citealt{Doyon:2023}). The remaining targets are observed with the Near Infrared Spectrograph (NIRSpec) using the Bright Object Time Series (BOTS) mode. The sample of 14 systems is primarily defined by the JWST Cycle 1 and 2 Guaranteed Time Observation Programs GTO 1201 and GTO 2759 (PI: Lafreni{\`e}re). These two programs include a number of rocky or likely-rocky planets orbiting M-dwarf stars, including TRAPPIST-1 \citep{Gillon:2017}, GJ 357 \citep{Luque:2019}, L 98-59 \citep{Kostov:2019, Pidhorodetska:2021}, LP 791-18 \citep{Crossfield:2019}, and L 231-32 \citep{Gunther:2019, VanEylen:2021}. These rocky planets will be used to constrain the mean molecular weight and bulk composition of terrestrial atmospheres. Several other programs with principle investigators within the NEAT collaboration are also contributing observations. These include the Cycle 1 GO program 2589 (PI: Lim) to observe TRAPPIST-1b and c with NIRISS and TRAPPIST-1g with NIRSpec and the Cycle 2 GO program 4098 (PI: Benneke) to observe planets around L 98-59, GJ 9827 \citep{Niraula:2017}, TOI-1685 \citep{Bluhm:2021, Hirano:2021},  GJ 3090 \citep{Almenara:2022}, and L 231-32 \citep{Gunther:2019, VanEylen:2021} with NIRISS and NIRSpec.

\subsection{NIRISS and NIRSpec Observations of TRAPPIST-1}\label{instrument_and_obs}
In this work, we focus on a subset of the TRAPPIST-1 observations obtained with NIRISS SOSS and NIRSpec BOTS as part of GTO 1201 and GO 2589. Our sample consists of NIRISS observations of two transits of TRAPPIST-1b, one transit of TRAPPIST-1c, and one transit of TRAPPIST-1f, and NIRSpec observations of two transits of TRAPPIST-1g. The TRAPPIST-1 observations are ongoing, with visits of TRAPPIST-1d and h not yet available at the time of this work. A total of 16.87 hr of monitoring time on target was obtained with NIRISS across all observations, consisting of 586 integrations taken at 105 s cadence (18 groups per integration). Similarly, a total of 9.91 hr of monitoring time on target was obtained with NIRSpec, consisting of 22,234 integrations taken at 1.6 s cadence with NIRSpec. Analysis of the NIRISS transit spectrum of TRAPPIST-1b is presented in \citet{Lim:2023}. Analysis of the NIRSpec transit spectrum of TRAPPIST-1g is presented in Benneke et al. (2023, submitted).

During the NIRISS observations, two large enhancements in the H$\alpha$ flux from the star indicative of flaring are observed during the second transit of TRAPPIST-1b and the transit of TRAPPIST-1f. A period of H$\alpha$ enhancement composed of two peaks is likewise detected just prior to the second transit of TRAPPIST-1g. We call the enhancements during the TRAPPIST-1g observations Flare 1 (F1) and Flare 2 (F2) throughout this work. The enhancement during the TRAPPIST-1f visit is called Flare 3 (F3) and the enhancement during the second visit of TRAPPIST-1b is called Flare 4 (F4). The event numbering is by decreasing flare energy in the TESS bandpass as described in \S \ref{basic_fl_properties} and given in Table \ref{tab:continuum_table}. A summary of these observations is presented in Table \ref{table:observations} and an overview of the H$\alpha$ time series are shown in Figure \ref{fig:activity_overview}.

NIRISS SOSS is designed to perform transit spectroscopy of exoplanets orbiting stars of 7$<$J$<$15 \citep{Doyon:2023}. The instrument acquires spectral time series observations at NIR wavelengths from 0.6-2.8~$\mu$m at medium resolution (R$\approx$700). NIRISS is a cross-dispersed spectrograph with three orders, although sufficient signal to noise (S/N) for science observations of most targets is only obtained for orders 1 and 2. The spectra in each order are defocused in the cross-dispersion direction to minimize pointing and flat-field systematics, leading to blending of the orders at several points along the trace \footnote{JWST User Documentation; jwst-docs.stsci.edu/jwst-near-infrared-imager-and-slitless-spectrograph}. 

Our NIRSpec BOTS observations are taken without a filter in the PRISM/CLEAR mode\footnote{JWST User Documentation; jwst-docs.stsci.edu/jwst-near-infrared-spectrograph} \citep{Boker:2023}. In our setup, NIRSpec BOTS obtains spectral time series of bright targets at NIR wavelengths from 0.6-5.3~$\mu$m at low resolution (R$\approx$100) and high throughput \citep{Birkmann:2022}. The BOTS mode of NIRSpec passes the spectra through the 1.6 in$^2$ S1600A1 aperture for maximum spectroscopic precision for planets around bright stars. The target is kept centered on the same pixels on the detector for increased stability.

\begin{table}
\centering
\caption{Overview of TRAPPIST-1 Observations}
\begin{tabular}{cccccccccc}
\hline
\hline
Program & Obs. & Planet & Visit & Date & $t_\mathrm{start}$ & $t_\mathrm{end}$ & $\Delta t_\mathrm{target}$ & $N_\mathrm{integrations}$ & Flares? \\
        &      &        &       & ISO  & UTC                & UTC              & hr                         &         &  \\
\hline
2589 &   1 & b & 1 & 2022-07-18 & 14:00:06 & 19:04:38 & 4.406 & 153 & no \\
2589 &   2 & b & 2 & 2022-07-20 & 02:15:42 & 07:20:14 & 4.406 & 153 & yes (F4) \\
2589 &   3 & c & 1 & 2022-10-28 & 18:14:52 & 23:29:50 & 4.58 & 159 & no \\
1201 & 101 & f & 1 & 2022-10-28 & 10:14:49 & 14:04:50 & 3.478 & 121 & yes (F3) \\
2589 & 6 & g$^\dag$ & 1 & 2022-07-17 & 04:41:12 & 09:58:17 & 4.953 & 11117 & no \\
2589 & 7 & g$^\dag$ & 2 & 2022-12-12 & 09:55:32 & 15:09:00 & 4.952 & 11117 & yes (F1\&2) \\
\hline
\hline
\end{tabular}
\label{table:observations}
{\newline\newline \textbf{Notes.} Description of the observations used in this work. Columns are the JWST Cycle 1 program ID, the archived observation number given in the JWST Visit Status Report, the TRAPPIST-1 planet observed, the visit sequence number, date, start and stop times listed for archived observations in the Visit Status Report, total time spent on target and usable for science observations, number of integrations in the time series, and whether any flaring was observed during the visit. A dagger denotes the instrument and mode as NIRSpec BOTS prism observations, while no dagger denotes the instrument and mode as NIRISS SOSS. The time on target is approximately 70\% of the total time listed in the Visit Status Report.}
\vspace{0.05cm}
\end{table}

\section{Extraction of Spectra and Light curves}\label{spec_reductions}
\subsection{NIRISS data reduction with supreme-SPOON}\label{supreme_SPOON}
We use the supreme-Steps to Process sOss ObservatioNs (\texttt{supreme-SPOON}) pipeline for our primary spectral reduction. \texttt{supreme-SPOON} begins with the raw, uncalibrated NIRISS SOSS FITS files and extracts one-dimensional spectra with differential calibration, correcting systematics on a group-by-group basis within each integration \citep{Feinstein:2023, Radica:2023, Coulombe:2023arXiv}. Correction of detector effects is largely consistent between \texttt{supreme-SPOON} and stage 1 of the \texttt{jwst} pipeline. Superbias correction is applied to each raw image within each group and integration, and saturation flags are applied to pixels with values below the noise floor or above the saturation limit. Following \citet{Feinstein:2023} and \citet{Radica:2023}, reference pixel correction is performed to further mitigate bias variations between groups and adjacent rows within individual frames. Jump detection is carried out after the other steps to avoid inducing false positives. Finally, the ramps are fit to produce \texttt{rateints} files. We do not correct for dark current due to the negligible dark current rate of the NIRISS observations \citet{Feinstein:2023}.

The Space Telescope Science Institute (STScI) JDox SOSS SUBSTRIP256 background model is used to correct for the zodiacal background \citep{Rigby:2023}. A region in the images without significant contamination from any of the orders is identified, and a scaling factor is defined from the median flux of the image region and the median flux of the comparison region of the background model and applied to the background model. The scaled background is then subtracted from each image. \texttt{supreme-SPOON} corrects for 1/f noise separately within each group prior to combining the data from all the groups in a given integration. The 1/f noise is a form of readout noise that changes with time due to the voltage amplification process. The presence of 1/f noise is mitigated using a difference image composed of the target frame and a median stack of all out-of-transit images in the same group and integration scaled to the flux in the target frame using the transit light curve. The median of each column in the difference image is then subtracted from the original column in the corresponding integration within the group. The y-positions of the trace are identified using the \texttt{edgetrigger} algorithm, which takes the derivative of the y-axis values and identifies the midpoint between the maximum and minimum values \citep{Radica:2022}. Lastly, the spectra are extracted as a differential flux-calibrated product for science analysis using a simple box aperture of 25 pixels since the transmission self-contamination level for TRAPPIST-1f is expected to be $<$20~ppm (with a similar value for TRAPPIST-1b), much smaller than the final uncertainties in our analysis \citep{Darveau-Bernier:2022}.

Photons of the same wavelength do not follow entirely vertical contours on the NIRISS detector. Rather, the wavelength maps follow contours that smoothly vary across multiple columns in patterns approximating cubic functions\footnote{github.com/spacetelescope/jdat\_notebooks/tree/main/notebooks/soss-transit-spectroscopy}. The curve of the trace intersects different parts of the wavelength maps at different y-pixel positions, inducing systematic offsets in the wavelength solution. We therefore cross-correlate the spectrum of TRAPPIST-1 with a PHOENIX model of the star from \citet{Wilson:2021} to determine these systematic offsets and correct the wavelengths.

\subsection{NIRISS data reduction with \texttt{transitspectroscopy}}\label{transitspectroscopy}
We use the \texttt{transitspectroscopy} pipeline \citep{Espinoza:2022} as a secondary spectral reduction to validate our primary reduction. \texttt{transitspectroscopy} (\texttt{ts}) is a spectral extraction and transit fitting package for NIRISS SOSS that starts from the \texttt{rateints} files produced by stage 1 of the \texttt{jwst} pipeline instead of the raw data \citep{Coulombe:2023arXiv}. First, \texttt{ts.trace\_spectrum} locates the trace for each order through cross-correlation of the image along the dispersion axis against a two-component Gaussian function with parameters $\mu_{G,1}$=-4.5, $\sigma_{G,1}$=3, $\mu_{G,2}$=4.5, $\sigma_{G,2}$=3. While all pixels in the columns are used when locating the order one trace, a y-tolerance of 20 pixels is used for order two. The order one trace is computed for columns $x\in$[4, 2043], and the order two trace is computed for $x\in$[700, 1625] to avoid order blending. Flux from the first order within the pixel radius is masked out before computing the second order trace. Each trace is then smoothed with cubic splines. We follow \citet{Feinstein:2023} in reporting the number of spline knots in order one $n_1$ and two $n_2$ and the partitions of the x-axis $x_1$ and $x_2$ of the spline fits: $n_1$=[4,2,3,4], $n_2$=[2, 2, 5], $x_1$=[[6, 1200-5], [1200, 1500-5], [1500, 1700-5], [1700, 2041]], and $x_2$=[[701, 850-5], [850, 1100-5], [1100,1624]].

A model provided by STScI in the JDox User Documentation is used to subtract the zodiacal background. A region in the images without significant contamination from any of the orders is identified and used to construct a scaling between the background model and the data. For the TRAPPIST-1b observations, this region is $x\in$[500, 800], $y\in$[190, 246]. For the TRAPPIST-1f observations a larger background region is available, $x\in$[10, 750], $y\in$[190, 246]. The ratio between the actual background and the model is measured for each pixel in the comparison region to generate a distribution of ratios. The scale factor for the TRAPPIST-1b observations is $f_S$=0.68 and is $f_S$=0.0064 for the TRAPPIST-1f observations with the larger background region. The bottom and top quartiles are clipped and the median of the remaining values is adopted as the scale parameter. For the TRAPPIST-1b observations, quartile three (0.5-0.75) is also clipped since most flux is below this range.

The pipeline computes a median stack of all background-subtracted images that were not obtained during the transit. The median reference frame is scaled by the normalized transit light curve for images to correct for the reduced flux. \texttt{transitspectroscopy} corrects 1/f noise after the groups in each integration have been combined since it does not work with the raw uncalibrated files. The dominant noise source left after background removal is 1/f banding. We select two regions close to the trace consisting of 15 pixels in the y-axis direction starting 20 pixels away from the top and bottom of the trace. The median of this region is computed on a per-column basis separately for the first and second orders and subtracted from the image. Cross-order contamination from this process has been demonstrated to not be significant \citep{Feinstein:2023}.

Spectra are extracted using \texttt{ts.spectroscopy.getSimpleSpectrum} for an aperture of 15 pixels for the first order and 30 pixels for the second order. Each spectrum is median-normalized, enabling a direct comparison of the variability observed for the flux in each column. The median and standard deviation of the resulting distribution of the normalized spectra are used to define a reference spectrum. The reference spectrum is needed to identify bad columns, such as those resulting from cosmic ray hits. The reference spectrum is scaled to the individual integrations and subtracted to compute residuals. Any 5$\sigma$ outlier is corrected based on the values adjacent to it.

We find the wavelength solution provided with \texttt{transitspectroscopy} is good to within a few nm. However, sub-pixel accuracy is desired for identifying emission lines, so we construct a correction factor for the wavelength solution. We perform a cross-correlation between the spectrum and the \citet{Wilson:2021} PHOENIX model of TRAPPIST-1. We divide the spectrum into sections of $\sim$0.2~$\mu$m and compute the wavelength offset in each section, observing they follow the trace. We fit a function that varies with wavelength in the same way the trace does, but with the y-axis scaled to the wavelength offsets instead of the detector position. We then verify the new solution reproduces the positions of well-known emission lines in the spectrum.

\begin{figure*}
	\centering
	{
		\includegraphics[width=0.98\textwidth]{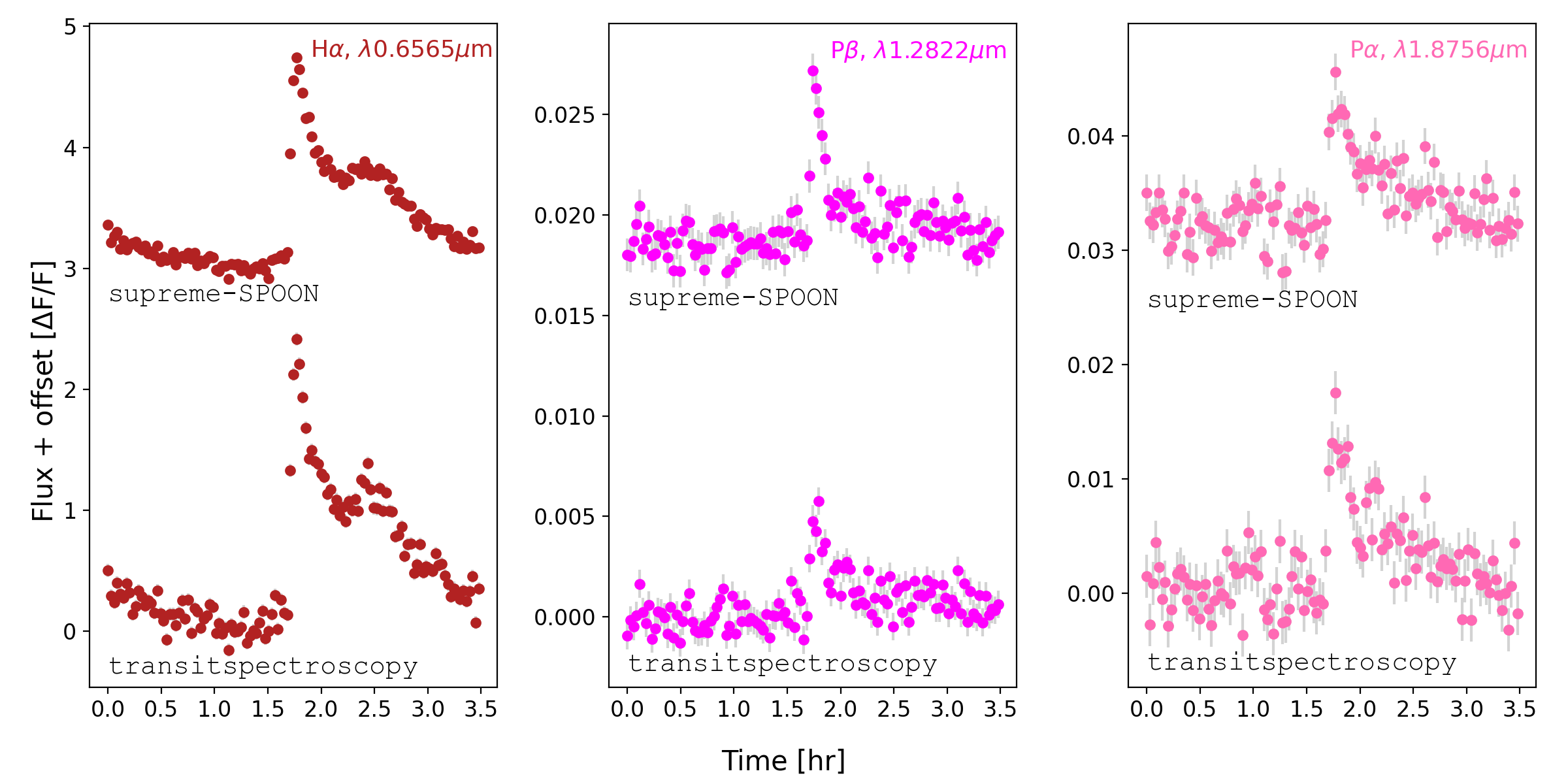}
	}
	\caption{Comparison of flare photometry from our implementation of the \texttt{supreme-SPOON} and \texttt{transitspectroscopy} pipelines for lines from the F3 flare at representative wavelengths. Both reductions produce qualitatively similar flare light curves. However, we find the \texttt{supreme-SPOON} reduction is more optimized for flare photometry as the point-to-point variability during the peak phase is reduced by $>$20\%. This effect is particularly noticeable in the second peak in the H$\alpha$ light curves and the P$\beta$ peak. }
	\label{fig:pipe_compare}
\end{figure*}

\subsection{NIRSpec BOTS data reduction}\label{custom_eureka}
NIRSpec prism data are reduced with a customized version of Stage 3 of the open-source \texttt{Eureka!} pipeline for the extraction and analysis of time series JWST observations (\citet{Bell:2022}, Benneke et al. 2023, submitted). The reduction starts from the ``uncal'' outputs from Stage 0 of the JWST pipeline (version 1.6.0). We perform data calibration in Stages 1 and 2 following the steps described for the analysis of the WASP-39 b NIRSpec PRISM observations with \textit{Eureka!} presented in \citet{Rustamkulov:2023}, except for how pixel saturation flags are set. We perform group scaling, and linearity correction using the standard JWST pipeline steps. Our customized routine first updates the saturation flags: rather than flagging a user-provided range of columns as saturated, we use the ``percentile'' method with a 50th percentile threshold: for each group, we flag as saturated entire columns which contain at least one pixel marked as saturated in the median frame of all integrations at this group. We then perform row odd-even by amplifier (ROEBA) correction of 1/f noise. Within each group, the median of the background pixels in the odd rows is subtracted from all odd rows, and the median of the even rows is subtracted from all even rows. Group-level background subtraction follows: for each column in each group, the median of the background pixels is subtracted from the entire column. Finally, ramps are fitted while ignoring 3$\sigma$ outliers. The flux calibration step is skipped since we flux calibrate directly from the PHOENIX model of TRAPPIST-1 as described below in \S \ref{flux_cal}. Stage 2 of the pipeline outputs calibrated the fitted slopes for each pixel and each integration ramp in the time series.

We perform custom spectral extraction with the \citet{Horne:1986} optimal extraction routine in Stage 3 of \textit{Eureka!}. Only columns 14 to 495 along the dispersion axis are included in the extraction due to the low throughput of NIRSpec in the excluded columns at the detector edges. The trace is located from the peak of a Gaussian fit to each column in a summed image across the integration. Similarly to Stage 1, we perform column-by-column median subtraction after masking all pixels with data quality flags and using only background pixels that are 8 or more pixels away from the central trace. Optimal median-weighted 1D extraction is then computed for the half-width aperture of 4 pixels in the cross-dispersion direction on either side of the trace to obtain the time series of spectra. The large (40--200\AA) wavelength bins do not require wavelength corrections. We only use the one NIRSpec reduction pipeline since we do not study NIR lines with the NIRSpec data and the shape of the continuum is robust across all flares.

\subsection{Flux calibration}\label{flux_cal}
We flux calibrate the extracted spectra of TRAPPIST-1 from each instrument and pipeline against a PHOENIX spectrum of TRAPPIST-1 whose reliability is verified with NIR photometry bands \citep{Wilson:2021}. The PHOENIX model is computed for the non-flaring stellar spectrum by linearly interpolating four PHOENIX model spectra computed at 2600 and 2700 K and log\textit{g} values of 5.0 and 5.5 to the values of 2628 K and log $g$=5.21 measured by \citet{Gonzales:2019}. Different values of the temperature and log $g$ result in different values in the absolute flux calibration, although uncertainties in both parameters are reported at the $\sim$1\% level in \citet{Gonzales:2019}. Other spectral energy distributions in the literature such as that of \citet{Pineda:2021} differ at the $\sim$10\% level based on the assumed stellar parameters.

For each set of observations, an averaged spectrum of all integrations that are both out of transit and out of flare is constructed for the color correction. The averaged spectrum and the PHOENIX model spectrum are both heavily binned to remove spectral features. For NIRISS, 200 wavelength points per bin are used; for NIRSpec, 90 points per bin are used. A smoothly-varying response curve results from dividing the resulting binned model by the data. The choice of bin size on systematic uncertainty of the flux calibrated spectrum is measured by varying the bin size and creating an alternate flux calibration. For NIRISS order one, the bin size is increased by 50\% in the alternate calibration, while the bin size is increased by 25\% for order two. For NIRSpec, the bin size is varied by 10\%. These values were selected because varying the bins within the specified ranges kept the bin scale large enough to blur out spectral features without altering the overall shape of the spectrum. Using these values, the mean systematic uncertainty of the order one NIRISS calibration is 7.8\%, with a 1$\sigma$ upper limit of 15\%, while the mean systematic uncertainty of the order two NIRISS calibration is 8.1\%, with a 1$\sigma$ upper limit of 22\%. The mean systematic uncertainty of the NIRSpec calibration in the wavelength range 0.6--3.5~$\mu$m is 3.6\%, with a 1$\sigma$ upper limit of 12\%. We use both the primary and secondary flux calibration to create two versions of the same flare spectra and compare the final values of the flare properties to ensure the impacts of this systematic uncertainty are not significant.

\begin{figure*}
	\centering
        \subfigure
	{
		\includegraphics[width=0.98\textwidth]{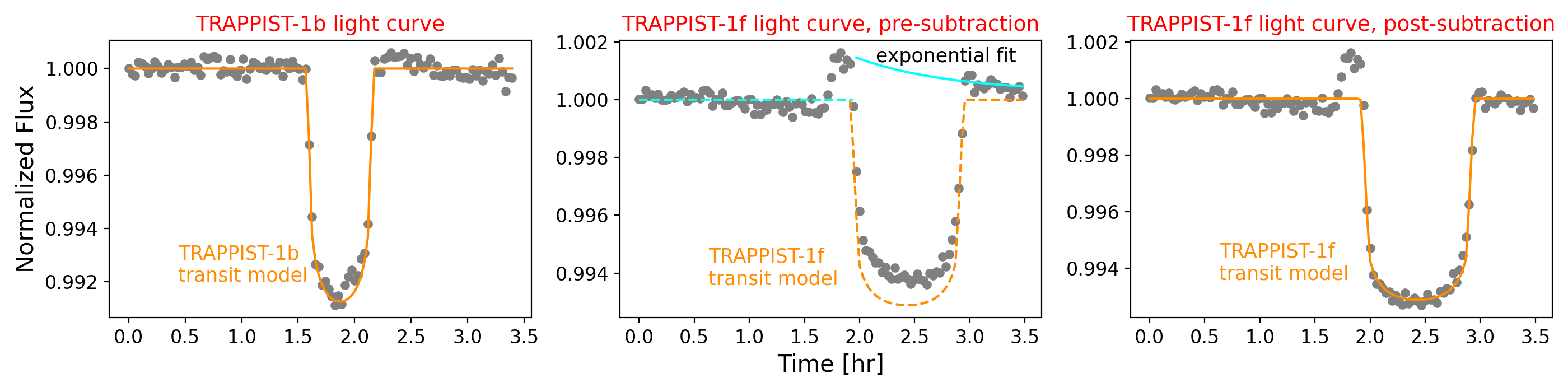}
	}
        \subfigure
	{
		\includegraphics[width=0.98\textwidth]{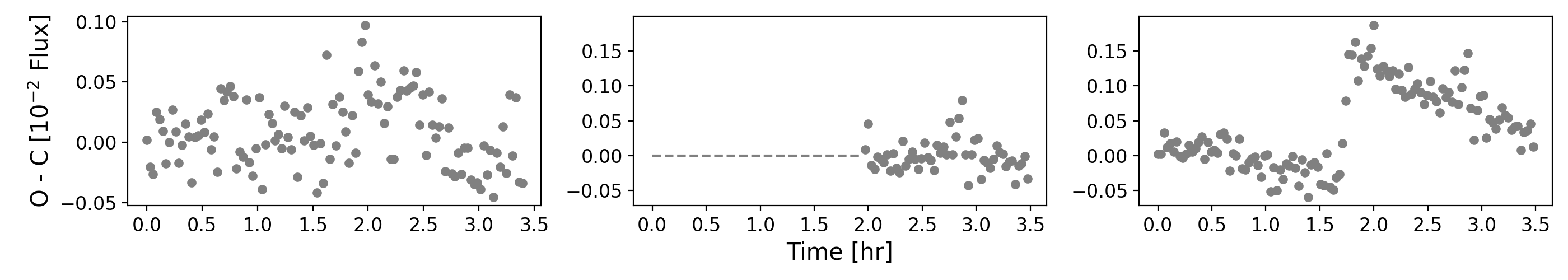}
	}
	\caption{Broadband light curves of the flare-affected transits of TRAPPIST-1b and f. The transit models are shown in orange, and the exponential fit to the integrations in both the transit and flare is shown in blue. The exponential fit is necessary to determine the shape of the transit light curve. As shown in the O-C diagrams at the bottom, we use an initial guess of $c_1$ and $c_2$ to fit the exponential, then subtract the exponential to obtain final values of $c_1$ and $c_2$. We note $c_1$ and $c_2$ represent the coefficients of the quadratic limb darkening equation. The first and third O-C diagrams are directly comparable and show the light curve with the final transit model removed. The middle O-C diagram shows an intermediate step in the determination of the TRAPPIST-1f transit model fit in which the initial guess on the transit model is subtracted prior to fitting and subtracting the exponential. The residuals of the middle panel therefore show the quality of the exponential fit.}
	\label{fig:transit_methods}
\end{figure*}

\subsection{Light curve extraction}\label{create_lcvs}
We create light curves for the flux within specific wavelength regions by iterating through the integrations, summing the counts for all wavelength points within the line. These light curves are useful for exploring the time evolution of line emission (e.g. the Balmer and Paschen lines) as described in \S \ref{identify_lines} and \ref{valid_lines_method}. We create light curves in each continuum bandpass by convolving the flux-calibrated spectra with the band response function and then summing the counts. We choose the TESS, J, H, and Ks bandpasses since they cover the range of wavelengths where most flare flux is observed (0.6--2.2~$\mu$m) and are widely recognized. We also note these bands allow us to connect our flares to the extensive flare literature from the TESS mission (e.g. \citealt{Gunther:2020,Feinstein:2022a,Howard_MacGregor:2022}) and flare observations from 2MASS \citep{Davenport:2012}. Photometric uncertainties are propagated from the individual flux uncertainties at each wavelength. Systematic uncertainties due to flux calibration are obtained by varying the calibration function as described above and re-computing the light curves. The light curves of the lines are not impacted by systematic color offsets, and the systematic color offsets of the continuum bands are small. Systematic uncertainties for the F3 and F4 flare light curves are at the 5\% level for the TESS band and 0.1\% level for the J, H, and Ks bands.

Transits are removed from the light curves prior to flare analysis using the \texttt{batman} package \citep{Kreidberg:2015} as shown in Figure \ref{fig:transit_methods}. The transit model is specified by fixed values of the orbital period, semi-major axis, inclination, and planetary radius taken from \cite{Gillon:2017}. The time of inferior conjunction is measured from the transit light curve since transit timing variations can change this value slightly between orbits. For TRAPPIST-1f where the flare covers the entire transit, this value is initially estimated as the center time of the transit and then verified after an exponential function is subtracted from the broadband transit light curve as described below. An eccentricity of zero and longitude of periastron of 90 degrees are assumed for each planet.

Quadratic limb darkening of the transit light curve is computed with \texttt{batman} as described in Equation 9 of \citet{Kreidberg:2015}. We reproduce the limb-darkened intensity in our Eq. \ref{eq_3}:
\begin{equation}
\label{eq_3}
    I(\mu) = I_0[1 - c_1(1-\mu) - c_2(1-\mu)^2]
\end{equation}
Here, $I_0$ is the 1D intensity profile prior to limb-darkening, $\mu=\sqrt{1-x^2}$ with the normalized radial coordinate $x$ given in units of stellar radius, and $c_1$ and $c_2$ are the quadratic limb darkening coefficients. Quadratic limb darkening coefficients are $c_1$=0.34 and $c_2$=0.26 for TRAPPIST-1b and $c_1$=0.17 and $c_2$=0.26 for TRAPPIST-1f. The TRAPPIST-1b coefficients are computed by fitting the broadband light curves of the transits, excluding in-flare integrations and setting the limb darkening coefficients as free parameters. The flare covers the entire transit of TRAPPIST-1f, so we first fit and remove an exponential representing the flare decay from the $\lambda>$1.1~$\mu$m broadband transit light curve using the \citet{Davenport:2014} flare template. The \citet{Davenport:2014} template is scaled to the data using a flux amplitude of 0.00148, width of 1.225 hr, and peak time during the 69$^\mathrm{th}$ integration. The coefficients were initially identified by manually varying the TRAPPIST-1b values for TRAPPIST-1f. The transit fit was then verified with a Monte Carlo analysis that minimized the residuals to the transit light curve. The errors on the fit are at the 10\% level and induce systematic uncertainties of 0.01\% in the resulting fluxes, an order of magnitude below the flux amplitudes of the flares as listed in Tables \ref{tab:line_table} and \ref{tab:continuum_table}. We note \citet{Lim:2023} find the best-fit \texttt{batman} transit parameters to the TRAPPIST-1b NIRISS data differ from the literature values. However, the 10$^3$--10$^4$ ppm signals of the flares are much less affected by uncertainties in the exact shape of the transit than are planetary signals of 100 ppm. This can be seen in the O-C diagrams of Figure \ref{fig:transit_methods}, where the shape of the flare shows a smooth decay after the transit is subtracted. The peak of each flare occurs during integrations outside of the transit or in the flat bottom, further minimizing the effect of uncertainties in the exact shape of the transit.

We compare the flare light curves between the \texttt{transitspectroscopy} and \texttt{supreme-SPOON} pipelines. Although neither reduction was initially designed for creating light curves of single lines, both pipelines produce qualitatively similar light curves for most lines. Example light curves are shown in Figure \ref{fig:pipe_compare} for the H$\alpha$, P$\alpha$, and P$\beta$ lines, since these span a range of wavelengths from 0.6565--1.8756~$\mu$m. The photometric variability in the lines is slightly lower for the \texttt{supreme-SPOON} reduction, which aids comparison of the flare light curve shapes. We note the point-to-point variability during the rapid peak phase of the flares is sometimes significantly lower for the \texttt{supreme-SPOON} reduction. The point-to-point variability during the flare peak is defined as the standard deviation of the difference between the flux of successive integrations. Lower amounts of scatter for \texttt{supreme-SPOON} are especially noticeable during the peak of the P$\beta$ flare light curve in the middle panel of Figure \ref{fig:pipe_compare} and the secondary flux increase in the H$\alpha$ flare light curve at $\sim$2.4 hr in the first panel of Figure \ref{fig:pipe_compare}. The point-to-point variability is typically 20--50\% greater for our implementation of \texttt{transitspectroscopy}, but is 270\% greater for the P$\beta$ peak and 100\% greater for the second H$\alpha$ peak. We speculate the difference may result from the group versus integration level reduction choices of the pipelines. Based on this comparison, we choose \texttt{supreme-SPOON} as our primary reduction for the forthcoming analysis.

\section{Characterizing the NIR line emission of the flares}\label{flare_line_methods}

\subsection{Creation of flare-only spectra}\label{create_flonly_spec}
The light curves are visually inspected to identify times during quiescence and times during flares. All flux calibrated spectra during non-flaring integrations are averaged together to create a quiescent reference spectrum. Likewise, spectra during integrations near the flare peak are averaged together to maximize the flare signal. A flare-only spectrum is created by subtracting the out-of-flare spectrum from the peak flare spectrum and propagating the errors of the component spectra. The peak spectrum of the TRAPPIST-1b observations occurs during the transit. In this case, the fractional depth of the transit light curve is used to correct the flux before subtracting the quiescent spectrum.

\subsection{Identifying flare lines}\label{identify_lines}
Candidate lines in the NIRISS data are first identified in an initial visual inspection of the flare-only spectra. To aid identification of lines, the central wavelengths for H$\alpha$, the Paschen series, and Brackett series in vacuum are overlaid on the spectrum. Other NIR flare lines in the literature (e.g. \citealt{Liebert:1999, Fuhrmeister:2008, Schmidt:2012, Kanodia:2022}) are also overlaid. While we overlay the Paschen series from P$\alpha$ at 1.8756~$\mu$m through the Paschen jump at 0.82~$\mu$m, we do not see any candidate lines exceeding the local noise beyond the P$\epsilon$ line at 0.9549~$\mu$m characterizing the 8--3 transition of hydrogen. We searched for Brackett emission up to Br$\gamma$, but only found a candidate line for Br$\beta$ at 2.6259~$\mu$m characterizing the 6--4 transition of hydrogen. Br$\alpha$ lies beyond the wavelength range of NIRISS at 4.0523~$\mu$m. We also searched for helium emission, including a helium I line at 0.7062~$\mu$m and the well-known helium infrared triplet (IRT) activity line at 1.0833~$\mu$m. We also searched for the Mg I and Fe I lines reported in Table 2 of \citet{Kanodia:2022}.

\subsection{Validating flare lines}\label{valid_lines_method}
Candidate lines are vetted in a two-step process designed to remove false positives. First, the significance of the peak $\sigma_\mathrm{line}$ is estimated compared to the surrounding continuum. The FWHM of each line is measured and a continuum region of 20 FWHM to either side of the line is defined. The mean $\mu_G$ and standard deviation $\sigma_G$ of the continuum region are measured and the probability that the maximum flux of the candidate line would be produced from a Gaussian distribution of fluxes with the same $\mu_G$ and $\sigma_G$ is computed. Twenty FWHM were selected to minimize the number of other potential peaks in the continuum region, to ensure consistency in the local noise properties, and to avoid arbitrary selection of continuum regions. 

Second, the time variation in the line is compared with the surrounding continuum. Light curves for continuum regions adjacent to and on either side of the candidate line are created. These light curves are visually examined to ensure the light curve fluxes at the target wavelength are higher than in the surrounding continuum. If the light curves of the line and continuum are similar, the candidate is classified as a false positive. This step is necessary because small continuum enhancements are present during the flare at wavelengths where line emission might be expected. Next, the significance of the flare light curve in the line $\sigma_\mathrm{lcv}$ is estimated compared to the out-of-flare regions in the light curve. The mean $\mu_G$ and standard deviation $\sigma_G$ of the out-of-flare portions of the light curve are measured and the probability that the maximum flux of the light curve would be produced from a Gaussian distribution of fluxes with the same $\mu_G$ and $\sigma_G$ is computed. We bin the light curves of flares that evolve more slowly than the instrument cadence prior to comparing the peak flux to the noise. The overall rise and decay of the F4 event occurs on $\sim$10--20 minute timescales. We therefore bin the light curves in each line into groups of ten, or 17.5 min per bin. Ca II IRTc is an exception, as the rise and decay of its light curve is captured with five points per bin, or 8.75 min per bin. We likewise bin the NIRSpec flares, F1 and F2, to 30 s cadence using 18 integrations per bin prior to computing the light curve significance.

\begin{figure*}
	\centering
        \subfigure
	{
		\includegraphics[width=0.98\textwidth]{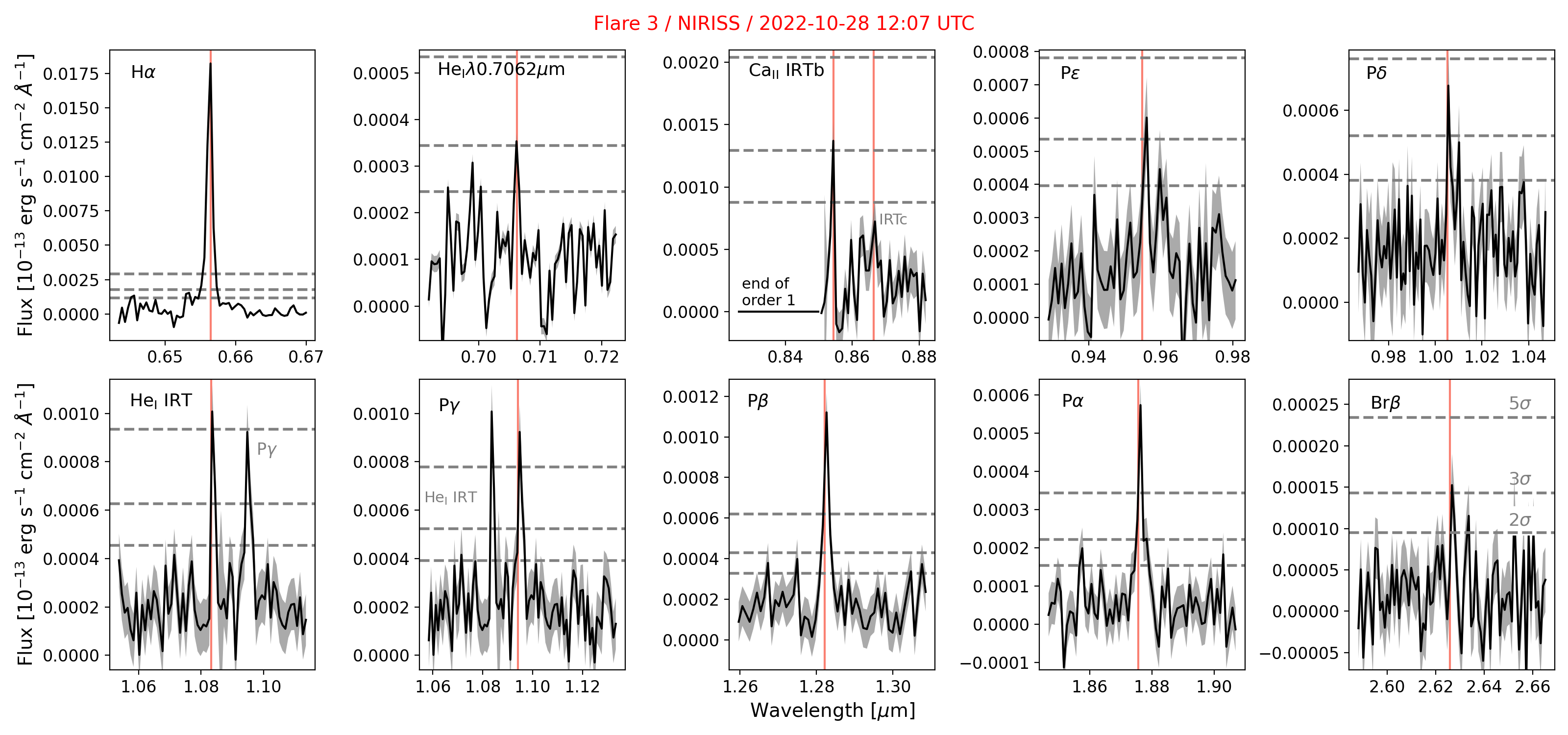}
	}
        \subfigure
	{
		\includegraphics[width=0.98\textwidth]{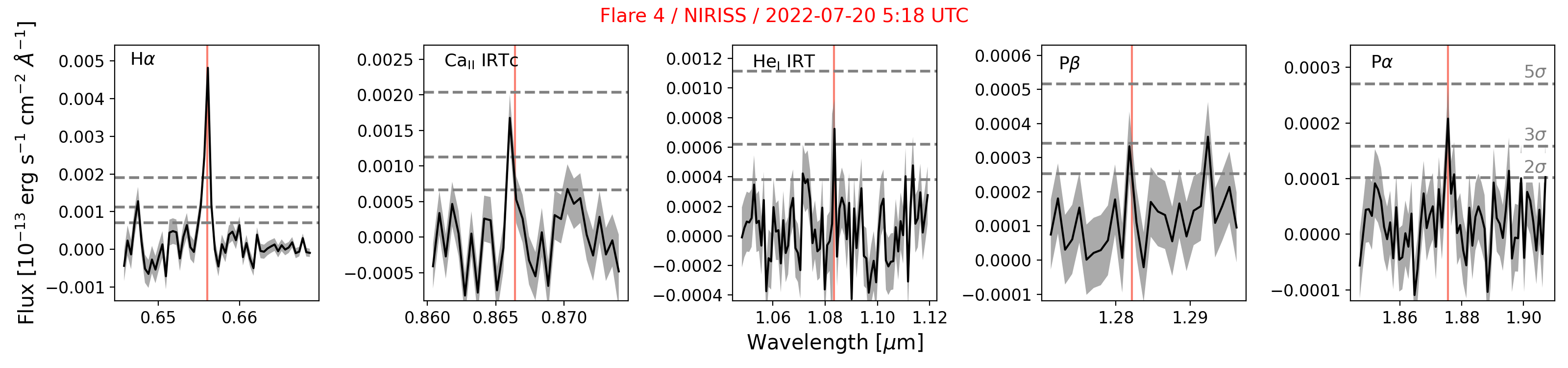}
	}
	\caption{Line emission is observed in the flare-only spectrum of the two NIRISS flares. Top: The F3 event is larger, with more candidate or confirmed lines during the flare peak. Bottom: The F4 event is the weakest flare for which lines beyond H$\alpha$ could be detected.}
	\label{fig:NIRISS_line_centers}
\end{figure*}

\begin{figure*}
	\centering
        \subfigure
	{
		\includegraphics[width=0.60\textwidth]{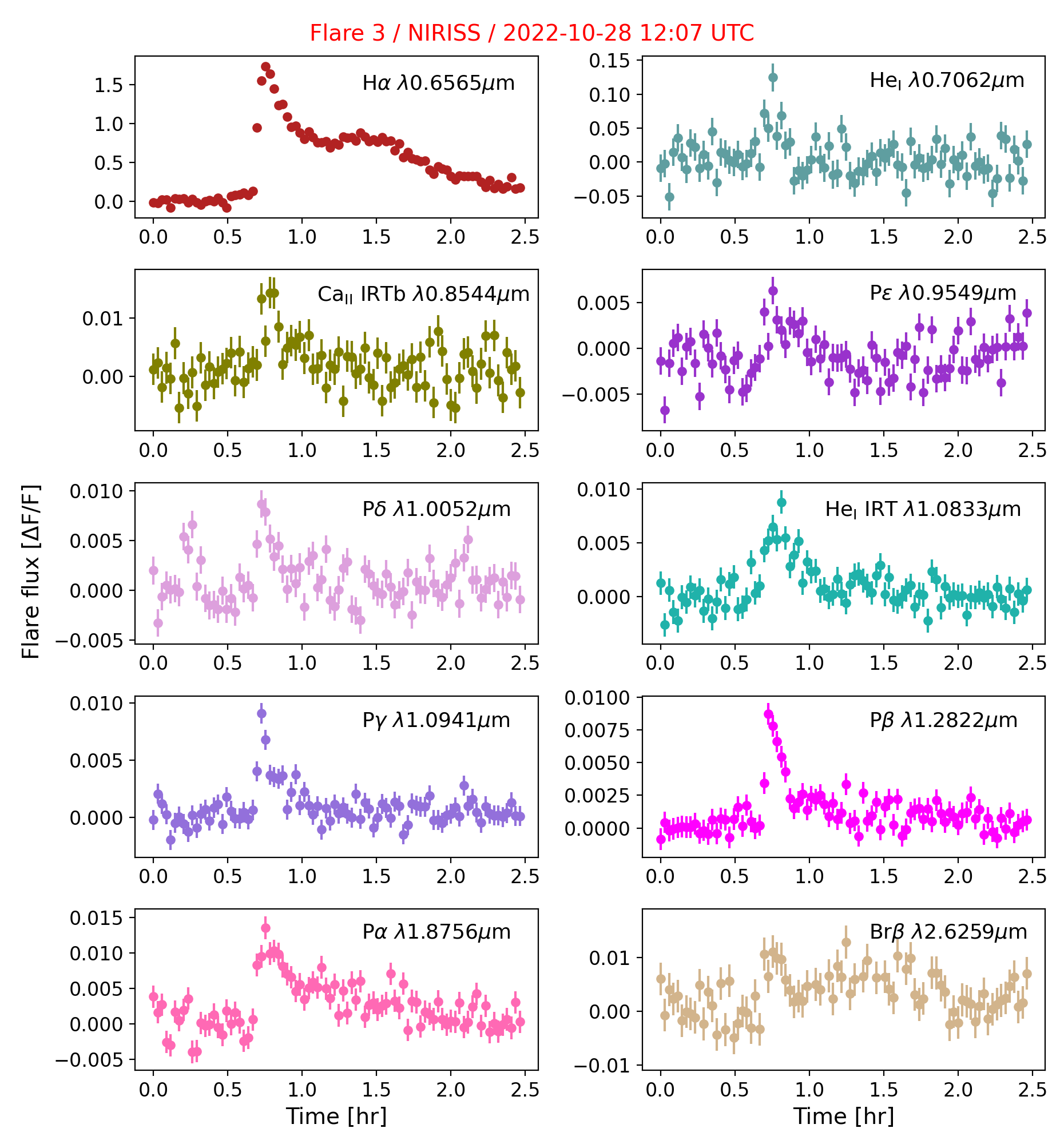}
	}
        \subfigure
	{
		\includegraphics[width=0.34\textwidth]{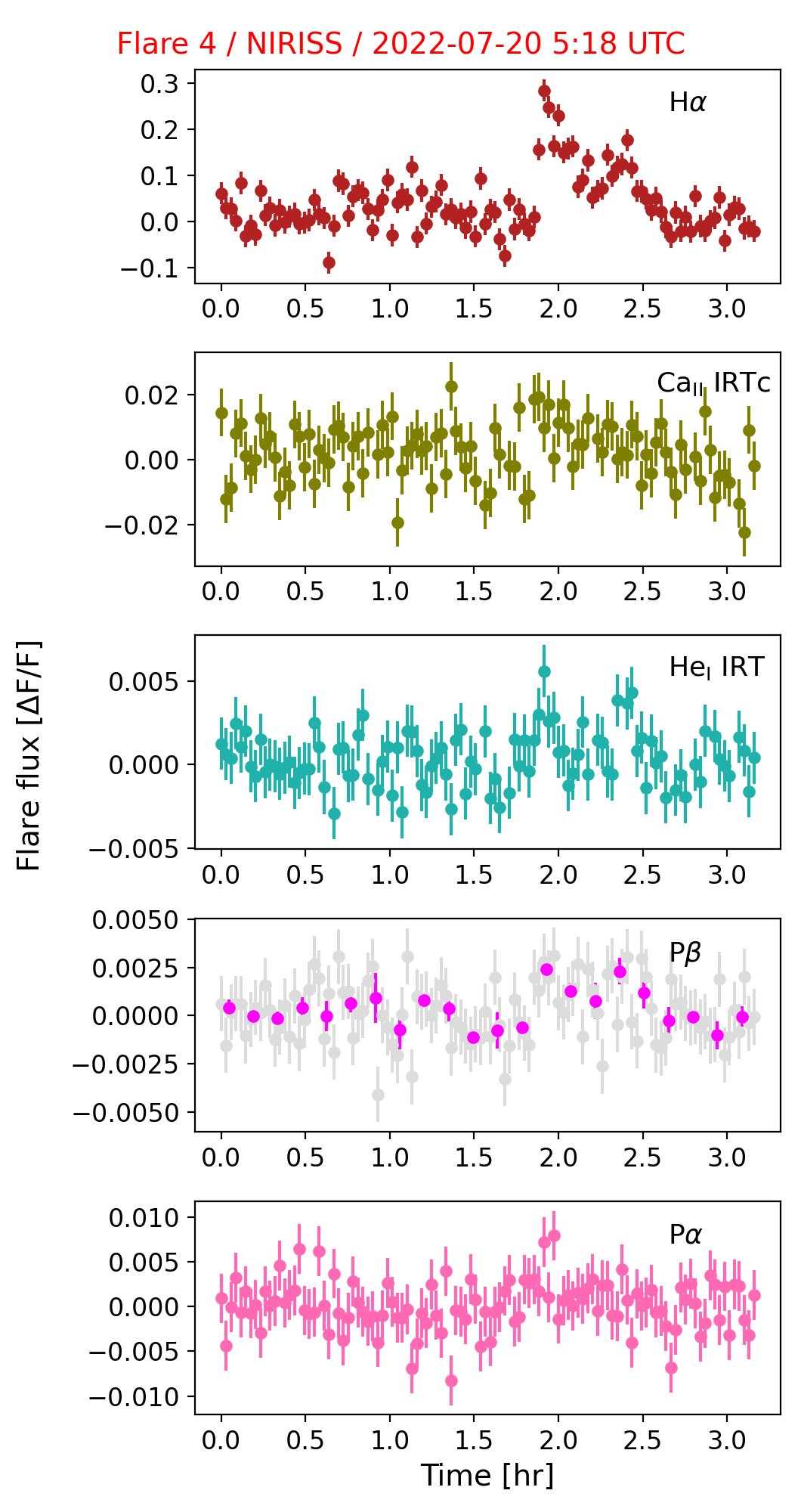}
	}
	\caption{Left: the light curves of the lines and candidate lines identified in the larger F3 flare. P$\alpha$ and Br$\beta$ are the first detections of flare emission from a main sequence star in these lines. Right: The light curves of the lines and candidate lines identified in the smaller F4 flare. $\Delta$F/F is the fractional flux as defined in \S \ref{basic_fl_properties}}
	\label{fig:jwst_overview}
\end{figure*}

For both the $\sigma_\mathrm{line}$ and $\sigma_\mathrm{lcv}$ estimates, Gaussian noise is assumed due to the small number of points in the continuum region and the out-of-flare portions of the time series during each 3--4 hour JWST observation, respectively. Random shuffling of points or bootstrapping are designed for scenarios in which higher-resolution spectra or longer time series are available. We consider a secure line detection to either exceed the 3$\sigma$ significance level for both the $\sigma_\mathrm{line}$ and $\sigma_\mathrm{lcv}$ measurements or to exceed 5$\sigma$ for one of these measurements and be clearly visible by eye in the light curve and spectrum. We allow for the second case because some lines such as the $\lambda$0.7062~$\mu$m He I line in the F3 event have lower values of $\sigma_\mathrm{line}$ of 2--3$\sigma$ but have a clearly visible flare light curve that is much stronger than the light curves for the continuum regions on either side of the line. We consider a formal but weak detection to reach 3$\sigma$ for either of the $\sigma_\mathrm{line}$ or $\sigma_\mathrm{lcv}$ measurements, and a candidate line to be above 2$\sigma$ but below 3$\sigma$ for both the $\sigma_\mathrm{line}$ and $\sigma_\mathrm{lcv}$ measurements. Each candidate or confirmed line in the spectra of the F3 and F4 events is shown in Figure \ref{fig:NIRISS_line_centers}. Likewise, the light curves of each line are shown in Figure \ref{fig:jwst_overview}.

\begin{figure*}
	\centering
	{
		\includegraphics[width=0.98\textwidth]{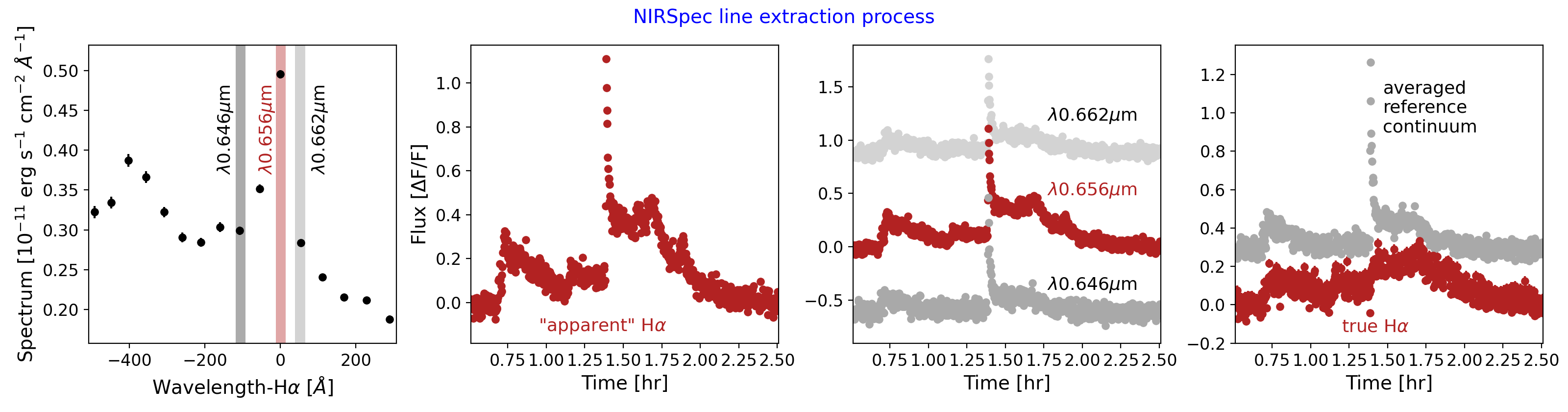}
	}
	\caption{Separation of H$\alpha$ from the continuum contribution in the same NIRSpec wavelength point. On the left, the wavelength point containing H$\alpha$ is highlighted in dark red, while two continuum reference wavelengths are highlighted in grey. In the second panel, the light curve at 0.6565~$\mu$m is displayed, presumably that of the H$\alpha$ line. In the third panel, light curves are extracted for each wavelength point and compared (with an offset applied in the plot). The average of the two reference light curves is divided from the target light curve to remove behavior common to all three wavelengths. On the right, the resulting corrected H$\alpha$ light curve is shown.}
	\label{fig:ref_lcv_method}
\end{figure*}

While the line widths are comparable to the wavelength bins of NIRISS, this is not true for NIRSpec. The NIRSpec wavelength bins vary from 42\AA\ at 0.6~$\mu$m to 201\AA\ at 1.55~$\mu$m, substantially diluting line emission. We isolate line emission from the continuum in the NIRSpec wavelength bins using reference light curves. The continuum remains essentially the same in the light curve of the flare at wavelengths immediately adjacent to the line, while the line emission is different. We therefore average the light curves in the continuum regions on either side of the line to estimate the contribution of the continuum in the target wavelength region. We normalize the light curve of the target wavelength region by the reference light curves and propagate the errors on the fluxes to produce a corrected light curve of the line. This process is illustrated for the H$\alpha$ line in Fig. \ref{fig:ref_lcv_method}. Our method indicates 61\% of the flux in both NIRSpec flares F1 and F2 is due to H$\alpha$ emission and 39\% is actually due to the underlying continuum. Light curves of the higher-order lines are much weaker than H$\alpha$, with 0.2\% enhancements above the continuum at peak for P$\alpha$, P$\delta$ and for the combined flux of the P$\gamma$ and the He I IRT lines. We note each of these excesses except for the P$\gamma$ and He I IRT emission are at the scale of the error bars and do not constitute detections. No evidence exists for other Paschen or Brackett lines in the NIRSpec data.

\section{Characterizing the NIR continuum of the flares}\label{flare_continuum_methods}

\begin{figure*}
	\centering
	{
		\includegraphics[width=0.98\textwidth]{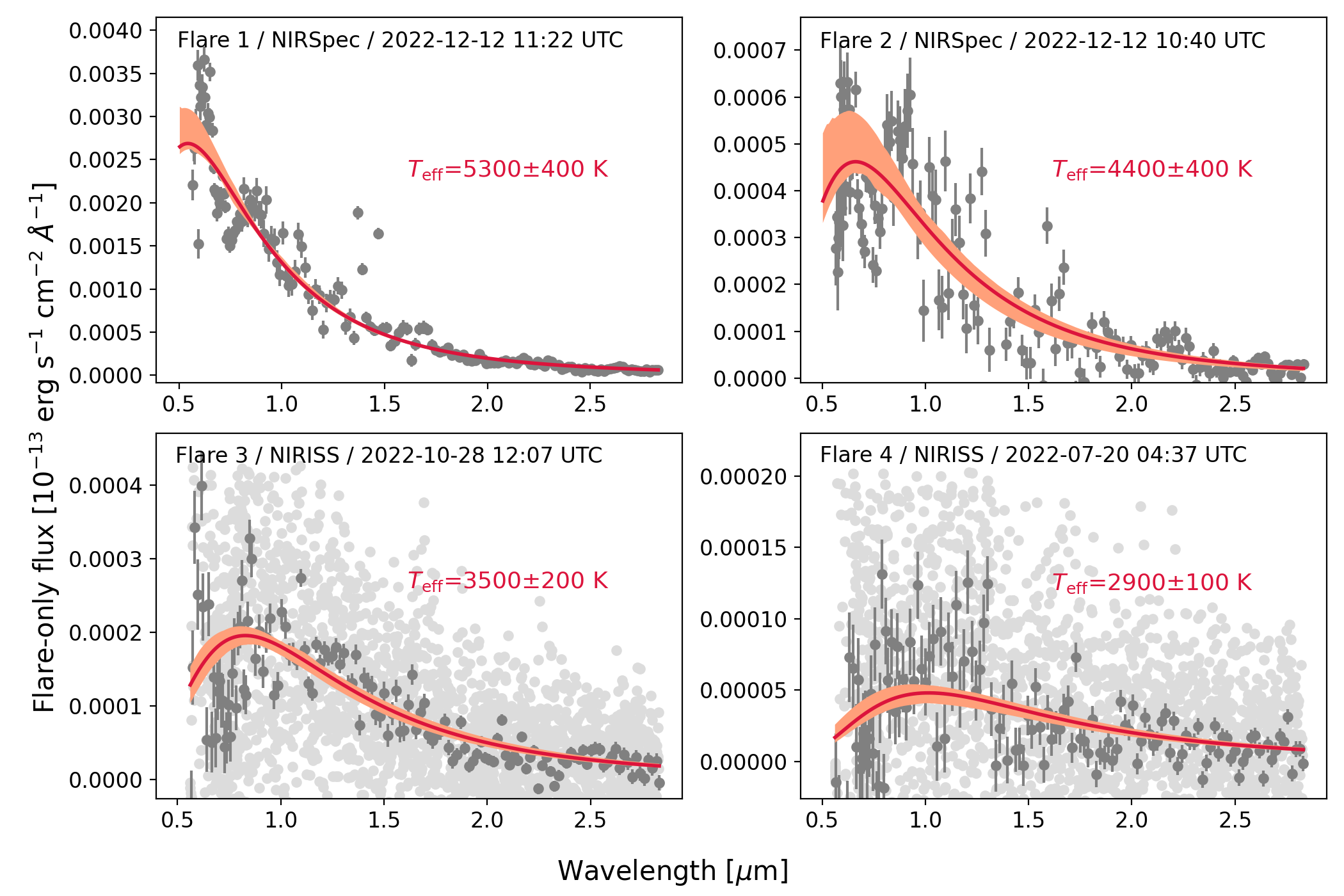}
	}
	\caption{Continuum is observed in each event. The continuum of the NIRSpec flares (top) is readily discernible without binning due to the low wavelength resolution. The wavelength points of the NIRISS flares (bottom) are binned to more clearly illustrate the flare spectrum, with the native resolution data shown in light gray and the binned spectrum in dark gray. The best-fit Planck function and uncertainties are displayed in red. The higher temperatures of the NIRSpec flares are a result of the higher cadence. Binning the NIRSpec flares to 105 s cadence produces nearly identical spectra to the NIRISS flares. For comparison, the effective temperature of the star is 2628 K. All spectra are plotted to the wavelength range of NIRISS to facilitate comparison.}
	\label{fig:bb_flares}
\end{figure*}

\subsection{Identifying flare continuum}\label{cand_lines}
The lines identified in \S \ref{identify_lines} are masked out of the flare-only spectra in order to observe any underlying continuum. The majority of the emission of a blackbody of $T_\mathrm{eff}$=5000--10,000 K is at wavelengths below 1~$\mu$m where the response curves of the instruments are lower. As a result, the flux calibrated flare-only spectrum is used to identify and characterize the continuum. We bin the NIRISS spectra by $N$=20 points per wavelength bin and visually inspect both the peak flare spectrum and the time evolution of the spectrum. We select 20 points per bin based on inspection of the bin size necessary to reduce scatter sufficiently to see the large scale variation in the spectrum. Binning is not necessary for NIRSpec due to the lower wavelength resolution.

An initial visual inspection reveals the spectrum of each flare appears consistent with that of a blackbody once the lines are masked out, as shown in Figure \ref{fig:bb_flares}. We fit the Planck function to the flares to determine if the NIR flare continuum is indeed consistent with the Rayleigh-Jeans tail of a blackbody. In addition to the apparent blackbody curve present from 0.6-3~$\mu$m, spectral features are also present on smaller scales. Increasingly deep spectral features and large error bars near 0.6~$\mu$m may represent systematics near the edge of the detector. Spectral features at longer wavelengths are unlikely to be systematics, however. Even though some features are present in both NIRISS and NIRSpec data, features are masked out separately to account for the different resolutions of the instruments. For NIRISS, we exclude regions from 0.723 to 0.761, 0.782 to 0.790, 1.370 to 1.395~$\mu$m as well as the region of the Paschen jump near the Paschen series limit from 0.820 to 0.867~$\mu$m. For NIRSpec, we exclude regions from 0.676 to 0.745, 1.351 to 1.410, and 1.453 to 1.527~$\mu$m. We note that the $\sim$1.5~$\mu$m feature is near the Brackett jump. The other features are in regions of high stellar flux and are likely astrophysical in origin.

We define new spectral flux errorbars to capture the empirical point-to-point variability as a function of wavelength, $\epsilon_\lambda$. The formal errorbars $\epsilon_\mathrm{formal}$ of the extracted spectra are often smaller than the scatter seen in the spectra. Our $\epsilon_\lambda$ errorbars are agnostic to the nature of this small-scale variability and improve the fit to the overall shape of the spectrum. We define $\epsilon_\lambda$ as the dispersion of the absolute value of the difference in flux between adjacent wavelength points in the spectrum. As shown in Fig.\ \ref{fig:bb_flares}, this approach does a good job representing the variation in the spectrum at smaller scales than the overall shape of the spectrum to be fit by the Planck function. Finally, each spectrum was reduced separately for both calibrations and compared against the other to account for systematic errors from flux calibration. The alternate calibration spectra are nearly identical to the primary reduction spectra.

\subsection{Fitting the Planck function to the flare continuum}\label{fitting_planck_fn}
The flare spectrum holds information on both the effective temperature $T_\mathrm{eff}$ and filling factor of the flare $X_\mathrm{eff}$, defined as $X_\mathrm{eff}$ = $a_\mathrm{fl}/a_{*}$ = $R_\mathrm{fl}^2$/$R_{*}^2$. Here, $R_\mathrm{fl}$ is the effective radius of the flare and $R_*$ is the stellar radius. The shape of the spectrum is determined by $T_\mathrm{eff}$, while the absolute luminosity of the flare is set by $X_\mathrm{eff}$. Following \citet{Hawley:2003}, we fit the spectrum for both variables simultaneously by scaling the flux received at the telescope $f_{\lambda, fl, \oplus}$ for the stellar distance $d_*$ and setting it equal to the monochromatic luminosity of the Planck function $F_{\lambda, fl, *}$ = 4$\pi^2 R_\mathrm{fl}^{2} B_{\lambda}(T_\mathrm{eff})$. We use values of the stellar radius and distance of $R_*/R_\odot$=0.1192$\pm$0.0013 \citep{Agol:2021} and $d_*$=12.43$\pm$0.02 pc \citep{Chambers:2016}. The flux of the flare at each wavelength is given from the Planck function as $B_\lambda(T_\mathrm{eff})$. We fit the resulting expression for the flare-only flux, shown in Eq. \ref{eq_1}:
\begin{equation}
\label{eq_1}
    f_{\lambda, fl, \oplus}(X_\mathrm{eff}, T_\mathrm{eff}) = X_\mathrm{eff} \frac{R_{*}^2}{d_{*}^2} \pi B_\lambda(T_\mathrm{eff}).  
\end{equation}
We fit the continuum with the expression given in Eq.\ \ref{eq_1} using a bootstrap approach. Fits are performed at the native wavelength resolution of NIRSpec and at the binned resolution for the NIRISS data using 200 bootstrap trials. In each trial, the fluxes of each wavelength point are varied within the empirical $\epsilon_\lambda$ errorbars and a 10--20\% region of the spectrum is randomly dropped to avoid over-fitting to any specific features. The best-fit and 1$\sigma$ range of the flux values for each wavelength bin are computed from the bootstrapped fits. Likewise, we record the best-fit and 1$\sigma$ range of the $T_\mathrm{eff}$ and $X_\mathrm{eff}$ values from the bootstrap. We also fit Eq.\ \ref{eq_1} to a flare-only spectrum produced with the alternate flux calibration to assess the effects of systematic errors arising from flux calibration. During the peak of the flares, the mean systematic offsets in temperature are 860, 410, 100, and 20 K for the F1, F2, F3, and F4 events, respectively. The mean systematic offsets of the filling factors between flux calibrations are 0.05, 0.06, 0.01, and $<$0.01\% for the F1, F2, F3, and F4 events, respectively. The systematic offsets are lower outside the peak times since most of the color correction occurs at the bluest wavelengths.

The continuum of each individual integration is fit to obtain time series for $T_\mathrm{eff}$ and $X_\mathrm{eff}$. For the smallest NIRISS flare (F4), light curves are binned in groups of 5 integrations to increase the signal to noise prior to fitting. Likewise, the native 1.6 s cadence of the NIRSpec data are binned to $\sim$30 s cadence with groups of 18 integrations. The flare spectrum is strongest during integrations with high count rates during the impulsive phase. We maximize the signal of the continuum by binning the points during the peak of each flare. For the NIRSpec flares (F1 and F2), 18 integrations per bin are sufficient to maximize the signal of the flare peak in a single point, while 4 integrations are used for the peak spectrum of the larger NIRISS flare (F3) and 5 integrations are used for the smaller NIRISS flare (F4). The continuum spectrum during the peak of each flare is displayed in Figure \ref{fig:bb_flares} at the native wavelength resolution of the instrument as well as the binned resolution for NIRISS. The best-fit blackbody models and uncertainties are overlaid on the spectrum of each flare.

\subsection{Validating continuum enhancements in the light curves}\label{valid_cont_methods}
\begin{figure*}
	\centering
        \subfigure
	{
		\includegraphics[width=0.31\textwidth]{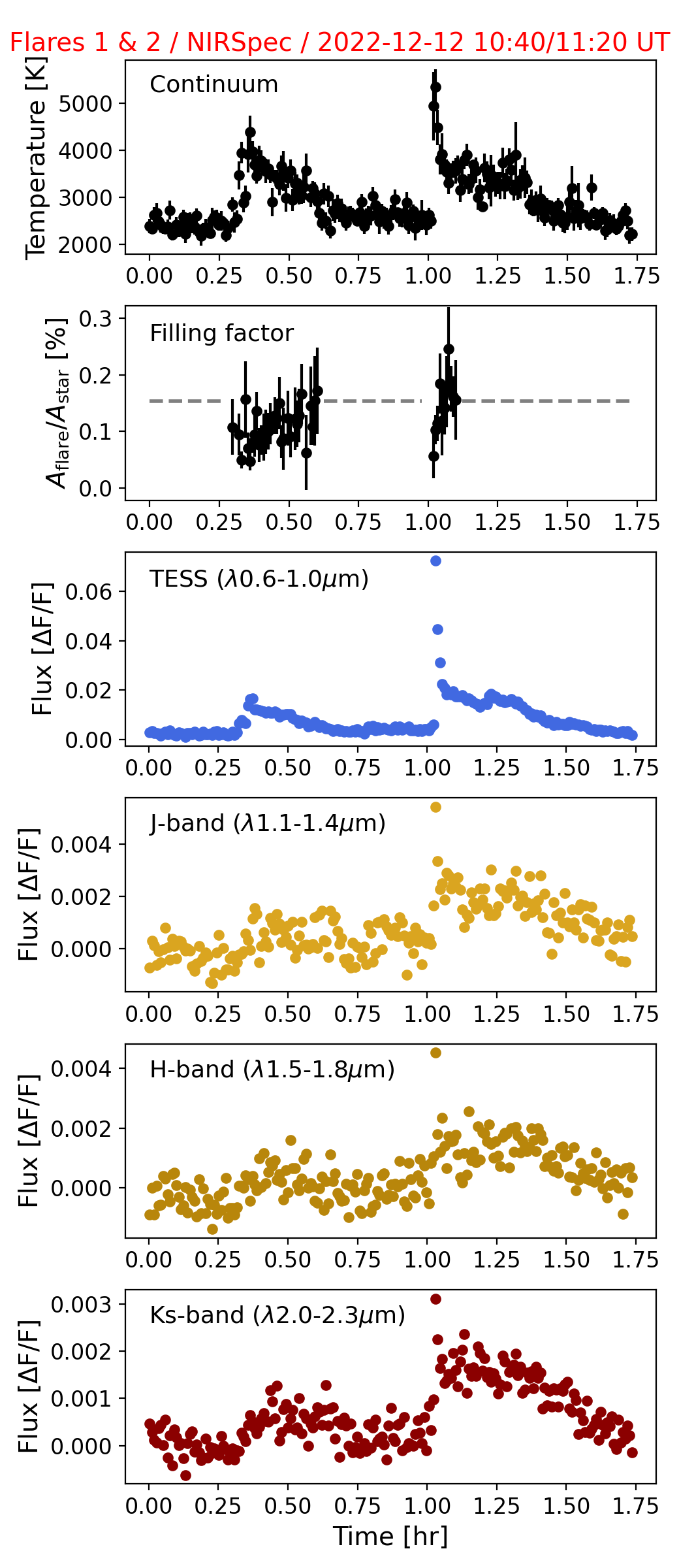}
	}
        \centering
        \subfigure
	{
		\includegraphics[width=0.31\textwidth]{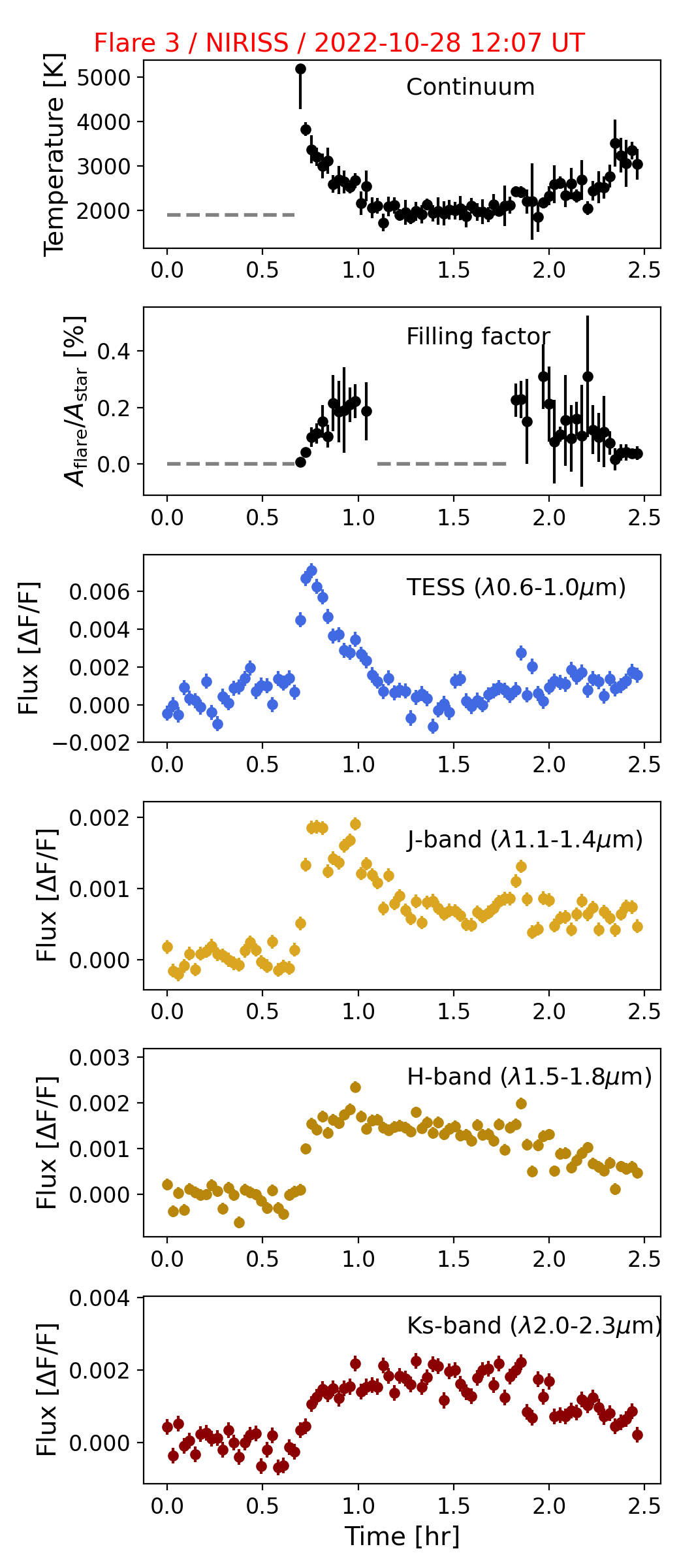}
	}
        \centering
        \subfigure
	{
		\includegraphics[width=0.31\textwidth]{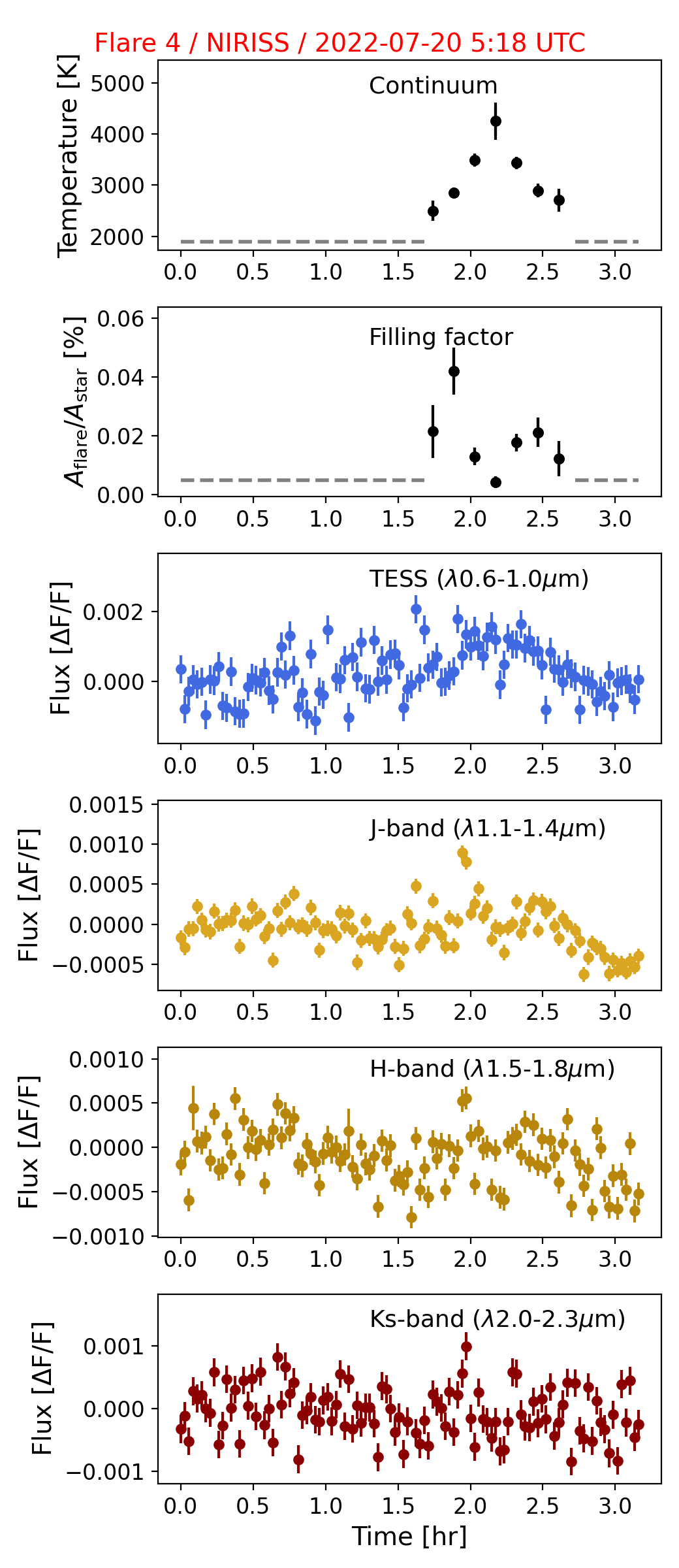}
	}
	\caption{Overview of the continuum properties of the flares. The F1 and F2 NIRSpec events are shown together in the left column of panels, while the F3 and F4 NIRISS flares are displayed in the middle and right columns. For each flare, the best-fit effective temperature and filling factor are shown for the integrations, as well as the broadband light curves for the TESS and 2MASS filters. The temperature is hottest near the beginning of the flares for the F1-F3 events, and cools throughout each event. It is difficult to compare the times of maximum temperature and peak flux for the F4 event due to the binning of integrations to increase the signal to noise. The flare area expands as the plasma cools, leading to an increase in the filling factor. The TESS band light curve evolves fastest, with the H and Ks band light curves evolving the slowest.}
	\label{fig:continuum_overview}
\end{figure*}

The continuum is sufficiently elevated during the flares to induce a flux increase in the broadband light curves. We therefore verify the significance of the light curve peaks seen in the TESS, J, H, and Ks bands using the $\sigma_\mathrm{lcv}$ method described in \S \ref{valid_lines_method}. The large scale structure of the flares evolves more slowly than the native cadence of the integrations. The TESS light curve is binned to 10 s cadence and the NIR bands are binned to 30 s cadence to increase the S/N by 2.5--4.3$\times$ prior to computing $\sigma_\mathrm{lcv}$ for NIRSpec. The TESS, J, H, and Ks light curves of both NIRISS flares (F3 and F4) are binned to 18 minute cadence using 10 integrations per bin to resolve only the large-scale structure in the flare light curve. The light curves are shown in Figure \ref{fig:continuum_overview}.

\section{Measurement of the flare parameters}\label{flare_params}
We measure the quiescent luminosity, start and stop times, flux amplitude, peak luminosity, and total energy of each flare using the light curves produced for each line and for the continuum bands as follows.

\subsection{Measuring quiescent luminosity for the lines and continuum bands}\label{quiescent_lumin}
We use the out of flare times to measure the quiescent luminosity $L_0$ in erg s$^{-1}$ for each line and for the continuum. For the lines, we sum the calibrated spectrum across all wavelength points identified within a given line and then scale the flux received by the detector to the flux emitted at the surface of the star using the stellar distance $d_{*}$=12.43$\pm$0.02 pc \citep{Chambers:2016}. We also multiply out the wavelength dependence using the effective wavelength coverage for each point in the line. We propagate the formal errors of the spectrum and stellar distance. Because most integrations during the TRAPPIST-1f visit are affected by flares, we use the quiescent integrations in the TRAPPIST-1b spectrum to measure $L_0$ for both sets of lines. The number of wavelength points within the lines of the stronger F3 flare are sometimes larger than for the smaller F4 flare, so we do not use the $L_0$ values from the F3 event directly. Rather, we visually compare the wavelength solution and number of points in each line between the spectra of the TRAPPIST-1b and f visits. We verify the same wavelength points are used for the quiescent spectrum that were used to measure the line emission in F3.

Quiescent luminosities are also measured for the TESS, J, H, and Ks continuum bands of each flare. The calibrated spectrum during each integration is scaled to the stellar distance and convolved with the response function of each band. Since the wavelength resolution of both NIRISS and NIRSpec changes significantly across each bandpass, we multiply the spectrum by the effective wavelength coverage of each point prior to summation. The fluxes of the convolved spectrum are summed and the uncertainties propagated to the final $L_0$ value. For both the lines and continuum bands, we re-compute the quiescence luminosity using the alternate flux calibration $L_{0,alt}$ and take the difference in luminosity computed with the two flux calibrations, $\Delta L_0$=$L_0$-$L_{0,alt}$, to assess systematic uncertainty. If $\Delta L_0$ is larger than the formal uncertainty, $\mid \Delta L_0 \mid$ is adopted as the final uncertainty value for $L_0$.

\subsection{Measuring flare amplitudes and energies from the fractional flux light curves}\label{basic_fl_properties}
The start and stop times of each flare are identified visually as excursions above the photometric noise. Fractional flux is computed as $\Delta$F/F=$\frac{F-F_0}{F_0}$ where $F_0$ is the median of the out-of-flare flux values \citep{Hawley:2014}. Small precursor events and small sympathetic events following the main flare are excluded from the median calculation. In the F3 NIRISS flare and the F1 and F2 NIRSpec flares, small flux enhancements during the gradual decay tails of flares that occurred prior to the beginning of the JWST observations are also excluded. The remaining out-of-flare flux values in the light curve are varied within their photometric uncertainties and the median flux is recomputed across 10,000 Monte Carlo (MC) trials. In each trial, 10\% of the light curve is dropped to reduce biases from periods of variability in the quiescent flux. Uncertainties in the median flux of the light curve $\sigma_{F_0}$ are computed as the 1$\sigma$ confidence interval of the resulting distribution.

The equivalent duration (ED) is defined as the area under the fractional flux light curve between the start and stop times in units of seconds \citep{Hawley:2014}. We measure the ED by summing the fractional flux values of each integration multiplied by the cadence. Uncertainty in the median out-of-flare flux $F_0$ propagates to the ED since lower values of $F_0$ produces higher normalized flux values. In order to fully account for this systematic source of uncertainty, the ED is recomputed for different $F_0$ values across 1000 MC trials. In each trial, the fractional flux is computed for a given median $F_{0,i}$ drawn from a Gaussian distribution of mean $F_0$ and standard deviation $\sigma_{F_0}$. The flux values in the resulting light curve are then varied within their photometric uncertainties and the ED is measured. Finally, the uncertainty in ED is computed as the 1$\sigma$ confidence interval of the distribution of ED values. 

The peak flux amplitude of each flare $A_\mathrm{fl}$ is defined as the maximum value of the flare peak in the fractional flux light curve, and the peak luminosity of the flare is defined as the quiescent luminosity of the line or band in erg s$^{-1}$ times the peak amplitude, $L_\mathrm{fl}$=$L_0\times A_\mathrm{fl}$ in units of erg s$^{-1}$. The total energy emitted by the flare in a given line or band in erg is defined as $E_{fl}$=$L_0\times ED$. Uncertainties in $L_0$ and ED are propagated to the luminosity and energy. The values of each flare parameter are given for the lines in Table \ref{tab:line_table} and for the continuum bands in Table \ref{tab:continuum_table}.

\begin{table}
\centering
\caption{Line emission properties of NIR flares from TRAPPIST-1}
\begin{tabular}{cccccccccccc}

\hline
\hline
Flare & Line & $\lambda$ & $\Delta\lambda$   & $\sigma_{line}$ & $\sigma_{lcv}$ & $t_\mathrm{rel}$ & $A_\mathrm{fl}$ & log $L_0$ & ED & log $E_\mathrm{line}$ & log $L_\mathrm{line}$ \\
      &      & $\mu$m     & $\AA$  &  &  & s & $\Delta$F/F & erg s$^{-1}$  & s         & erg & erg s$^{-1}$ \\
\hline
F1$^\dag$ & H$\alpha$ & 0.6565 & 55 & 5.1$\sigma$ & 8.6$\sigma$ & 0.0 & 0.34$\pm$0.03 & 26.68$\pm$0.02 & 540$\pm$9 & 29.42$\pm$0.02 & 26.21$\pm$0.04 \\
F2$^\dag$ & H$\alpha$ & 0.6565 & 55 & 8.7$\sigma$ & 5.6$\sigma$ & 0.0 & 0.22$\pm$0.02 & 26.68$\pm$0.02 & 240$\pm$5 & 29.07$\pm$0.02 & 26.02$\pm$0.05 \\
\hline
F3 & H$\alpha$ & 0.6565 & 9 & 32$\sigma$ & 22$\sigma$ & 0.0 & 1.74$\pm$0.04 & 26.27$\pm$0.02 & 4400$\pm$100 & 29.91$\pm$0.02 & 26.51$\pm$0.02 \\
F3 & He I & 0.7062 & 12 & 2.1$\sigma$ & 5.3$\sigma$ & -3.4 & 0.125$\pm$0.02 & 25.86$\pm$0.001 & 43$\pm$13 & 27.49$\pm$0.14 & 24.96$\pm$0.07 \\
F3 & Ca II IRTb & 0.8544 & 14 & 2.4$\sigma$ & 4.4$\sigma$ & - & 0.014$\pm$0.003 & 26.81$\pm$0.03 & 13.2$\pm$2.2 & 27.93$\pm$0.08 & 24.97$\pm$0.10 \\
F3 & P$\epsilon$ & 0.9549 & 28 & 2.3$\sigma$ & 3.6$\sigma$ & -3.4 & 0.008$\pm$0.002 & 27.59$\pm$0.02 & 7.8$\pm$2.3 & 28.48$\pm$0.13 & 25.49$\pm$0.10 \\
F3 & P$\delta$ & 1.0052 & 28 & 4.0$\sigma$ & 4.8$\sigma$ & -1.7 & 0.009$\pm$0.002 & 27.54$\pm$0.01 & 7.2$\pm$2.7 & 28.4$\pm$0.2 & 25.48$\pm$0.08 \\
F3 & He I IRT & 1.0833 & 19 & 5.1$\sigma$ & 8.8$\sigma$ & 0.0 & 0.007$\pm$0.001 & 27.65$\pm$0.001 & 8.4$\pm$0.9 & 28.57$\pm$0.05 & 25.46$\pm$0.09 \\
F3 & P$\gamma$ & 1.0941 & 27 & 6.1$\sigma$ & 10.1$\sigma$ & -1.7 & 0.009$\pm$0.001 & 27.84$\pm$0.01 & 7.3$\pm$1.1 & 28.70$\pm$0.07 & 25.8$\pm$0.04 \\
F3 & P$\beta$ & 1.2822 & 19 & 10.2$\sigma$ & 11.6$\sigma$ & -1.7 & 0.009$\pm$0.001 & 27.81$\pm$0.01 & 10.2$\pm$0.9 & 28.82$\pm$0.04 & 25.75$\pm$0.05 \\
F3 & P$\alpha$ & 1.8756 & 18 & 8.4$\sigma$ & 7.2$\sigma$ & 0.0 & 0.014$\pm$0.002 & 27.29$\pm$0.01 & 22.8$\pm$2.2 & 28.65$\pm$0.04 & 25.42$\pm$0.05 \\
F3 & Br$\beta$ & 2.6259 & 21 & 3.0$\sigma$ & 3.6$\sigma$ & -3.4 & 0.011$\pm$0.004 & 26.84$\pm$0.001 & 24.7$\pm$3.8 & 28.23$\pm$0.07 & 24.88$\pm$0.14 \\
\hline
F4 & H$\alpha$ & 0.6565 & 8 & 12$\sigma$ & 8.5$\sigma$ & 0.0 & 0.28$\pm$0.02 & 26.52$\pm$0.02 & 320$\pm$20 & 29.03$\pm$0.03 & 25.97$\pm$0.04 \\
F4 & Ca II IRTc & 0.8665 & 6 & 4.1$\sigma$ & 2.6$\sigma$ & -1.7  & 0.02$\pm$0.01 & 27.03$\pm$0.01 & 23$\pm$7 & 28.39$\pm$0.14 & 25.37$\pm$0.17 \\
F4 & He I IRT & 1.0833 & 9 & 3.3$\sigma$ & 3.6$\sigma$ & 0.0 & 0.006$\pm$0.001 & 27.35$\pm$0.001 & 3.8$\pm$1.2 & 27.93$\pm$0.14 & 25.11$\pm$0.12 \\
F4 & P$\beta$ & 1.2822 & 12 & 2.7$\sigma$ & 2.3$\sigma$ & - & 0.003$\pm$0.001 & 27.34$\pm$0.001 & 3.4$\pm$1.2 & 27.87$\pm$0.16 & 24.86$\pm$0.20 \\
F4 & P$\alpha$ & 1.8756 & 6 & 3.8$\sigma$ & 3.1$\sigma$ & - & 0.008$\pm$0.003 & 26.82$\pm$0.01 & 4.2$\pm$1.7 & 27.44$\pm$0.19 & 24.75$\pm$0.16 \\
\hline
\end{tabular}
\label{tab:line_table}
{\newline\newline \textbf{Notes.} Properties of the line emission of each flare. For each flare and line, the central wavelength, FWHM of the line, significance of the spectral line relative to the spectral continuum to either side, significance of the flare peak in the light curve, time of the peak relative to the H$\alpha$ peak time in min, amplitude in fractional flux, quiescent luminosity in erg s$^{-1}$, total flare energy in erg, and maximum luminosity in erg s$^{-1}$ are given. A dagger denotes the flare was observed with NIRSpec, while no dagger denotes it was observed with NIRISS. For multi-peaked flares, the first peak is used to compute $t_\mathrm{rel}$. A machine readable version of the table is available.}
\end{table}

\begin{table}
\centering
\caption{Continuum emission properties of NIR flares from TRAPPIST-1}
\begin{tabular}{ccccccccccc}

\hline
\hline
Flare & Band & $\lambda$ & $\sigma_{lcv}$ & $t_\mathrm{rel}$ & $A_\mathrm{fl}$ & log $L_0$ & ED & log $E_\mathrm{band}$ & log $L_\mathrm{band}$ & T$_\mathrm{eff}$ \\
      &      & $\mu$m    &                & min              & $\Delta$F/F     & erg s$^{-1}$ & s  & erg & erg s$^{-1}$ & K \\
\hline
F1$^\dag$ & TESS & 0.787 & 66$\sigma$ & -1.5 & 0.1061$\pm$0.0006 & 29.37$\pm$0.02 & 37.0$\pm$0.4 & 30.94$\pm$0.02 & 28.40$\pm$ 0.02 & 5300$\pm$400 \\
F1$^\dag$ & J & 1.235 & 13$\sigma$ & -1.5 & 0.0066$\pm$0.0001 & 29.55$\pm$0.004 & 10.6$\pm$0.2 & 30.58$\pm$0.01 & 27.37$\pm$0.01 & 5300$\pm$400 \\
F1$^\dag$ & H & 1.662 & 11$\sigma$ & -1.5 & 0.0058$\pm$0.0001 & 29.53$\pm$0.001 & 8.8$\pm$0.2 & 30.48$\pm$0.01 & 27.30$\pm$0.01 & 5300$\pm$400 \\
F1$^\dag$ & Ks & 2.159 & 14$\sigma$ & -1.5 & 0.0041$\pm$0.0003 & 29.24$\pm$0.002 & 8.3$\pm$0.2 & 30.15$\pm$0.01 & 26.85$\pm$0.03 & 5300$\pm$400 \\
\hline
F2$^\dag$ & TESS & 0.787 & 11$\sigma$ & -1.0 & 0.0178$\pm$0.0004 & 29.37$\pm$0.02 & 15.3$\pm$0.1 & 30.56$\pm$0.02 & 27.62$\pm$0.02 & 4400$\pm$400 \\
F2$^\dag$ & J & 1.235 & 5.4$\sigma$ & 0.0 & 0.0027$\pm$0.0001 & 29.55$\pm$0.004 & 3.8$\pm$0.2 & 30.13$\pm$0.02 & 26.98$\pm$0.02 & 4400$\pm$400 \\
F2$^\dag$ & H & 1.662 & 5.5$\sigma$ & 1.9 & 0.0024$\pm$0.0001 & 29.53$\pm$0.001 & 3.4$\pm$0.2 & 30.07$\pm$0.02 & 26.92$\pm$0.02 & 4400$\pm$400 \\
F2$^\dag$ & Ks & 2.159 & 8$\sigma$ & 3.4 & 0.0022$\pm$0.0002 & 29.24$\pm$0.002 & 3.4$\pm$0.2 & 29.77$\pm$0.02 & 26.58$\pm$0.05 & 4400$\pm$400 \\ 
\hline
F3 & TESS & 0.787 & 9.4$\sigma$ & 0.0 & 0.0069$\pm$0.0005 & 29.39$\pm$0.02 & 9.0$\pm$0.8 & 30.35$\pm$0.04 & 27.23$\pm$0.04 & 3500$\pm$200 \\
F3 & J & 1.235 & 9.0$\sigma$ & 0.0 & 0.0018$\pm$0.0001 & 29.56$\pm$0.002 & 5.6$\pm$0.2 & 30.31$\pm$0.01 & 26.81$\pm$0.03 & 3500$\pm$200 \\
F3 & H & 1.662 & 5.7$\sigma$ & 0.0 & 0.0014$\pm$0.0001 & 29.54$\pm$0.001 & 7.3$\pm$0.2 & 30.4$\pm$0.01 & 26.7$\pm$0.04 & 3500$\pm$200 \\
F3 & Ks & 2.159 & 4.7$\sigma$ & 3.4 & 0.0012$\pm$0.0002 & 29.23$\pm$0.001 & 7.5$\pm$0.4 & 30.11$\pm$0.02 & 26.33$\pm$0.08 & 3500$\pm$200 \\
\hline
F4 & TESS & 0.787 & 4.8$\sigma$ & 0.0 & 0.002$\pm$0.0004 & 29.4$\pm$0.01 & 3.2$\pm$0.4 & 29.9$\pm$0.06 & 26.7$\pm$0.1 & 2900$\pm$100 \\
F4 & J & 1.235 & 3.3$\sigma$ & 1.7 & 0.0012$\pm$0.0001 & 29.56$\pm$0.001 & 1.2$\pm$0.1 & 29.6$\pm$0.04 & 26.64$\pm$0.04 & 2900$\pm$100 \\
F4 & H & 1.662 & 2.5$\sigma$ & 1.7 & 0.0009$\pm$0.0001 & 29.54$\pm$0.001 & 0.9$\pm$0.1 & 29.5$\pm$0.05 & 26.5$\pm$0.1 & 2900$\pm$100 \\
F4 & Ks & 2.159 & 3.1$\sigma$ & 3.4 & 0.0012$\pm$0.0002 & 29.23$\pm$0.001 & 0.4$\pm$0.2 & 28.8$\pm$0.2 & 26.3$\pm$0.1 & 2900$\pm$100 \\
\hline
\end{tabular}
\label{tab:continuum_table}
{\newline\newline \textbf{Notes.} Properties of the continuum emission of each flare. For each flare and band, the central wavelength of the filter, significance of the flare peak in the light curve, time of the peak relative to the H$\alpha$ peak time in min, amplitude in fractional flux, quiescent luminosity in erg s$^{-1}$, total flare energy in erg, maximum luminosity in erg s$^{-1}$, and effective blackbody temperature at peak in K are given. A dagger denotes the flare was observed with NIRSpec, while no dagger denotes it was observed with NIRISS. For NIRSpec flares, $t_\mathrm{rel}$ is computed in 30 s bins for NIRSpec to reduce scatter around the peak time. For multi-peaked flares, the first peak is used to compute $t_\mathrm{rel}$. A machine readable version of the table is available.}
\end{table}

\section{Results on the NIR characteristics of the flares}\label{results}
Small-scale variability in the H$\alpha$ line exists throughout the spectroscopic time series. We identify a sample of four flares of sufficient size for multi-wavelength characterization, F1--F4. These flares produce detectable levels of continuum and line emission in the NIR and emit approximately $E_\mathrm{H\alpha}$=10$^{29}$ erg and $E_\mathrm{TESS}$=10$^{30}$ erg. Two overlapping flares F1 and F2 are detected with NIRSpec, which we consider as separate events since the peak phases of the flares are distinct and therefore should represent two heating episodes in the stellar atmosphere with different continuum formation temperatures. We likewise identify two flares in the NIRISS data during the TRAPPIST-1f visit and the second TRAPPIST-1b visit, F3 and F4, respectively. Among the flares, F3 is a truly exceptional event in which the star increased in brightness by a factor of 2.74 in the H$\alpha$ line and overlapped the ingress and flat bottom of the TRAPPIST-1f transit. The majority of the NIR lines investigated in this work occurred during the F3 event.

\subsection{Flare Line Discoveries and Properties}\label{result_lines}
Seven lines are securely detected at NIR wavelengths of 0.7--3~$\mu$m from the F3 event and two lines are securely detected from F4 (Fig. \ref{fig:jwst_overview}). A number of hydrogen lines are securely detected for the F3 event, which are P$\alpha$, P$\beta$, P$\gamma$, P$\delta$, and Br$\beta$. Two helium lines are securely detected from F3, He I $\lambda$0.7062~$\mu$m and He I $\lambda$1.0833~$\mu$m IRT. Securely detected lines for the F4 event include the He I IRT and P$\alpha$ lines. Lines that exceed 3$\sigma$ in either the $\sigma_\mathrm{line}$ or $\sigma_\mathrm{lcv}$ measurements and are considered weakly detected are the Ca II IRTb and P$\epsilon$ lines for the F3 event and Ca II IRTc and P$\beta$ lines for the F4 event. The H$\alpha$ line is strongly detected in all four flares. The NIR lines are connected to the optical by the H$\alpha$ detections, as H$\alpha$ is the primary line used to assess flares from low-mass stars \citep{Kanodia:2022}.

\subsubsection{The Balmer, Paschen, and Brackett lines}\label{hydrogen_results}
\begin{figure*}
	\centering
        \subfigure
	{
		\includegraphics[trim=30 10 30 10, width=0.48\textwidth]{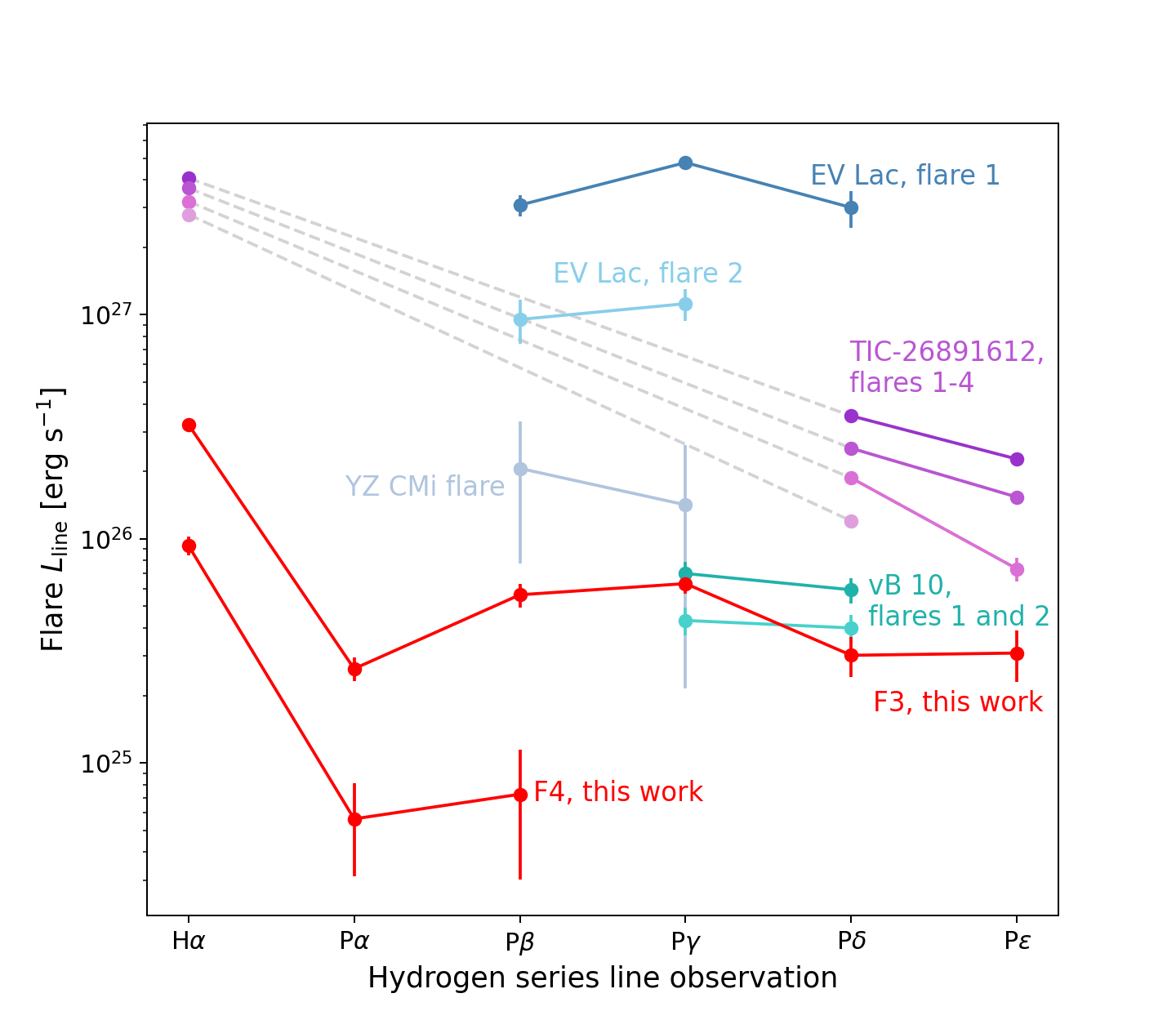}
	}
        \subfigure
	{
		\includegraphics[trim=30 10 30 10, width=0.48\textwidth]{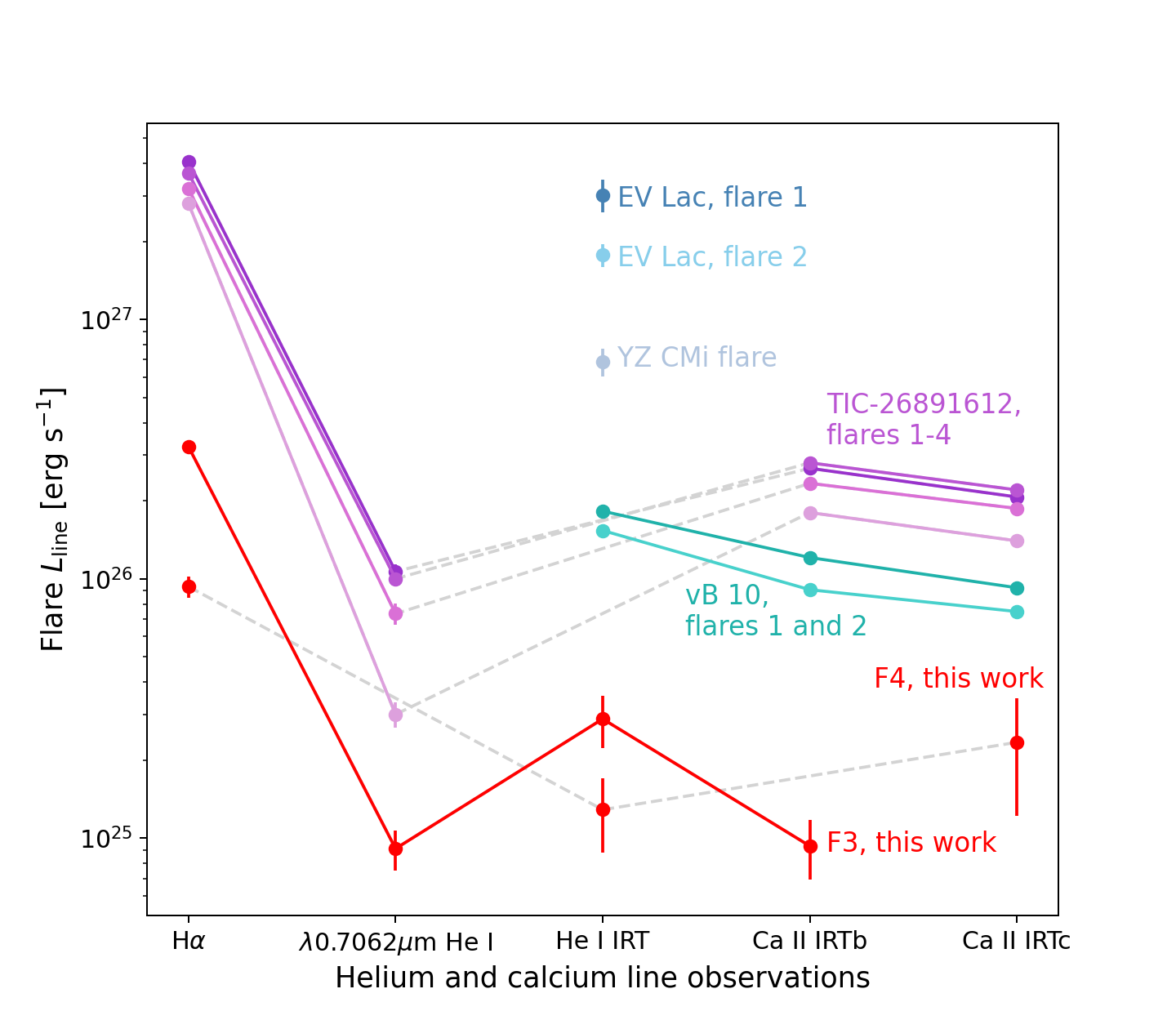}
	}
	\caption{Left: Luminosity emitted in each hydrogen line during the flare peak. Other flares with multiple NIR lines detected simultaneously are displayed for reference \citep{Liebert:1999, Schmidt:2012, Kanodia:2022}. Both our flares and the strongest EV Lac flare indicate that higher P$\gamma$ luminosities might be a typical feature of NIR flares. Right: The same as the left panel, but for the helium and calcium lines.}
	\label{fig:line_energy_partition}
\end{figure*}

In general, a decrement in line energy is expected for the higher order transitions within each series. The Paschen decrement has been confirmed in several flares at and above the 6$\rightarrow$3 P$\gamma$ transition \citep{Kanodia:2022}. However, \citet{Schmidt:2012} reported a higher peak luminosity for P$\gamma$ than $P\beta$ for two large flares from EV Lac, which raises the intriguing possibility that the lower energy states may not be energetically favored. As shown in the left panel of Figure \ref{fig:line_energy_partition}, we observe a reversed Paschen decrement for P$\alpha$-P$\gamma$ before the luminosities decrease again from P$\gamma$ to P$\delta$-P$\epsilon$. P$\gamma$ appears to sit at the peak of the energy partition curve in Figure \ref{fig:line_energy_partition}, but it is worth emphasizing that the P$\beta$ and P$\gamma$ luminosities are within 1$\sigma$. Likewise, the P$\delta$ and P$\epsilon$ line luminosities are the same within errors, consistent with the Paschen decrement observed for P$\gamma$-P$\epsilon$ in the literature (Fig. \ref{fig:line_energy_partition}). The wings of He I IRT line 107$\AA$ away are a potential source of line contamination for P$\gamma$. However, the majority of the flux at 1.0941~$\mu$m in Figure \ref{fig:NIRISS_line_centers} is clearly associated with the P$\gamma$ line since the two lines are 4 FWHM apart. Flares with multiple Paschen lines in the literature are plotted for comparison to F3 and F4. These include two flares from the M4 dwarf EV Lac and one flare from the M4 dwarf YZ CMi \citep{Schmidt:2012}, two flares from the M8 dwarf vB 10 \citep{Kanodia:2022}, and four flares from the M9.5 dwarf TIC 26891612 \citep{Liebert:1999}. The largest flare from EV Lac shows a similar increase in the P$\gamma$ luminosity relative to P$\beta$ and P$\delta$ seen in our flares. This is significant for two reasons. First, the similar energy partition of the Paschen series in flares from both an M4 and an ultra-cool dwarf strengthens the case this behavior could be typical. Second, the resolution of the spectra in \citet{Schmidt:2012} is much higher than NIRISS. The separation between the He I IRT and P$\gamma$ lines is fully resolved in Figure 1 of \citet{Schmidt:2012}.

\begin{figure*}
	\centering
        \subfigure
	{
		\includegraphics[trim=10 10 10 10, width=0.48\textwidth]{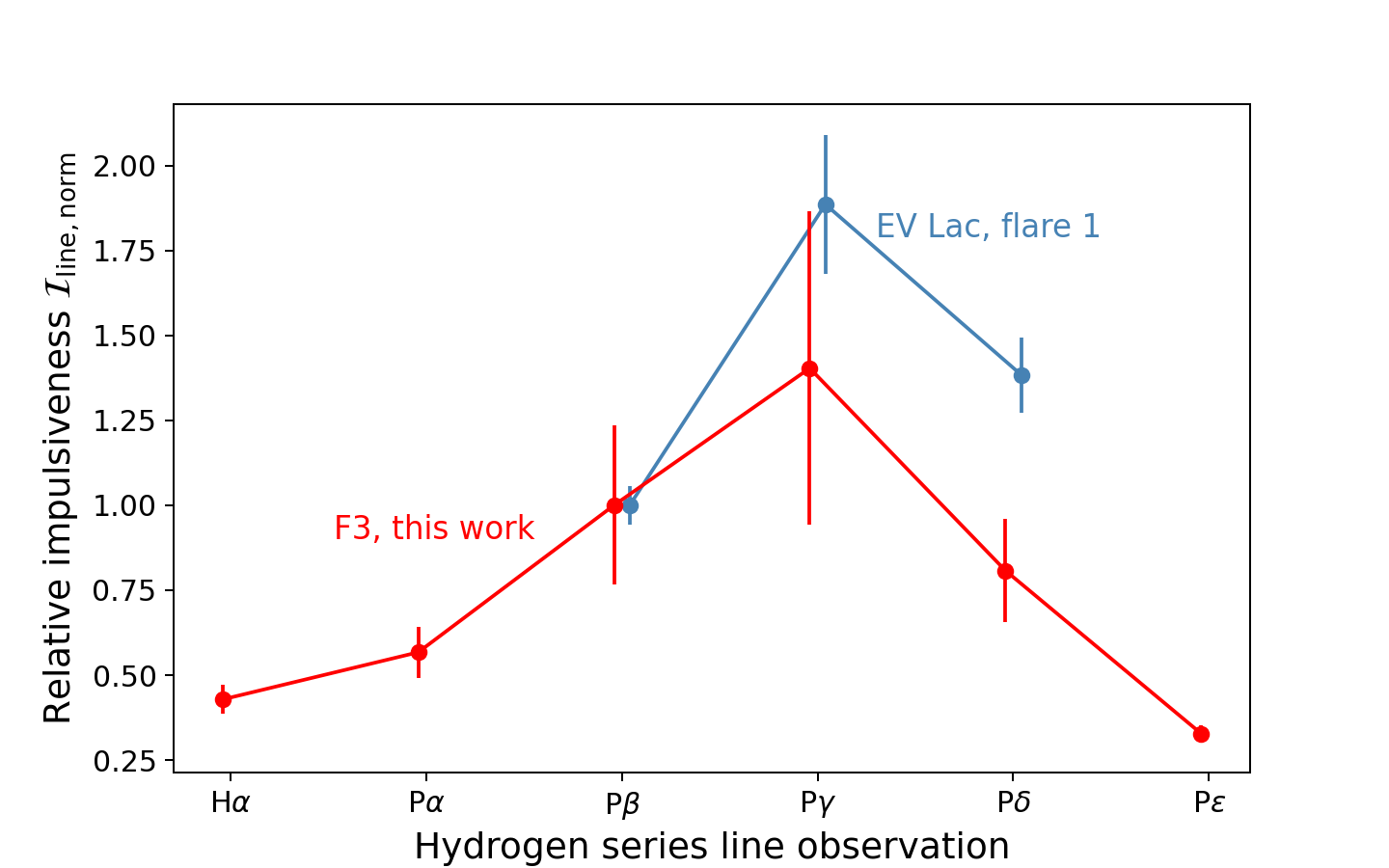}
	}
        \subfigure
	{
		\includegraphics[trim=10 10 10 10, width=0.48\textwidth]{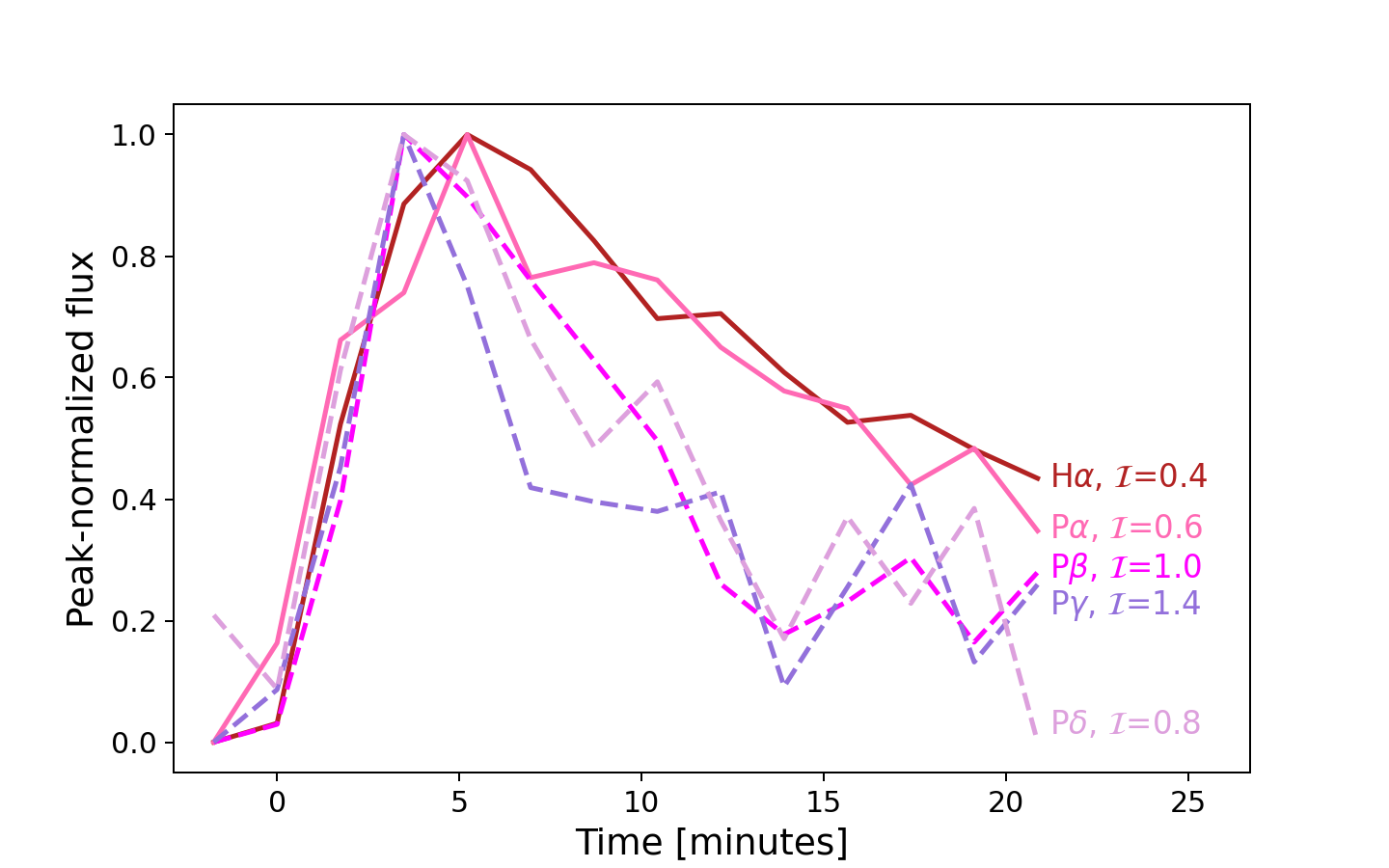}
	}
	\caption{Left: The relative impulsiveness of the flare light curve for each hydrogen line in our F3 event (red) and the largest EV Lac flare from \citet{Schmidt:2012}. More impulsive flares have light curves that are spiky, while less impulsive flares have light curves that are gradual. To facilitate comparisons with different flare light curves of different amplitudes and FWHM times, we normalize amplitudes to one and FWHM times by the FWHM of the P$\beta$ line of each flare prior to computing the impulsiveness index $\mathcal{I}_\mathrm{line,norm}=A_\mathrm{fl,line}/FWHM_\mathrm{line}$. The impulsiveness of both flares reaches a maximum at P$\gamma$, suggesting that its excitation period is brief and intense relative to the lower and higher order transitions. Right: Light curves of the F3 lines are peak-normalized to show changes in the light curve shape across the Paschen series, described by the line-normalized impulse $\mathcal{I}_\mathrm{line,norm}$. P$\alpha$ and H$\alpha$ are shown with solid lines to emphasize their similar shapes, while the other lines are shown with dashed lines.}
	\label{fig:line_impulse_lcvs}
\end{figure*}

The shape of the flare light curve also changes from P$\alpha$ to P$\epsilon$. The change in shape can be seen most clearly when the flare light curves are normalized to one, as in the right panel of Figure \ref{fig:line_impulse_lcvs}. Here, the peaks and FWHM times of the H$\alpha$ and P$\alpha$ light curves evolve slower than do those of P$\beta$ and P$\gamma$. We quantify this change using the impulsiveness index of \citet{Kowalski:2013}, defined as the peak amplitude of a flare in fractional-flux units over its FWHM time in minutes, $\mathcal{I}_\mathrm{fl}$=$A_\mathrm{fl}$/$\Delta t_\mathrm{FWHM}$ \citep{Kowalski:2013}. Spiky flares have a high impulsiveness index while more gradual flares have a low impulsiveness index. We modify this definition and define the line-normalized impulsiveness index $\mathcal{I}_\mathrm{line,norm}$ to account for changes in the flare luminosity at different wavelengths. For example, the H$\alpha$ and P$\alpha$ light curves are nearly identical when their peaks are normalized to one (Fig. \ref{fig:line_impulse_lcvs}), but the 100$\times$ greater flux in H$\alpha$ gives it a 100$\times$ higher impulsiveness index than P$\alpha$. We therefore normalize the flare light curves to one before computing $\mathcal{I}_\mathrm{line,norm}$. In a similar fashion, different flares last for different amounts of time. The different evolution timescales of different flares are captured by the FWHM time $\Delta t_\mathrm{FWHM}$ and set the y-scale of the impulsiveness index. To facilitate comparison between different flares, we normalize the FWHM times of the light curves for each line in our F3 event and for the larger EV Lac flare from \citet{Schmidt:2012} by the FWHM time of the P$\beta$ line. P$\beta$ is chosen as the lowest-order line available in both data sets. Finally, we compute $\mathcal{I}_\mathrm{line,norm}$ for the flares once the amplitudes are set to one and the widths are in FWHM units rather than minutes. The results are plotted for our F3 event and the large EV Lac flare from \citet{Schmidt:2012} in the left panel of Figure \ref{fig:line_impulse_lcvs}. We see that the impulsiveness increases from P$\alpha$ through P$\gamma$ before decreasing again from P$\gamma$ through P$\epsilon$.

\begin{figure*}
	\centering
        \subfigure
	{
		\includegraphics[width=0.48\textwidth]{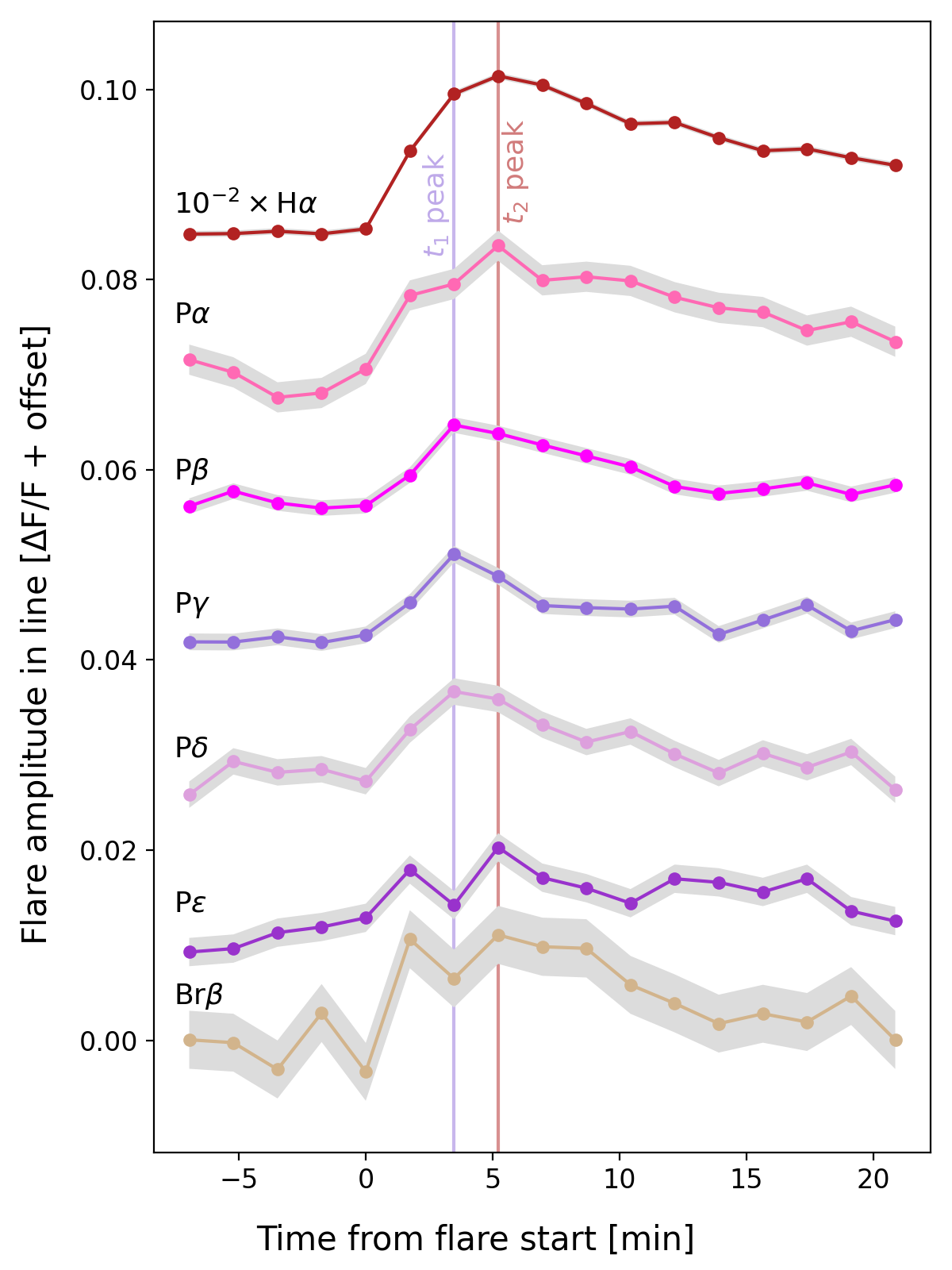}
	}
        \subfigure
	{
		\includegraphics[width=0.48\textwidth]{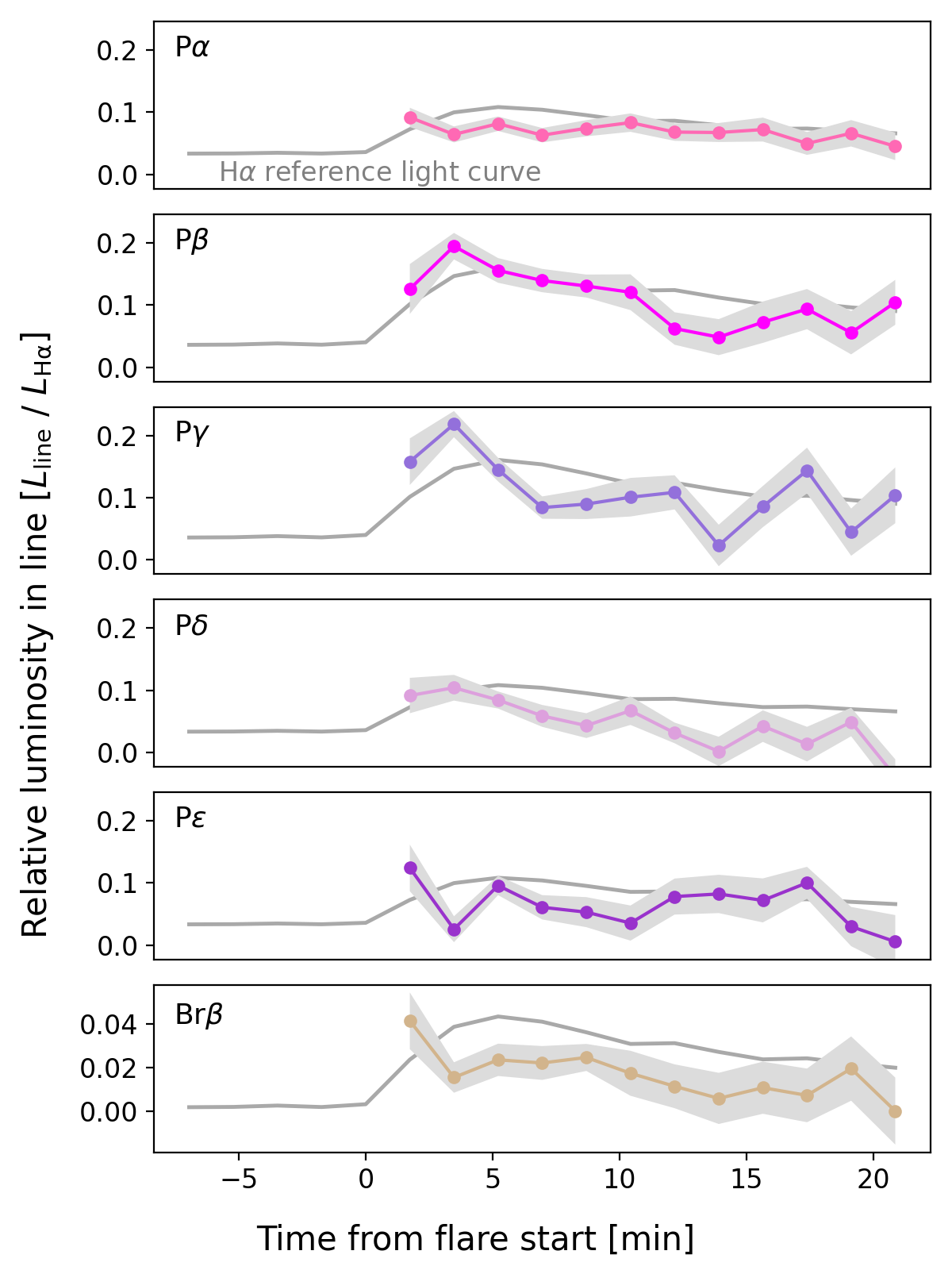}
	}
	\caption{Left: Relative timing of hydrogen line emission during the F3 flare. The faster rise of the higher order P$\beta$-P$\delta$ lines likely reflects a shorter period of excitation compared to the H$\alpha$ and P$\alpha$ lines. Note H$\alpha$ has been scaled down by 100$\times$ to facilitate comparison with the weaker lines. Right: Relative luminosity of the hydrogen lines during the F3 flare, normalized by the luminosity of H$\alpha$. P$\beta$ and P$\gamma$ exhibit both the highest peak luminosity and fastest decay. The H$\alpha$ light curve is shown for reference.}
	\label{fig:relative_line_results}
\end{figure*}

Next, we look at the rise time for each line, or how long it takes after the flare start time for the light curve to reach its maximum value \citep{Howard_MacGregor:2022}. In the left panel of Figure \ref{fig:relative_line_results}, the P$\alpha$-$P\delta$ lines peak during one of two integrations. The lower order lines peak one integration later than the higher order lines up through P$\delta$. The P$\epsilon$ and Br$\gamma$ light curves have two peaks, although the signal to noise is lower and hence the difference in peak time may not be significant. The first peak of the P$\epsilon$ and Br$\gamma$ light curves appears to occur an integration earlier than P$\gamma$-P$\delta$ and two integrations prior to H$\alpha$ and P$\alpha$, while the second peak is coincident with H$\alpha$ and P$\alpha$. In the right panel, the relative evolution of the luminosities is shown. We normalize the luminosity light curves by the H$\alpha$ light curve to search for departures from the dominant transition of hydrogen. The $L_\mathrm{line}/L_\mathrm{Ha}$ ratios of the Paschen lines are $\sim$0.1 for most integrations, while the Br$\gamma$ ratio is $\sim$0.02. The P$\beta$ and P$\gamma$ lines diverge most strongly from this value during the flare peak compared to the other lines. P$\epsilon$ and Br$\beta$ show similar luminosity curves.

Insight into the reversed Paschen decrement, rise times, and impulsiveness of the lines may be gained from a similarly reversed decrement for the Balmer series. \texttt{RADYN} models of the relative intensities of the Balmer series find H$\alpha$ to be weaker than H$\beta$-H$\gamma$, with a peak in intensity at H$\gamma$ \citep{Kowalski:2022a}. This is because the effective temperature of the source contribution function of the H$\gamma$ line is $\sim$1000 K higher than for H$\alpha$, producing more overall flux. On the Sun, the region of the chromosphere where H$\alpha$ is formed is optically thicker than for H$\gamma$, producing greater collisional pressure broadening of the H$\alpha$ line in flares. However, the height from the photosphere at which H$\alpha$ escapes (i.e. at an optical depth of $\tau\approx1$) is greater than for the optically thin H$\gamma$ region \citep{Kowalski:2022a}. If P$\alpha$ is similarly formed at a lower effective temperature in an optically thick region and reaches $\tau\approx$1 at a greater height from the photosphere than P$\gamma$, a reversed decrement would also be expected. Furthermore, this scenario naturally results in the increased impulsiveness of the P$\beta$-P$\gamma$ light curves relative to P$\alpha$ and H$\alpha$. As the precipitating electron beam decays during the flare, fewer electrons would interact with the plasma in the optically thin layer where P$\beta$-P$\gamma$ are formed than the optically thick layer where P$\alpha$ is formed \citep{Kowalski:2017a}. The faster drop in excitation of the P$\beta$-P$\gamma$ lines then produces more impulsive peaks compared to the P$\alpha$ and H$\alpha$ lines where excitation continues for a longer time.

\subsubsection{Helium and calcium lines}\label{helium_results}
The strongest helium line is the He I IRT, which appears as a single line at the resolution of NIRISS. The He I IRT has been observed during a number of solar (e.g. \citealt{You_Oertel:1992, Penn_Kuhn:1995, Zeng:2014, Kerr:2021}) and stellar \citep{Schmidt:2012, Fuhrmeister:2020, Kanodia:2022} flares. In solar flares, He I IRT emission results from recombination after extreme UV (EUV) photoionization, collisional excitation, and collisional ionization in the chromospheric footprints of flare loops \citep{Ding:2005}. The IRT occurs alongside soft X-ray and EUV emission in moderate to large solar flares \citep{Ding:2005}, making it valuable as a diagnostic of the X-ray and EUV radiation environment of exoplanets during transmission spectroscopy \citep{Fuhrmeister:2020}. We also detect the He I $\lambda$0.7062~$\mu$m line seen previously in the \citet{Liebert:1999} and \citet{Fuhrmeister:2008} stellar flares. The 0.7062~$\mu$m He I line is the strongest helium line in the NIR after the IRT, explaining why this is the only other He I line we detect \citep{Liebert:1999}.

In the right panel of Figure \ref{fig:line_energy_partition}, we compare the helium and calcium line luminosity values of the F3 and F4 events against literature values. The H$\alpha$ luminosity is also included as a reference point. Our flares have the lowest luminosity in the He IRT observed to date, 5$\times$ lower than the \citet{Kanodia:2022} flares from vB 10 and 60$\times$ lower than the \citet{Schmidt:2012} flares from EV Lac. The low luminosity of our F3 and F4 flares enables us to confirm that line ratios hold across a range of flare sizes. The four flares from the M9.5 dwarf TIC 26891612 have H$\alpha$ to $\lambda$0.7062~$\mu$m He I ratios  L$_\mathrm{H\alpha}$/L$_\mathrm{He~I}$ of 38.1$\pm$2.5, 36.7$\pm$2.5, 43.6$\pm$4.1 and 93$\pm$11, ordered from the largest flare to the smallest. Our F3 flare has a ratio of L$_\mathrm{H\alpha}$/L$_\mathrm{He~I}$=35.5$\pm$2.6, consistent with the larger three flares from TIC 26891612. For another example, \citet{Kanodia:2022} estimate the H$\alpha$ luminosity of the vB 10 flares from the ratio L$_\mathrm{H\alpha}$/L$_\mathrm{He~I~IRT}$=6.66 since the spectral range of the HPF does not observe both lines. We measure L$_\mathrm{H\alpha}$/L$_\mathrm{He~I~IRT}$ values of 11.2$\pm$2.6 and 7.2$\pm$2.4 for F3 and F4, respectively. If these ratios hold for the larger flares observed from vB 10, the H$\alpha$ emission associated with its flares may be slightly higher than expected.

\begin{figure*}
	\centering
        \subfigure
	{
		\includegraphics[trim=0 -15 0 15, width=0.48\textwidth]{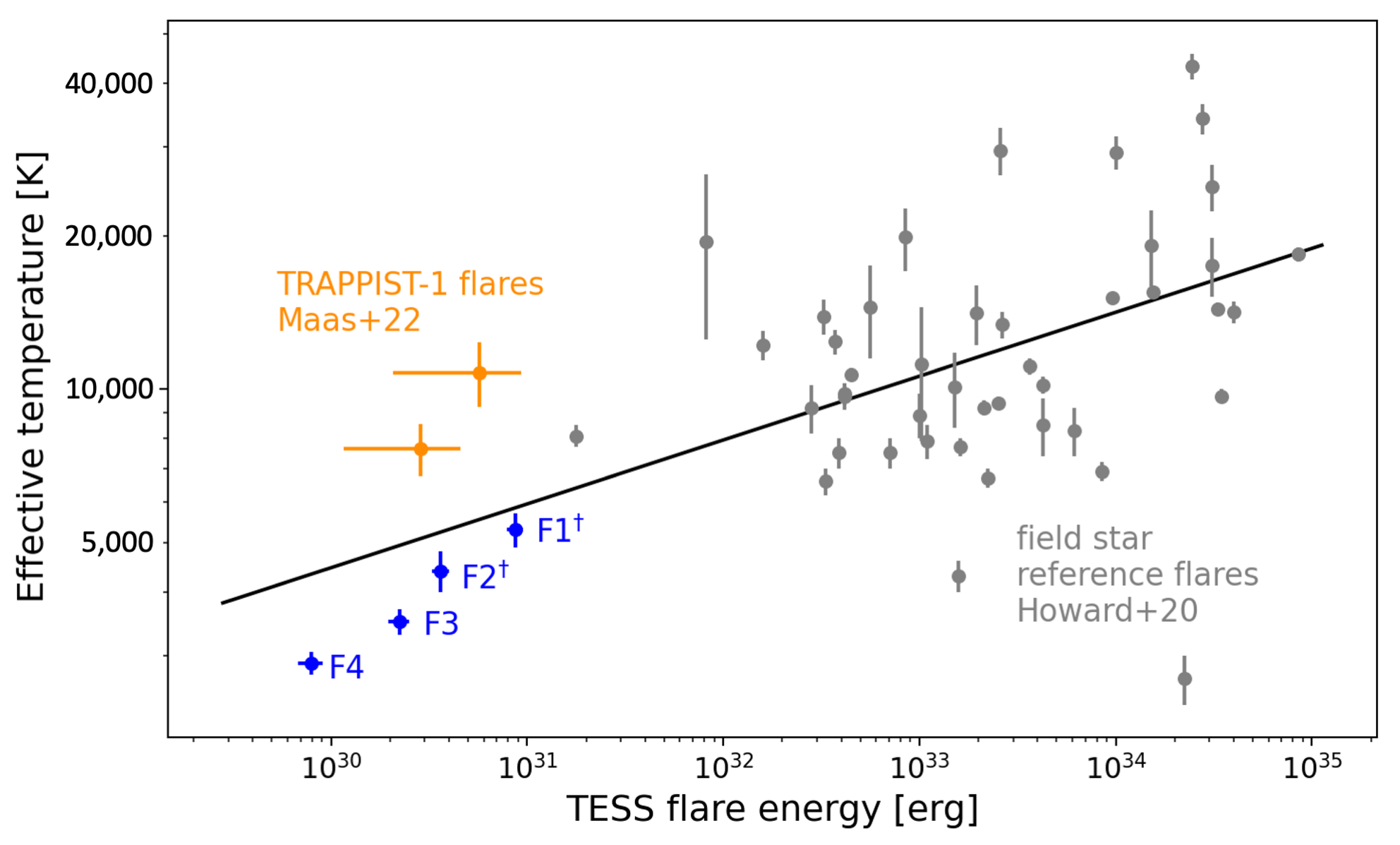}
	}
        \subfigure
	{
		\includegraphics[width=0.48\textwidth]{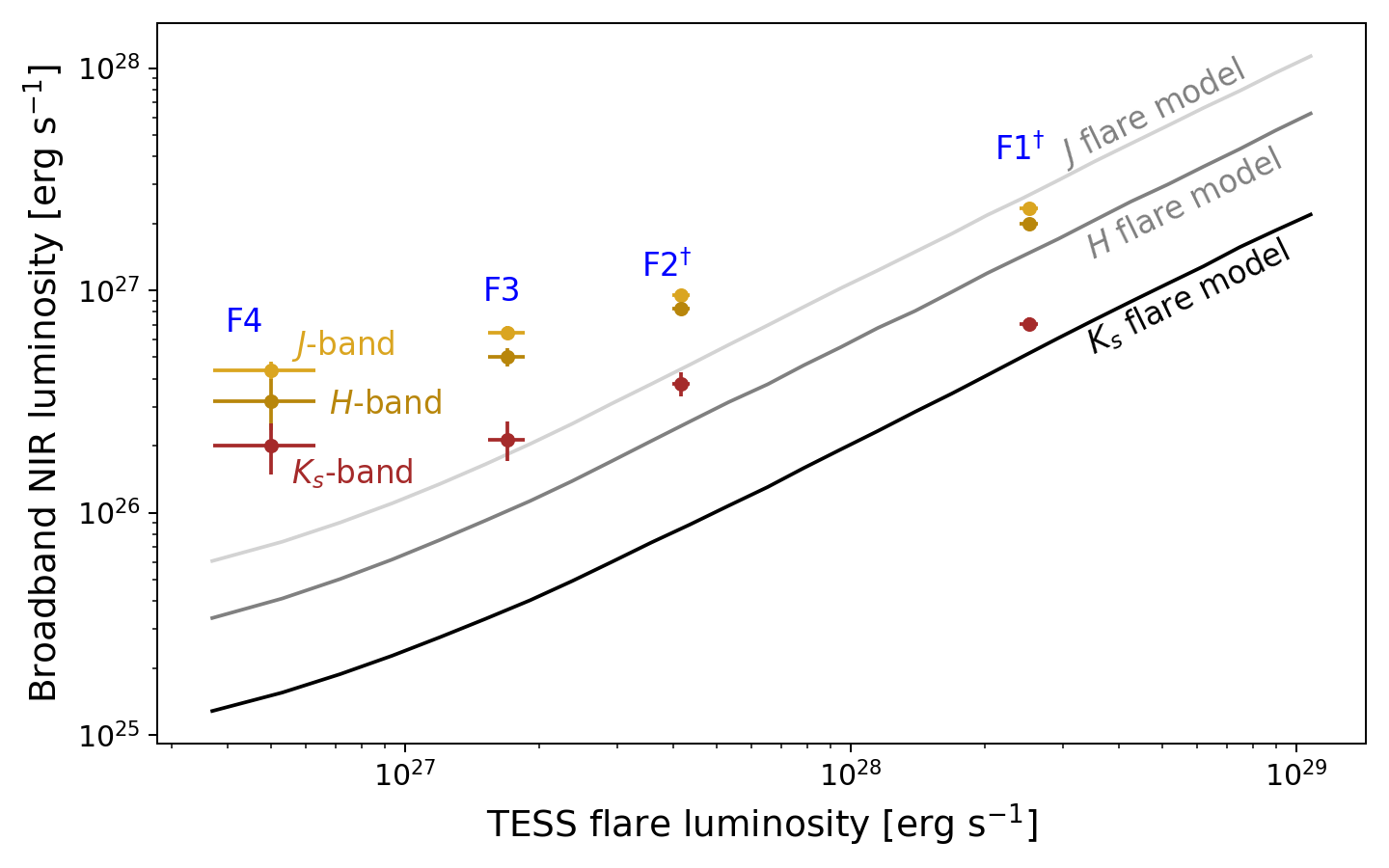}
	}
	\caption{Left: Effective temperature of the flares during the peak phase $T_\mathrm{eff}$ as a function of TESS band energy $E_\mathrm{TESS}$. The $T_\mathrm{eff}$ values are lower than the peak temperatures often observed in earlier M-dwarf flares \citep{Kowalski:2013}, as well as those reported in a sample of TRAPPIST-1 flares by \citet{Maas:2022}. For reference, we display the distribution of peak flare temperatures as a function of TESS energy for early and mid M-dwarfs observed by Evryscope and TESS \citep{Howard:2020}. The $E_\mathrm{TESS}$-$T_\mathrm{eff}$ scaling relationship from \citet{Howard:2020} is also shown in black and qualitatively agrees with the lower temperatures of the TRAPPIST-1 flares. Right: Scaling relations from the TESS band to the 2MASS broadband luminosities during the flare peaks. Model predictions from \citet{Davenport:2012} are shown for reference.}
	\label{fig:continuum_scaling}
\end{figure*}

\begin{figure*}
	\centering
        \subfigure
	{
		\includegraphics[width=0.48\textwidth]{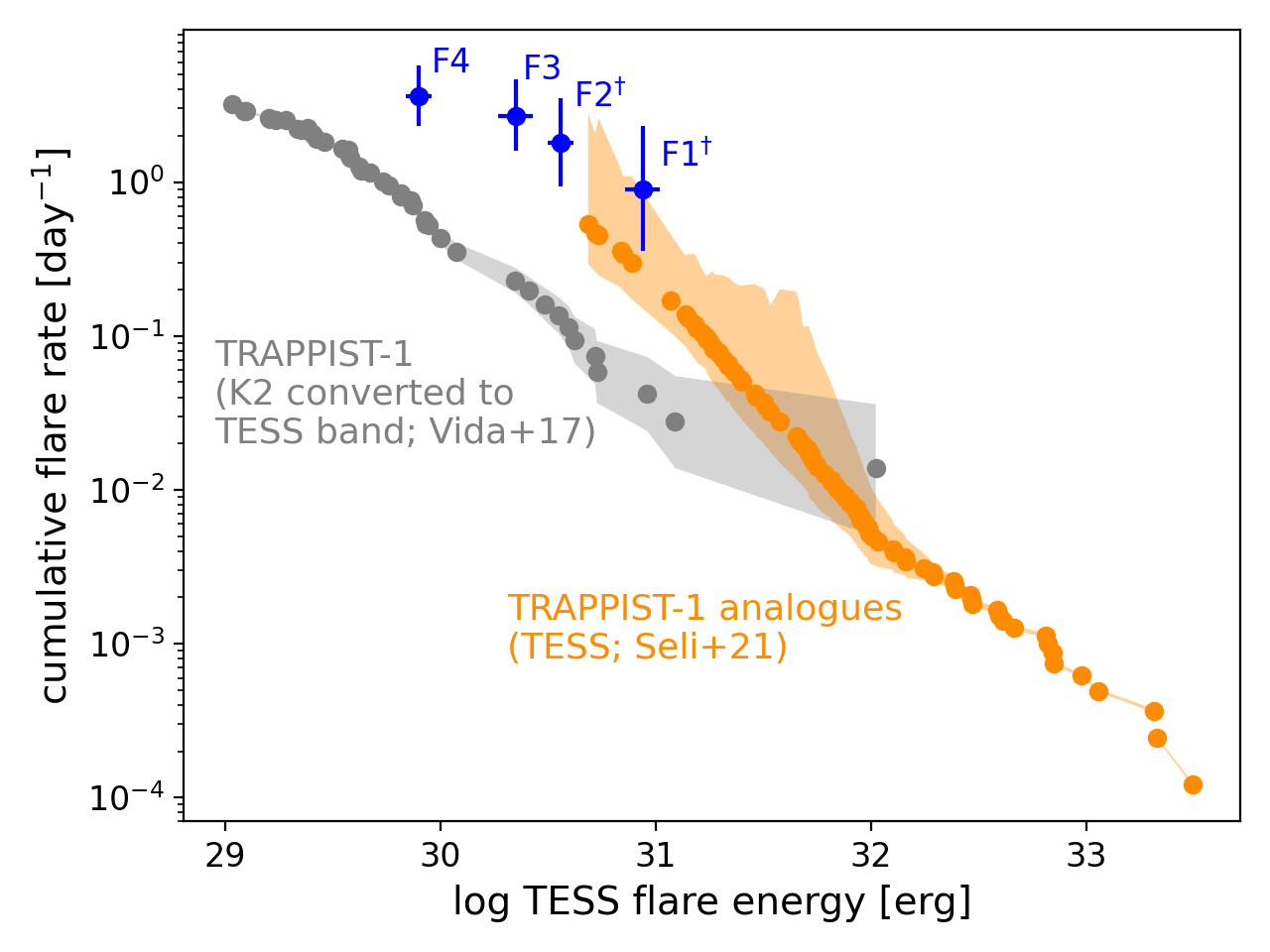}
	}
        \subfigure
	{
		\includegraphics[width=0.48\textwidth]{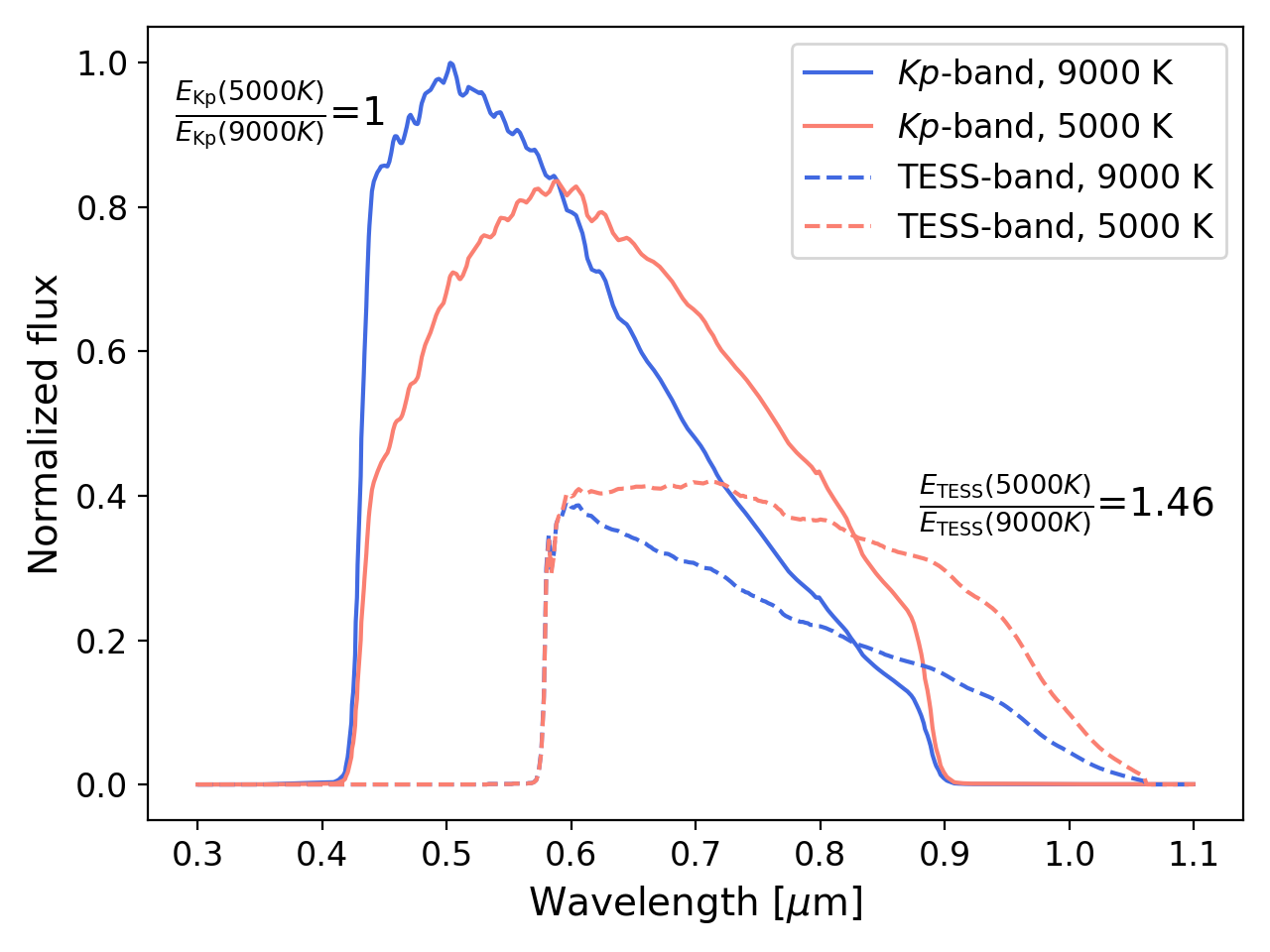}
	}
	\caption{Left: We compare the JWST measured flare rate of TRAPPIST-1 in the TESS band (0.6--1~$\mu$m; blue points) against two previous flare frequency distributions of the star reproduced from \citet{Seli:2021}. The K2 flare rate of TRAPPIST-1 from \citet{Vida:2017} is shown in gray and scaled to the TESS bandpass by \citet{Seli:2021}. The flare rate of TRAPPIST-1 analog stars observed by the TESS mission in its full-frame images is shown in orange and is higher than that predicted from K2. The flare rate measured by JWST agrees with the flare rate measured from the TRAPPIST-1 analogs, suggesting flares are likely to contaminate more transits than suggested by the K2 data. NIRSpec flares are noted with a dagger. Right: Cooler flares lead to an under-estimated TESS-band energy when scaling from the \textit{Kepler} band using a 9000 K blackbody. Both 9000 K and 5000 K flare spectra emit the same energy in the \textit{Kepler} band, but the TESS band energies are higher for the cooler flare by 46\%.}
	\label{fig:tess_ffd_lc}
\end{figure*}

\subsection{Results for the NIR continuum of the flares}\label{continuum_results}
Continuum is clearly observed in the flare peak from 0.6 to 3.5~$\mu$m in the F1 and F2 NIRSpec flares and from 0.6 to 2.8~$\mu$m in the F3  NIRISS flare as shown in Figure \ref{fig:bb_flares}. The F4 flare is weaker, with continuum present out to at least $\sim$2.5~$\mu$m and possibly further. The NIRSpec data reduction does not produce fluxes beyond 3.5~$\mu$m. The mean and 1$\sigma$ errors of the flare-only flux density do not drop to zero even at the longest wavelengths of each spectrum in Fig.\ \ref{fig:bb_flares}. The mean continuum flux density of the F1 and F2 peaks for 3-3.5~$\mu$m at the distance of Earth is 3.9$\pm$0.14$\times$10$^{-18}$ and 1.6$\pm$0.14$\times$10$^{-18}$ erg s$^{-1}$ cm$^{-2}$ $\AA^{-1}$, respectively. The mean continuum flux density of the F3 and F4 peaks for 2.5-2.8~$\mu$m at the distance of Earth is 2.5$\pm$0.23 $\times$10$^{-18}$ and 7.3$\pm$1.4$\times$10$^{-19}$ erg s$^{-1}$ cm$^{-2}$ $\AA^{-1}$, respectively.

\subsubsection{Properties of the spectral continuum}\label{continuum_properties}
The Planck function fits the spectra of the flares reasonably well as shown in Figure \ref{fig:bb_flares}. We confirm that the continuum is indeed the dominant source of the flare spectrum at most wavelengths outside the hydrogen and helium lines. As a further check, we divide the blackbody model spectrum by the flare spectrum to see if they are consistent with one. The ratios of the F1, F2, F3, and F4 flares are 1.04$\pm$0.27, 1.05$\pm$0.53, 1.05$\pm$0.42 and 0.82$\pm$0.81, respectively. The best-fit blackbody temperature for each flare peak is considerably lower than the canonical 9000 K blackbody temperature. Cooler flare temperatures were previously observed in broadband photometry of two flares from TRAPPIST-1 by \citet{Maas:2022}. Although these temperatures are lower than 9000 K, they are broadly consistent with the temperatures of small flares in the flare energy to temperature scaling relationship from \citet{Howard:2020}. Furthermore, the spectral range of our flares is redder than the wavelengths at which flare temperatures are usually measured. Two blackbody components are regularly observed in optical flares, with a hot $>$10,000 K blackbody component that decays quickly and a cooler $\sim$5000 K blackbody that decays slowly. The hot component is thought to trace prompt emission at the flare footprints due to the initial episode of particle acceleration, while the cool component is thought to trace emission in larger flare structures heated by the earlier prompt emission \citep{Kowalski:2016}. The cooler component dominates the flux at longer wavelengths and may impact our derived temperatures.

\subsubsection{Properties of the broadband emission}\label{broadband_properties}
Peak luminosity decreases with flare size, effective temperature, and longer wavelengths as listed in Table \ref{tab:continuum_table}. Within each flare, the peak luminosity drops across the broadband filter measurements from the TESS band to the J, H, and Ks bands as shown in the right panel of Figure \ref{fig:continuum_scaling}. A drop in luminosity was also observed in a sample of flares observed in Stripe 82 of the Sloan Digital Sky Survey (SDSS) and the Two Micron All Sky Survey (2MASS). We reproduce scaling relations for the J, H, and Ks bands from \citet{Davenport:2012} and plot them alongside our observations. An earlier search for flare emission in the 2MASS bands by \citet{Tofflemire:2012} did not produce any detections and is not considered here. We note the 2MASS flares in \citet{Davenport:2012} were observed for only a single point in time and that their $\textit{i}$ band is similar to the TESS band. We linearly extrapolate an M8 relation from the M4-M6 dwarf relations in \citet{Davenport:2012} after verifying the flare amplitudes vary nearly linearly with stellar mass from 0.1--0.2M$_\odot$. Our flares show a shallower drop in the NIR bands relative to TESS than do the \citet{Davenport:2012} flares. The discrepancy could be due to changes in the effective temperature of the flares with stellar mass and the difference between the peak luminosity measured for our flares and the decay phase luminosity observed for many of the 2MASS flares.

The shape of the broadband light curves changes from the TESS band through the Ks band. As shown in Figure \ref{fig:continuum_overview}, the peaks of the flares are impulsive in the TESS band and less pronounced in the 2MASS bands. The H and Ks light curves have a particularly noticeable period of elevated emission in the decay phase. The gradual decay observed at the longest wavelengths is consistent with the longer decay phase of the secondary emission of the cooler blackbody component in the two-temperature flares observed in optical spectra \citep{Kowalski:2016}. The rise time of the TESS and 2MASS band flares varies between events. The bands peak at the same time in the F1 flare, but longer wavelength bands peak at later times for the F2-F4 flares. 

\subsubsection{The flare rate of TRAPPIST-1}\label{flare_rate}
Four flares observed in the TESS bandpass observed over a total of 26.8 hr place strong constraints on the true flare rate of the star. We measure the cumulative flare frequency distribution (FFD) for the TESS band energies to determine whether the flare rate derived from K2 observations of the star or TESS flares observed from M8 TRAPPIST-1 analog stars is more accurate. We also desire to place the NIR properties of our flares into the context of how often a flare of that size occurs from the star. The cumulative FFD is measured from the number of flares of energy $E_\mathrm{TESS}$ or greater observed across the total monitoring time. Uncertainties on the cumulative occurrence rate for flares of each energy are measured with a binomial 1$\sigma$ confidence interval. We find flares equivalent in size to our F4 event occur at a rate of 3.6$\substack{+2.1 \\ -1.3}$ flare d$^{-1}$, and flares equivalent in size to our F3 event occur at a rate of 2.7$\substack{+2.0 \\ -1.1}$ flare d$^{-1}$. Flares the size of our largest event, F1, occur at a rate of 0.9$\substack{+1.4 \\ -0.5}$ flare d$^{-1}$. The F4, F3, and F1 flares therefore occur in our data at 5.7, 11.7, and 21$\times$ the rates predicted by scaling from the K2 flares, respectively. As shown in the left panel of Fig. \ref{fig:tess_ffd_lc}, our FFD excludes the flare rate estimated by scaling the K2 flares into the TESS bandpass, but agrees with the FFD computed from the TRAPPIST-1 analog stars of \citet{Seli:2021}. These results suggest that the underlying spectral energy distribution of the flares in the \textit{Kepler} and TESS wavelength regimes may differ and complicate scaling relations between wavelengths. Assuming a 9000 K temperature for a 5000 K flare when scaling from the \textit{Kepler} to the TESS bandpass under-estimates the TESS band energy by 46\% (right panel in Figure \ref{fig:tess_ffd_lc}). We also find the true rate of flares that may contaminate transit spectroscopy observations is at the higher end of the literature range.

\section{Radiative hydrodynamic modeling of the flares}\label{x_rel_results}
\begin{figure*}
	\centering
        \subfigure
	{
		\includegraphics[trim=10 0 0 10, width=0.33\textwidth]{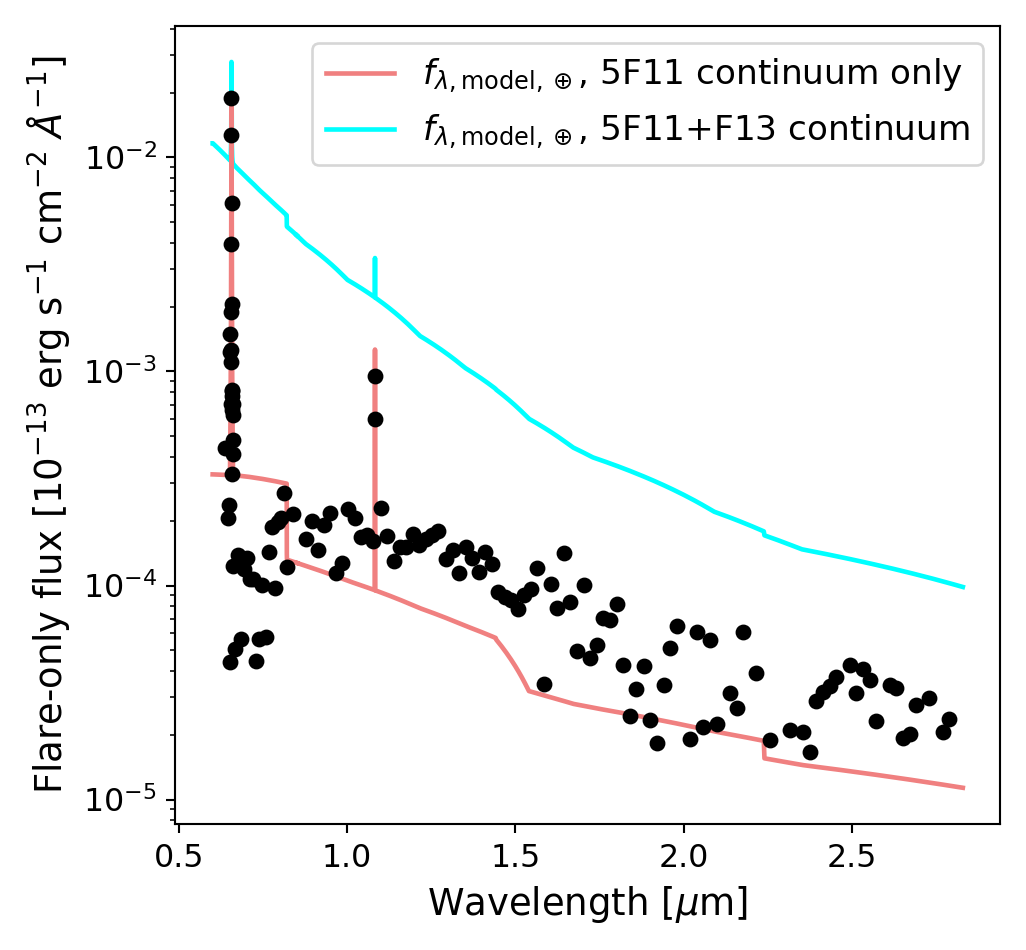}
	}
        \subfigure
	{
		\includegraphics[width=0.62\textwidth]{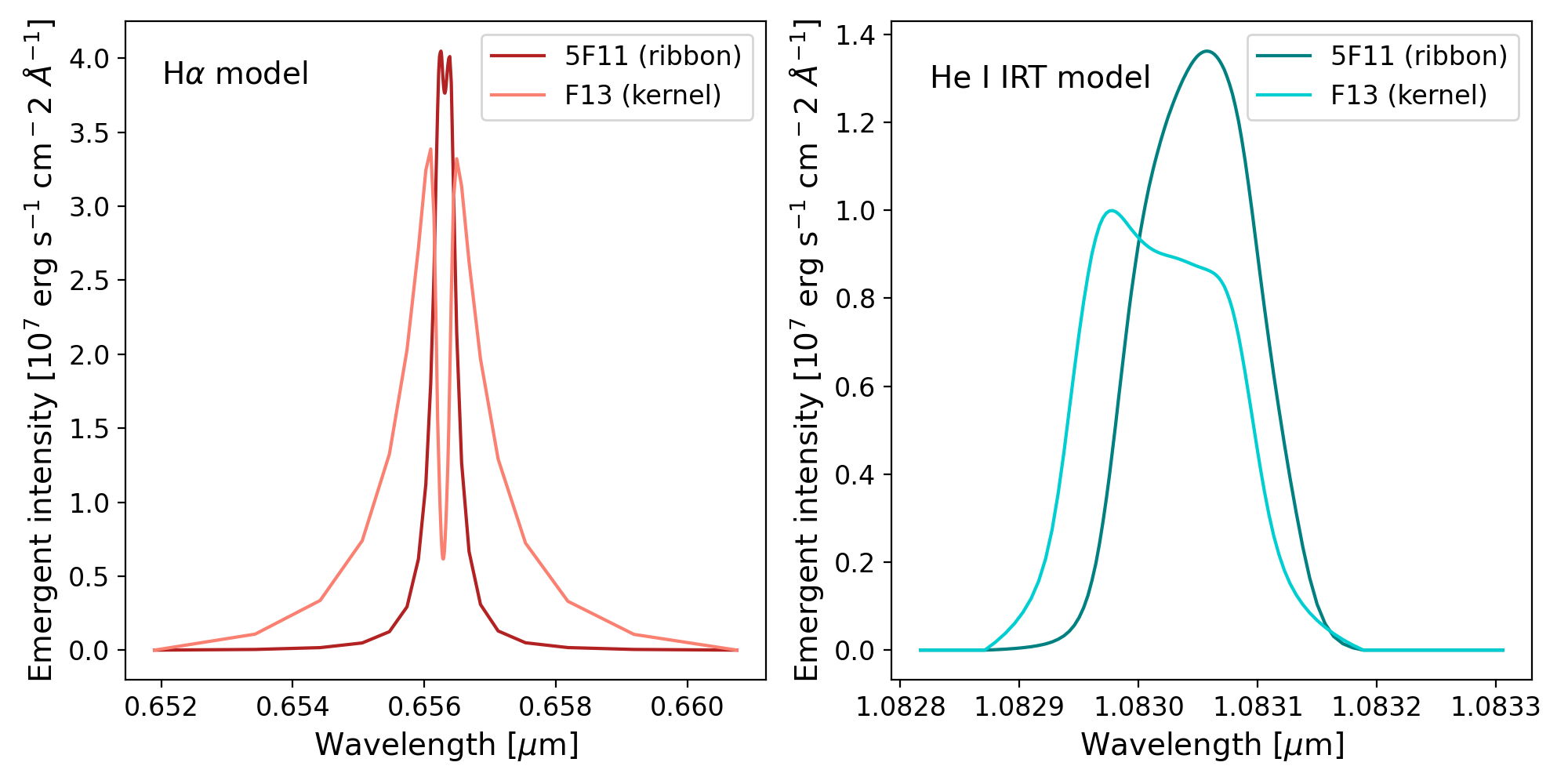}
	}
	\caption{Left: The best-fit model for the observed strengths of the H$\alpha$ and He I IRT lines in our F3 flare, with and without continuum from the F13 kernel. Continuum from the 5F11 ribbon is included in both cases. Middle and Right: The F13 kernel flux is pressure broadened to a greater degree than the 5F11 ribbon flux, especially for H$\alpha$. As a result, the 5F11 flux dominates the fit to the line centers and the F13 flux dominates the wings.}
	\label{fig:radyn_model}
\end{figure*}

Non-local thermodynamic equilibrium (non-LTE) radiative hydrodynamic models are able to match NUV/optical spectra of stellar flares (e.g. \citealt{Kowalski:2016, Kowalski:2017b, Kowalski:2022b, Brasseur:2023}), but have not been compared with NIR observations. The \texttt{RADYN} code \citep{Carlsson_Stein:1995, Carlsson_Stein:1997} solves the coupled non-LTE radiative transport and hydrodynamics equations for the stellar atmosphere during nonthermal electron beam heating in flares \citep{Allred:2006, Allred:2015}. A grid of \texttt{RADYN} models exploring a broad range of injected electron energies has recently become available \citep{Kowalski:2022b}, enabling us to test if the line and continuum emission of our flares is consistent with non-LTE models. The \texttt{RADYN} models each inject a non-thermal beam of electrons into the plasma according to a power-law distribution of energies given by the index of accelerated electrons $\delta$, varying the maximum flux and low-energy cutoff. The models assume apex coronal conditions of a starting atmosphere with an effective temperature of $\sim$3600 K, a magnetic loop half-length of 10$^9$ cm, log $g$ of 4.75, ambient electron density of 3$\times$10$^{10}$ cm$^{-3}$, a gas temperature of 5 MK, and a uniform cross-sectional area. The response of the atmosphere is computed across a 1 s increase and 10 s decrease of the beam intensity following the pulsed injection profile described in \citet{Aschwanden:2004}. The spectra are then averaged across the 10 s period to produce a fiducial spectrum of the flare. The hydrogen lines were calculated using the pressure broadening profiles of \citet{Tremblay_Bergeron:2009} as described in \citet{Kowalski:2022a}. Further details are given in \citet{Kowalski:2017b} and Kowalski et al. (2023, in prep). Model outputs in the spectral range of our observations are given as emergent radiative fluxes at the stellar surface in erg s$^{-1}$ cm$^{-2}$ $\AA^{-1}$ and include line profiles for the H$\alpha$ and He I IRT lines and continuum emission through 2.83~$\mu$m.

In addition to stellar flares, solar flares have also been observed at NIR wavelengths \citep{Simoes:2017, Penn:2016}. Both a high-energy and low-energy electron beam are needed to match observations of solar and stellar flare spectra \citep{Kowalski:2022b, Brasseur:2023}. The high-energy beam penetrates deep into the chromosphere where it thermally ionizes the plasma and induces pressure broadening of the hydrogen lines. Observations of solar flares indicate high-energy beams are coincident with the compact flare kernel responsible for the majority of the continuum emission \citep{Kawate:2016}. The low-energy beam heats plasma in the upper layers of the chromosphere and produces strong, narrow hydrogen lines and a faint continuum. On the Sun, this emission is thought to occur in the elongated flare ribbons. We adopt the mF13-85-3 (henceforth F13) and m5F11-25-4 (henceforth 5F11) \texttt{RADYN} models for our high-energy and low-energy beams, respectively. The mF13-85-3 model injects an electron beam with a maximum flux of 10$^{13}$ erg s$^{-1}$ cm$^{-2}$, a low-energy cutoff of 85 keV, and $\delta$=3. The m5F11-25-4 model injects an electron beam with a maximum flux of 5$\times$10$^{11}$ erg s$^{-1}$ cm$^{-2}$, a low-energy cutoff of 25 keV, and $\delta$=4. These two models are representative of the allowed range of electron beam properties in \texttt{RADYN} \citep{Brasseur:2023}.

We subtract the continuum from the H$\alpha$ and He I IRT model spectra to obtain the flux due to the lines \citep{Kowalski:2022b}. A linear superposition of the F13 and 5F11 line fluxes is defined with the filling factors as free parameters as in \citet{Kowalski:2022b}, shown in Eq. \ref{eq_2}:
\begin{equation}
\label{eq_2}
    f_{\lambda, \mathrm{model}, \oplus} = (X_\mathrm{F13}S_{\lambda, \mathrm{F13}} + X_\mathrm{5F11}S_{\lambda, \mathrm{5F11}}) \frac{R_{*}^2}{d_{*}^2} 
\end{equation}
Here, $f_{\lambda, \mathrm{model}, \oplus}$ is the flux density of the combined model at the distance of Earth, $X_\mathrm{F13}$ and $X_\mathrm{5F11}$ are the filling factors, and $S_{\lambda,\mathrm{F13}}$ and $S_{\lambda,\mathrm{5F11}}$ are the emergent flux densities of the models. The relative filling factor $X_\mathrm{rel}$=$X_\mathrm{5F11}$/$X_\mathrm{F13}$ describes how much larger the effective area of the flare ribbon is compared to the kernel. We bin the combined model $f_{\lambda, \mathrm{model}, \oplus}$ to the wavelength resolution of NIRISS and fit the spectra of the H$\alpha$ and He I IRT lines shown in the middle and right panels of Figure \ref{fig:radyn_model} to those of our F3 flare. The result is shown in the left panel of Figure \ref{fig:radyn_model}. The fit produces reasonable filling factors of $X_\mathrm{5F11}$=0.0038 and $X_\mathrm{F13}$=0.0018 for an $X_\mathrm{rel}$ of 2.1. These filling factors are typical of large flares and consistent with the filling factors derived from the simple blackbody fits in \S \ref{fitting_planck_fn}. Our $X_\mathrm{rel}$ of 2.1 is comparable to the $X_\mathrm{rel}$=2.28 measured from the H$\gamma$ line for the Great Flare of AD Leo \citep{Kowalski:2022b}. The models do not reproduce the continuum well, suggesting X-ray/extreme-UV backwarming \citep{Hawley_Fisher:1992} or other processes may be more important than in optical spectra. The continuum of the 5F11 ribbon component is within a factor of $\sim$3 of the observed spectrum, but the F13 continuum from the kernel is 5-30$\times$ higher. The 5F11 model also dominates the fit to the line centers, indicating ribbons are a key source of flare emission in the NIR. Fitting the F4 flare spectrum produces $X_\mathrm{rel}\approx$3, but the S/N of the fit is low. Since our empirical blackbody fits trace the continuum more closely than the model, we employ our blackbody fits to decontaminate the transit spectra.

\section{Correcting flare contamination in transit spectra using our models}\label{planet_results}
\begin{figure*}
	\centering
	{
		\includegraphics[width=0.98\textwidth]{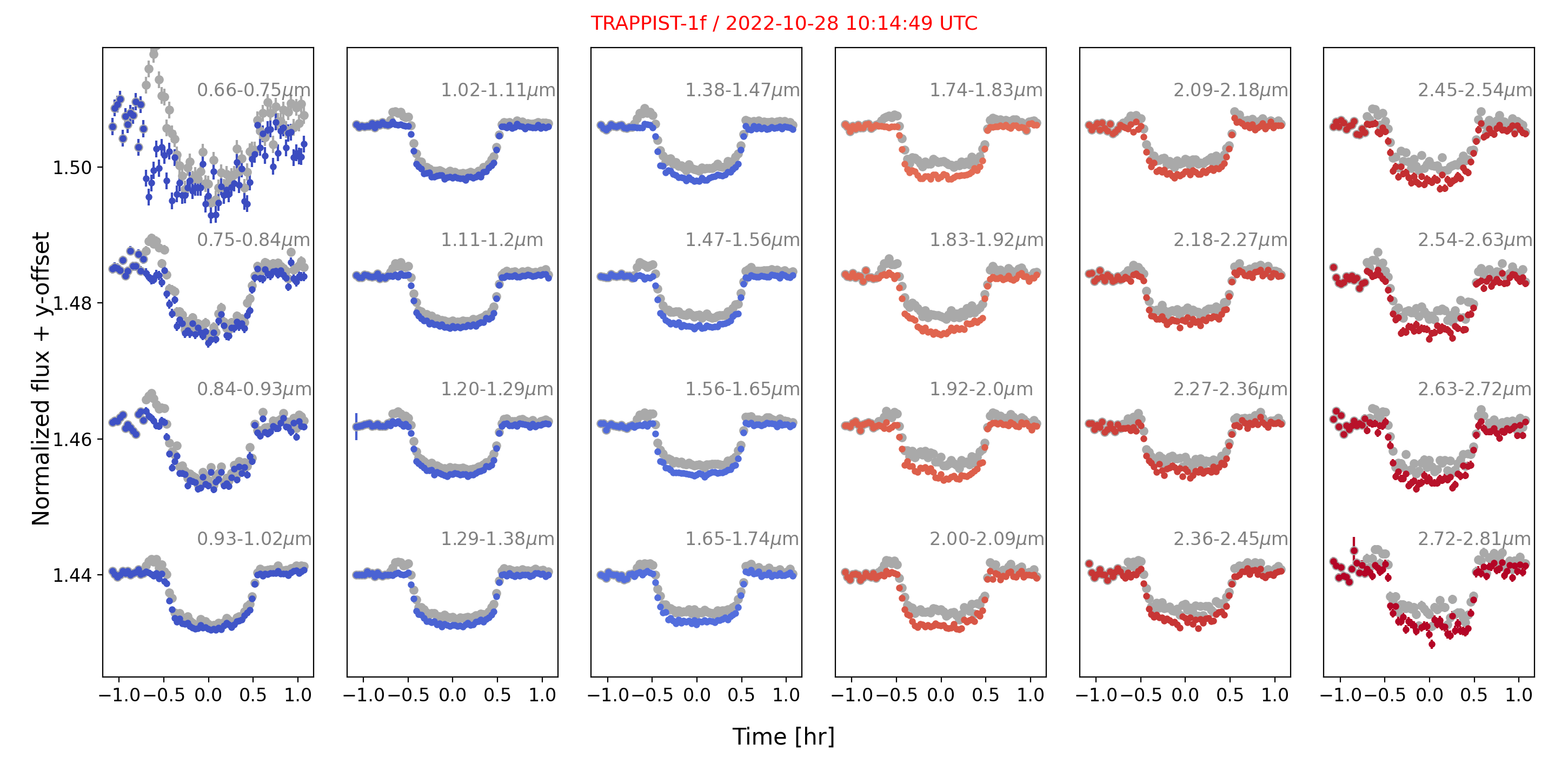}
	}
	\caption{Correction of flare contamination across a range of wavelengths for the transit spectra of TRAPPIST-1f. The F3 flare peaks just prior to ingress. The original NIRISS order 1 and 2 spectra of the uncorrected transit are shown for reference in gray, while the corrected spectra are color-coded by wavelength. While the F3 flare is clearly seen in the uncorrected spectra, it is much weaker in the corrected spectra after the best-fit blackbody model from Fig. \ref{fig:bb_flares} is subtracted and the activity lines are masked.}
	\label{fig:bb_correction_T1f}
\end{figure*}

\begin{figure*}
	\centering
	{
		\includegraphics[width=0.98\textwidth]{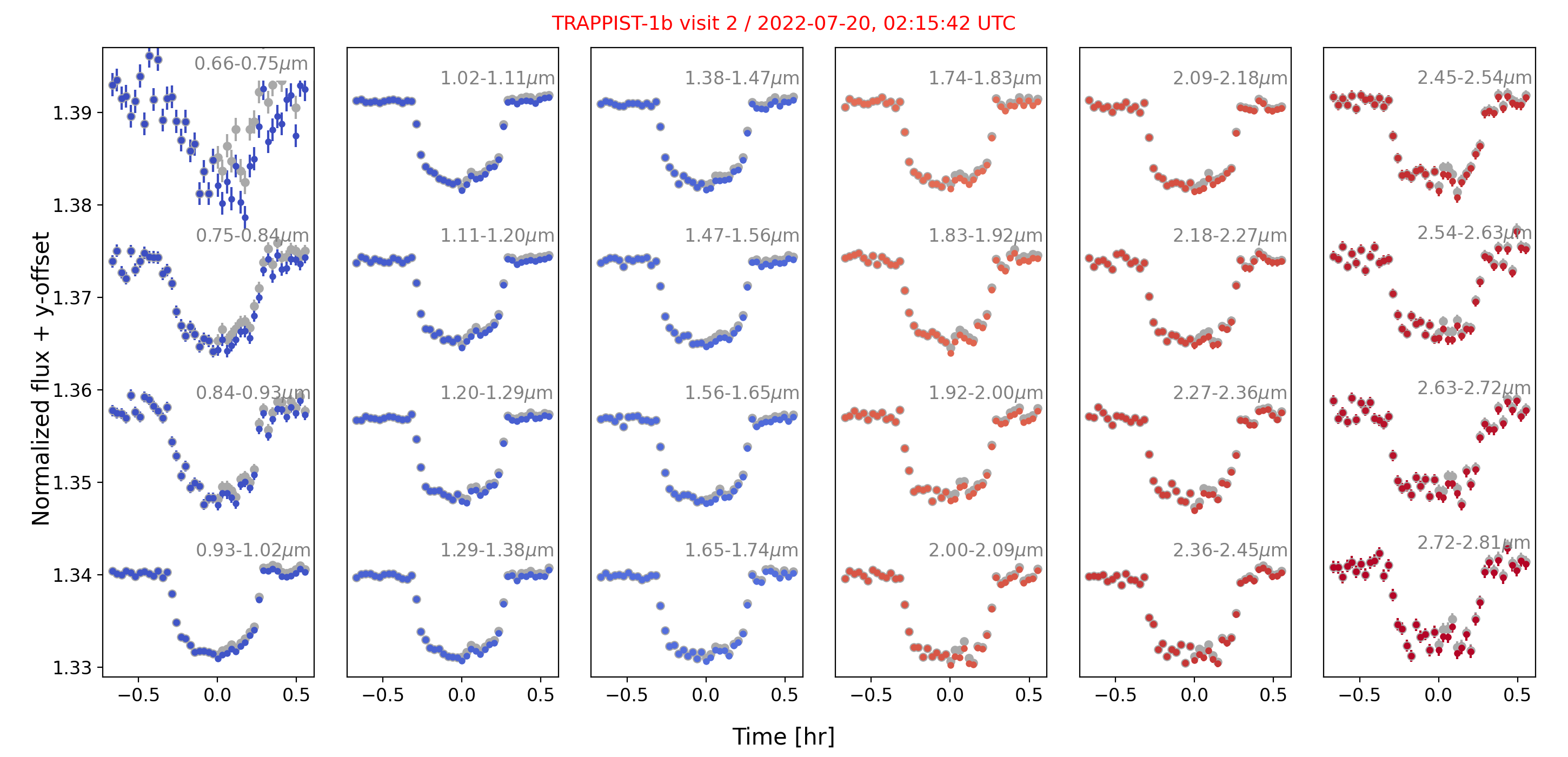}
	}
	\caption{Correction of flare contamination across a range of wavelengths for the transit spectra during the second visit on TRAPPIST-1b. The F4 flare peaks during the flat bottom and continues through egress. The original NIRISS order 1 and 2 spectra of the uncorrected transit are shown for reference in gray, while the corrected spectra are color-coded by wavelength. The F4 flare is much smaller than the F3 event, leading to a less pronounced difference between the uncorrected and corrected spectra during the flare.}
	\label{fig:bb_correction_T1b}
\end{figure*}

We correct for flare contamination in the transit spectra of TRAPPIST-1f and b during the F3 and F4 events, respectively. The best-fit blackbody model with parameters $T_\mathrm{eff}$ and $X_\mathrm{eff}$ from the flare-only spectrum is subtracted from the original spectral time series. Wavelength points with candidate lines are also masked out. We do not perform blackbody fits for integrations outside of the flare start and stop times as the lack of continuum outside these times can lead to an unconstrained fit. We also do not correct for contamination in the TRAPPIST-1g transit from the F1 and F2 flares as the transit occurs far into the decay phase of the flares where the continuum signals are weakest. The large F3 flare during the transit of TRAPPIST-1f shows the clearest improvement across all wavelengths in the spectrum. As shown in Figure \ref{fig:bb_correction_T1f}, the flare peaks just prior to ingress in the uncorrected data, but is almost completely absent in the corrected spectrum. The smaller F4 flare during the flat bottom and egress of the TRAPPIST-1b transit shows some improvement in Figure \ref{fig:bb_correction_T1b}, but the continuum fits do not remove the flare as cleanly.

The contamination correction does not have a minimum wavelength bin size, but is illustrated for representative wavelengths across the full spectral range of NIRISS. Corrections to the first-order spectra are clearer than the second order due to the higher precision in the first order. For both transits, the blackbody fits perform best at wavelengths from approximately 1.0--2.4~$\mu$m and worse at wavelengths close to 0.6 and 2.83~$\mu$m. In the F3 flare, the blackbody that best fits the spectrum of the flare from 0.6--2~$\mu$m slightly overshoots the flux for $\lambda>$2.0~$\mu$m. The extreme Rayleigh-Jeans tail does not constrain the flare temperature \citep{Fuhrmeister:2008}, so we use the global fit for the determination of $T_\mathrm{eff}$ and $X_\mathrm{eff}$. However, we fit a second blackbody to the spectrum from 1.7~$\mu$m--2.83~$\mu$m in order to decontaminate the spectral time series in the Rayleigh-Jeans tail beyond 1.7~$\mu$m for F3 and 2.0~$\mu$m for F4 where the global fit slightly over-estimates the flux.

\begin{figure*}
	\centering
        \subfigure
	{
		\includegraphics[width=0.48\textwidth]{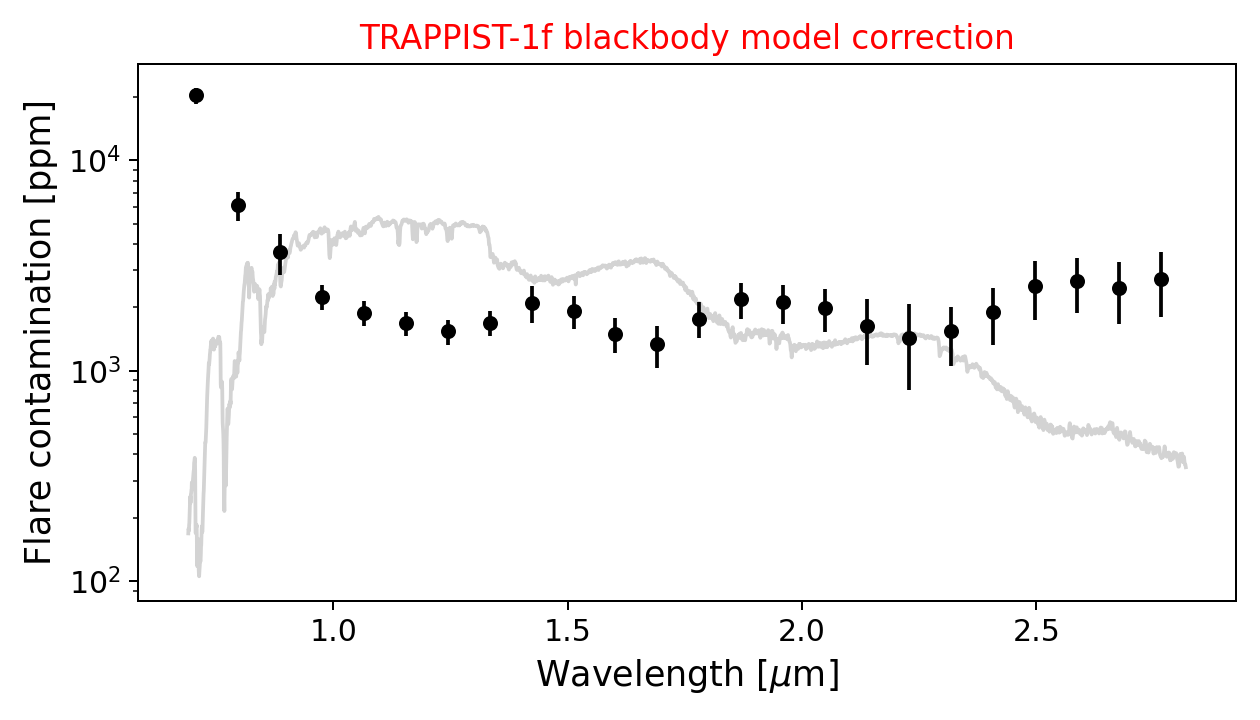}
	}
        \subfigure
	{
		\includegraphics[width=0.48\textwidth]{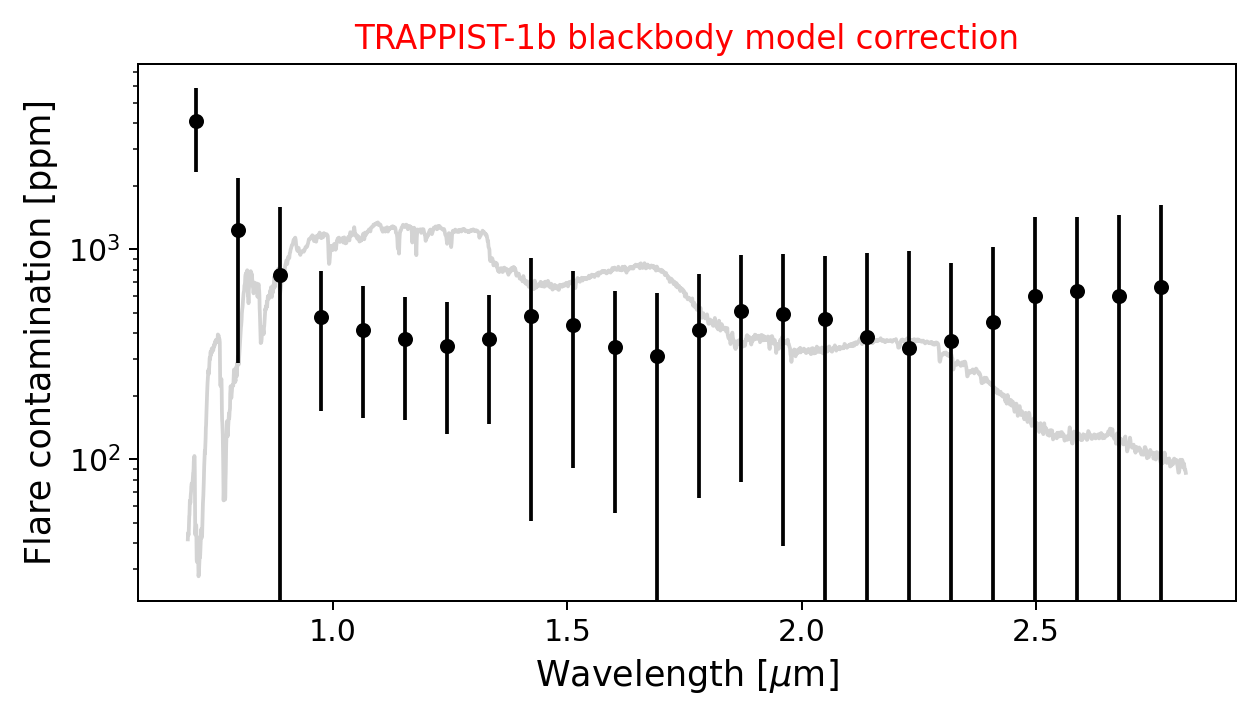}
	}
        \subfigure
	{
		\includegraphics[width=0.48\textwidth]{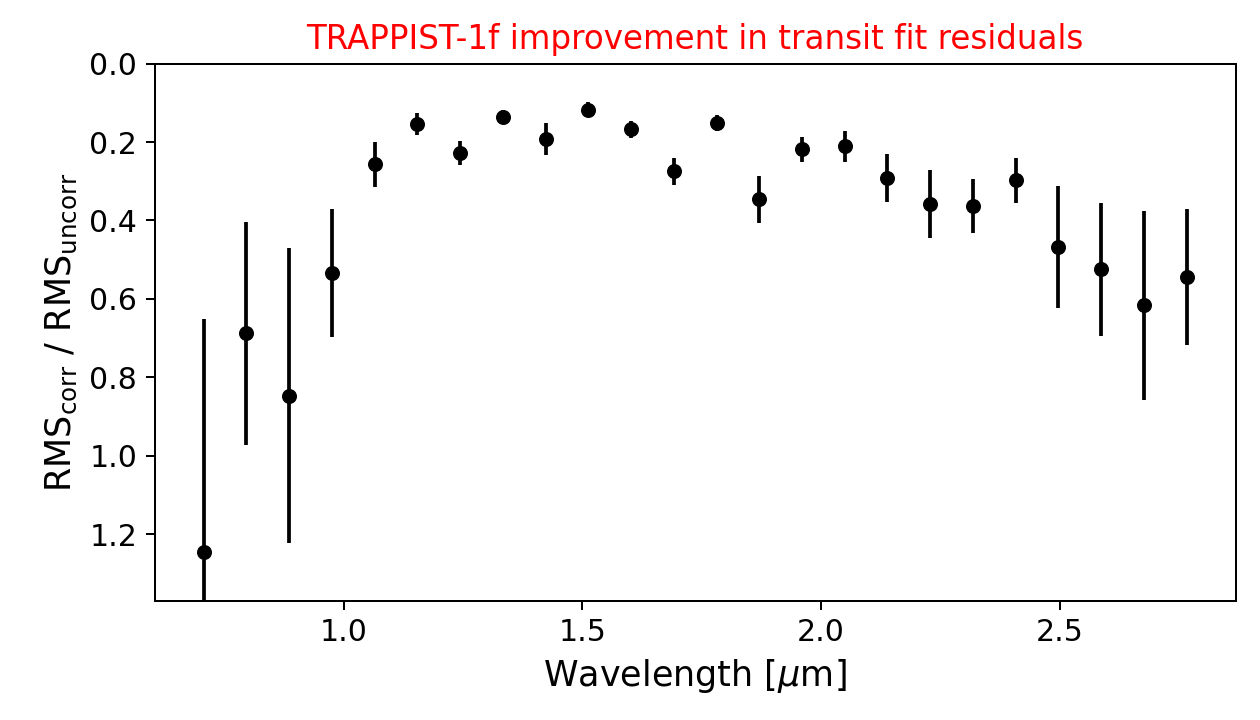}
	}
        \subfigure
	{
		\includegraphics[width=0.48\textwidth]{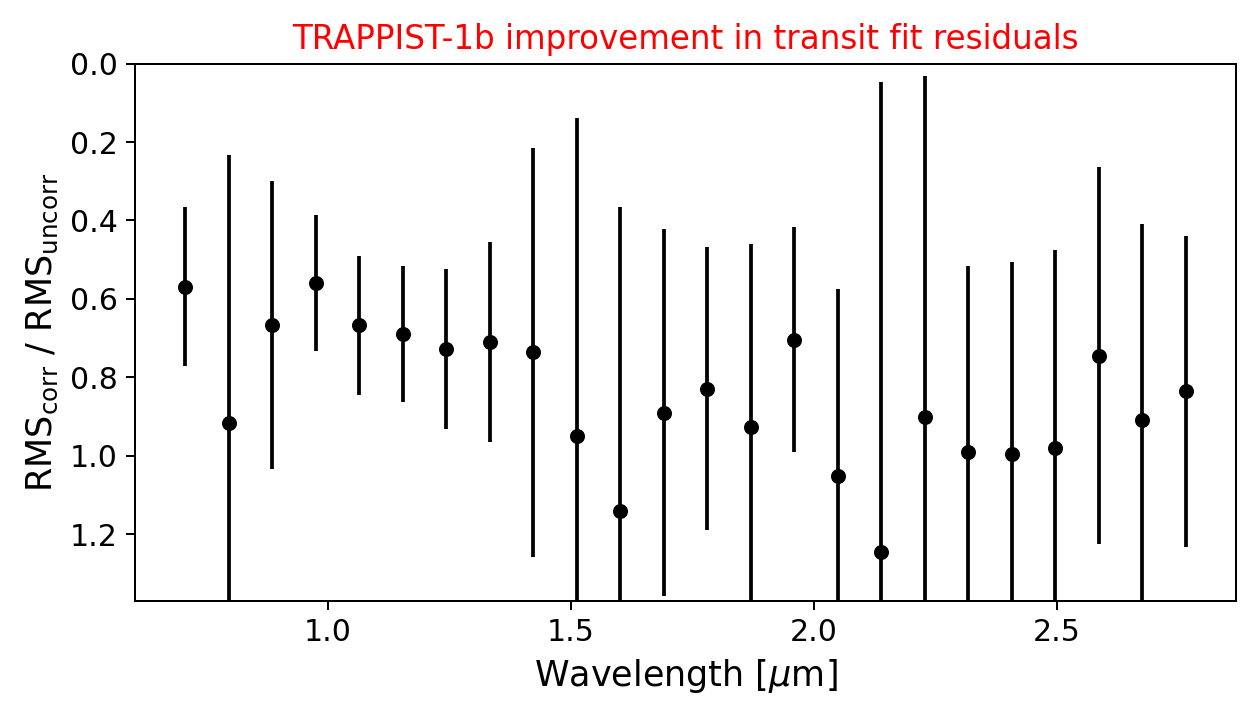}
	}
	\caption{Fractional change in the transit depth as a function of wavelength after subtracting the flare contamination model for TRAPPIST-1f (top left) and b (top right). The spectrum of the star is shown in gray for reference. The larger F3 flare induces contamination at the 2100 ppm level at 2~$\mu$m, while the smaller F4 flare causes contamination of $\sim$500 ppm at 2~$\mu$m. TRAPPIST-1's flare rate predicts that flares of these sizes occur multiple times per day. The improvement in the RMS of the transit spectrum before and after applying the transit correction is shown for TRAPPIST-1f (bottom left) and b (bottom right).}
	\label{fig:contamination_level}
\end{figure*}

We explore the efficacy of the contamination correction by computing the fractional change to the transit depth between the original and corrected transit spectrum. The comparison is performed for the integration during which the flare is brightest in white light. The results are shown for the TRAPPIST-1f and b transits in Figure \ref{fig:contamination_level}. The wavelength dependence of the correction varies inversely with the contrast of the flare against the stellar spectrum. The peak contamination in the TRAPPIST-1f transit light curves due to the larger F3 event ranges from 20000$\pm$1700 ppm at 0.7~$\mu$m, 2100$\pm$400 ppm at 2~$\mu$m and 2700$\pm$900 ppm at 2.8~$\mu$m, assuming the best-fit blackbody. The contamination in the TRAPPIST-1b transit light curve from the smaller F4 flare ranges from 4100$\pm$1800 ppm at 0.7~$\mu$m, 500$\pm$450, and 650$\pm$960 ppm at 2.8~$\mu$m (i.e. consistent with zero).

We also quantify the improvement in the transit signal as a function of wavelength. Residuals are computed for the difference between the transit model and the original and flare-corrected light curves. The root mean square (RMS) of the transit light curves is computed for the corrected and uncorrected transit light curves and the ratio RMS$_\mathrm{corr}$/RMS$_\mathrm{uncorr}$ is plotted for the TRAPPIST-1f and b transits in the bottom panels of Figure \ref{fig:contamination_level}. The transit improvement RMS$_\mathrm{corr}$/RMS$_\mathrm{uncorr}$ of the F3 flare is better than 0.5 for wavelengths between 1.0--2.4~$\mu$m, with no measurable improvement near 0.6~$\mu$m. The best improvement in F3 is observed at 1.12--2.1~$\mu$m, where $\sim$80\% of the flare contamination is removed. The improvement of F4 is not well constrained. The wavelength bins with the smallest errorbars (e.g. 0.9--1.2~$\mu$m) are each consistent with values from 10--60\%.

\section{Discussion and Conclusions}\label{discuss_conclude}
We have characterized the combination of line and continuum emission of stellar flares at NIR wavelengths of 1--3.5~$\mu$m for the first time. Several spectral features of major carbon and oxygen bearing species are produced in this range, such as the $\sim$100~ppm CO$_2$ features at 2.0 and 2.8~$\mu$m. However, the most significant CO$_2$ feature expected in the atmospheres of the TRAPPIST-1 planets exists at 4.3~$\mu$m \citep{Lustig-Yaeger:2019}. We estimate our larger F3 flare would have produced a 300--1600~ppm signal at 4.3$\mu$m through extrapolation of the best-fit blackbody spectrum. Such a signal would be larger than the 160 ppm CO$_2$ feature at 4.3~$\mu$m.

The strongest lines in the largely unexplored wavelength range of 1--3.5~$\mu$m are the Paschen $\alpha$ to $\delta$ lines, Brackett $\beta$ and the He I IRT. We report the first detection of P$\alpha$ and Br$\beta$ in a stellar flare from the NIRISS spectra of our F3 event. All lines from 0.7--2.8~$\mu$m are at least an order of magnitude weaker than H$\alpha$, with the exception of P$\beta$ and P$\gamma$ which are 0.7 dex weaker. We discover that the intensity of the Paschen lines does not decrease sequentially for higher transitions. Instead, intensity increases from P$\alpha$ to P$\gamma$ and then decreases out to P$\epsilon$. Future work is needed to model the conditions that induce increased flux for the less energetically favored states.

The NIRISS and NIRSpec observations of four flares from TRAPPIST-1 reveal that flare continuum exists out to at least 3.5~$\mu$m, and is well-described at these wavelengths by a blackbody with an effective temperature of $\sim$5000 K. Note the $\sim$3000 K peak temperatures of the NIRISS flares are due to the longer integration times. These temperatures are much lower than the 9000 K blackbody typically assumed when estimating the amount of flare emission at any given wavelength \citep{Osten_Wolk:2015}, but comparable to those seen in several red optical flare spectra \citep{Kowalski:2016}. We confirm the continuum is the dominant source of flare contamination. Indeed, the continuum gives rise to signals of 10$^3$ ppm at wavelengths of 1--3~$\mu$m for even small flares of $E_\mathrm{TESS}$=10$^{30}$ erg. Multiple flares of this size occur each day, requiring care in the reduction and interpretation of the transit spectra. Smaller flares of $E_\mathrm{TESS}$=10$^{29}$ erg occur an order of magnitude more frequently, but are expected to lead to lower levels of contamination of 30--150 ppm at 4.3~$\mu$m. For context, an $E_\mathrm{TESS}$=10$^{30}$ erg flare produces a mean contamination level of $\sim$1900 ppm from 1--3~$\mu$m, while an $E_\mathrm{TESS}$=10$^{29}$ erg flare produces a mean contamination level of $\sim$200 ppm from 1--3~$\mu$m.

In this work, we focus on contamination of the transit spectra by the flares. However, photochemistry due to the high-energy radiation of the flares may dominate the molecular abundances detectable in the planetary atmospheres with JWST \citep{Tsai:2023, Batalha:2023}. Several flares have been observed simultaneously in the TESS bandpass and at X-ray/UV wavelengths (e.g. \citealt{MacGregor:2021, Paudel:2021, Jackman:2022}). \citet{Jackman:2023} use TESS and GALEX flare rates of field age late M-dwarfs to show that far UV (FUV) and near UV flare energies exceed those predicted from the TESS energies using pseudo-continuum templates of $\sim$9000 K by factors of 30.6$\pm$10.0 and 6.5$\pm$0.7, respectively. Underestimation of the FUV energies may be even higher for a 5000 K blackbody than the assumed 9000 K blackbody. Similarly, the He I IRT emission in flares observed by JWST can be used to estimate the soft X-ray and EUV emission by scaling from the multi-wavelength properties of solar flares of comparable size to those in this work \citep{Ding:2005}. A C3.9 solar flare observed simultaneously in the GOES 1-8$\AA$ soft X-ray band and in the $\lambda$1.083~$\mu$m He I IRT reached a luminosity ratio of $L_\mathrm{SXR}$/$L_\mathrm{He~I~IRT} \approx$440 \citep{Zeng:2014}. We obtain this ratio by multiplying the emergent flux in the He I IRT line by the area of the flare extraction regions and propagating the flux to 1 au to compare with the GOES X-ray flux. If a similar ratio holds for our flares, the F3 and F4 flares would have emitted 1.3$\times$10$^{28}$ and 5.7$\times$10$^{27}$ erg s$^{-1}$ in the 1--8~$\AA$ GOES band, respectively. The F3 and F4 events would therefore be large X45-X20 flares, which are associated with extreme coronal mass ejections from the Sun \citep{Youngblood:2017}. The flat spectrum of TRAPPIST-1b and c in recent MIRI observations by \citet{Greene:2023} and \citet{Zieba:2023} are consistent with no-atmosphere scenarios that could result from high fluxes of stellar energetic particles \citep{Tilley:2019}. These results are also consistent with the spectrum of TRAPPIST-1b observed with NIRISS SOSS in \citet{Lim:2023}, which showed no evidence of an atmosphere on the planet.

ow effective temperatures may not necessarily imply favorable conditions for pre-biotic chemistry or survivable levels of UV flux for surface life. Low effective temperatures have been previously reported for the decay phase of TRAPPIST-1 flares by \citet{Maas:2022}, who consider the impacts of these low temperatures for the habitability of planets around ultracool dwarfs (UCDs). While hot flares emit a greater portion of their energy at NUV wavelengths, cool flares do not emit nearly as much high-energy radiation. Since photochemistry in both exoplanet atmospheres \citep{Ranjan:2020, Chen:2021} and biochemistry \citep{Ranjan:2017, Abrevaya:2020} is sensitive to light at wavelengths of 200--300 nm, lower flare temperatures could mean UV surface conditions are more favorable for life than often assumed \citep{Kowalski:2022b}.

However, pre-biotic chemistry on planets around M-dwarfs requires UV radiation from flares due to the lack of quiescent UV emission, assuming the RNA world hypothesis \citep{Ranjan:2017, Rimmer:2018}. Cooler flares from UCDs would produce less energy for origin of life scenarios for systems like TRAPPIST-1. Furthermore, neither the \citet{Maas:2022} study nor our own sufficiently probes the wavelengths for the hot blackbody component in the two-temperature flare scenario \citep{Kowalski:2016}. The hotter blackbody of 10,000--12,000 K is dominant below 0.5~$\mu$m, while the shortest broadband filter used in the \citet{Maas:2022} study is $g$ (400--550 nm). Likewise, our spectra only extend down to 0.6~$\mu$m. \citet{Maas:2022} observe the flare temperatures decrease with color across $g$-$r$, $g$-$i$, and $g$-$z_s$, which could be due to the increasing prominence of the cooler component at longer wavelengths.

Our work illustrates the utility of JWST transit spectroscopy for stellar astrophysics in addition to exoplanets. The spectral time series of M-dwarfs obtained as a by-product of exoplanet programs provide a rich dataset for the characterization of stellar activity. Multiple programs targeting TRAPPIST-1 and other high priority exoplanet systems have and will continue to obtain data on the host stars, substantially increasing the temporal baseline of the stellar variability of M-dwarfs at NIR wavelengths. At the same time, we find that understanding the line and continuum emission of the flares enables the amount of contamination to be assessed and even mitigated in the transit spectra. Along these lines, Flagg et al. (2023, in prep.) are applying flare mitigation based on this work for specific features in the transmission spectra such as for major carbon and oxygen bearing species. We recommend masking the lower-order Paschen and Brackett series lines and fitting blackbody models to the continuum to decontaminate transit spectra during flares. As in \citet{Lim:2023}, we also recommend caution when interpreting features in the transit spectrum if excess H$\alpha$ emission is observed.

As it becomes available, archival JWST transit spectroscopy of early and mid M-dwarf flare stars such as LTT1445A and L 98-59 will enable us to explore whether stellar mass and age impact the NIR properties of the flares. The flare rates of the host stars of $\sim$3 flare d$^{-1}$ are sufficiently high to contaminate transits \citep{Howard:2022, Brown:2022}. Our results suggest mitigation of the flare contamination may be possible and increase the scientific impact of the transit observations. A survey of the flaring TESS Objects of Interest found 25.7$\substack{+12 \\ -7}$\% of potentially-rocky TESS planets suitable for JWST transit spectroscopy orbit flare stars. Beyond JWST, flares have been observed during transits of planets from the far-UV \citep{Feinstein:2022b} to the mid-IR \citep{Gillon:2017}. Based on flares observed in the FUV during two transits of AU Mic b, \citet{Feinstein:2023} estimate flare contamination from 1--2~$\mu$m will occur at the 200 and 70 ppm level for AU Mic b and c, respectively. Likewise, \textit{K2} observations of a flare during the transit of TRAPPIST-1h \citep{Luger:2017} and \textit{Spitzer} observations of five flares at 4.5$\mu$m overlapping several transits of TRAPPIST-1 planets \citep{Gillon:2017, Ducrot:2020} induce contamination as described in \citet{Davenport:2017}. Retrievals of atmospheric species at these and other wavelengths may also benefit from improved flare modeling.

\section{Acknowledgements}\label{acknowledge}
We would like to thank the anonymous referee who graciously gave their time and effort in writing a thoughtful and helpful report.

WH thanks Zach Berta-Thompson for helpful discussions on reducing NIRISS transit data, Eileen Gonzales for a good discussion on UCDs, Brad Barlow for a good discussion on flare light curves, Steven Cranmer for helpful discussions on solar flares, David Wilson for discussing SEDs, and Laura Crosskey for a helpful framing on completing such a long paper.

WH acknowledges partial funding support through the TESS Cycle 4 GO 4132. This material is also based upon work supported by the National Aeronautics and Space Administration under award No. 19-ICAR19\_2-0041. Support for this work was also provided by NASA through the NASA Hubble Fellowship grant HST-HF2-51531 awarded by the Space Telescope Science Institute, which is operated by the Association of Universities for Research in Astronomy, Inc., for NASA, under contract NAS5-26555.

DJ is supported by NRC Canada and by an NSERC Discovery Grant.

J.D.T was supported for this work by NASA through the NASA Hubble Fellowship grant HST-HF2-51495.001-A awarded by the Space Telescope Science Institute, which is operated by the Association of Universities for Research in Astronomy, Incorporated, under NASA contract NAS5-26555.

This work is based on observations made with the NASA/ESA/CSA James Webb Space Telescope. The data were obtained from the Mikulski Archive for Space Telescopes at the Space Telescope Science Institute, which is operated by the Association of Universities for Research in Astronomy, Inc., under NASA contract NAS 5-03127 for JWST. The specific observations analyzed can be accessed via \dataset{10.17909/b0ee-8j61}. These observations are associated with programs GTO 1201 and GO 2589. The authors acknowledge the Lim team for developing their observing program with a zero-exclusive-access period.

\texttt{astropy} \citep{astropy:2013,astropy:2018}, \texttt{jwst} v1.6.0, v1.6.2, 1.8.2 \citep{bushouse_2022_7041998, bushouse_2022_7325378}, \texttt{supreme-SPOON} \citep{Feinstein:2023, Radica:2022}, \texttt{transitspectroscopy} \citep{Espinoza:2022}, \texttt{RADYN} \citep{Allred:2015,Kowalski:2017b}

\bibliography{References.bib}

\end{document}